\documentclass[final,5p]{elsarticle}
\usepackage{enumitem}
\usepackage{caption}
\usepackage{array}
\usepackage{xspace}
\usepackage{mathtools}
\usepackage{stfloats}
\usepackage{multirow}
\usepackage{aasmacros}
\usepackage[switch]{lineno}
\usepackage[english]{babel}
\usepackage{xcolor}
\usepackage[colorlinks=True, linkcolor=red]{hyperref}
\usepackage{placeins}
\usepackage{subcaption}

\newcommand{\Deg}{$^{\circ}$}
\newcommand{\lnA}{$\langle$$\mathrm{lnA}$$\rangle$\xspace}
\newcommand{\Bave}{$\langle$B$\rangle$\xspace}
\newcommand{\Brms}{B$_{\mathrm{RMS}}$\xspace}

\renewcommand{\paragraph}[1]{\noindent \textbf{#1}}

\newcommand{\thickhline}{\noalign {\global\arrayrulewidth=1.5pt} \hline \noalign {\global\arrayrulewidth=0.4pt}}

\newcommand{\QGSJet}{QGS\-Jet\-II-04}
\newcommand{\Sibyll}{Si\-byll\-2.3c}
\newcommand{\orcidlink}[2]{%
    \href{https://orcid.org/#2}{#1}%
}

\begin{document}
\setlist[enumerate,1]{left=10pt, label=(\alph*)}
\journal{The Astrophysical Journal (ApJ)}

\begin{frontmatter}

\title{Combined Fit of Spectrum and Composition for FR0 Radio-galaxy-emitted Ultra-high-energy Cosmic Rays with Resulting Secondary Photons and Neutrinos}

\author[1]{\orcidlink{Jon Paul Lundquist}{0000-0002-4245-5092}\corref{cor1}}
\ead{jlundquist@ung.si}
\author[1]{\orcidlink{Serguei Vorobiov}{0000-0001-8679-3424}}
\author[2,3]{\orcidlink{Lukas Merten}{0000-0003-1332-9895}}
\author[4]{\orcidlink{Anita Reimer}{0000-0001-8604-7077}}
\author[4]{\orcidlink{Margot Boughelilba}{0000-0003-1046-1647}}
\author[4]{\orcidlink{Paolo Da Vela}{0000-0003-0604-4517}}
\author[5]{\orcidlink{Fabrizio Tavecchio}{0000-0003-0256-0995}}
\author[5]{\orcidlink{Giacomo Bonnoli}{0000-0003-2464-9077}}
\author[5]{\orcidlink{Chiara Righi}{0000-0002-1218-9555}}

\cortext[cor1]{Corresponding author}

\address[1]{Center for Astrophysics and Cosmology (CAC), University of Nova Gorica, Nova Gorica, Slovenia}
\address[2]{Institute for Theoretical Physics IV, Faculty for Physics and Astronomy, Ruhr University Bochum, Universitätsstraße 150, 44801 Bochum, Germany}
\address[3]{Ruhr Astroparticle and Plasma Physics Center (RAPP Center), Bochum, Germany}
\address[4]{Universität Innsbruck, Institut für Astro- und Teilchenphysik, Technikerstraße 25, 6020 Innsbruck, Austria}
\address[5]{Astronomical Observatory of Brera, Via Brera 28, 20121 Milano, Italy}

\begin{abstract}
This study comprehensively investigates the gamma-ray dim population of Fanaroff-Riley Type 0 (FR0) radio galaxies as potentially significant sources of ultra-high-energy cosmic rays (UHECRs, E~$>$~10$^{18}$~eV) detected on Earth. While individual FR0 luminosities are relatively low compared to the more powerful Fanaroff-Riley Type 1 and Type 2 galaxies, FR0s are substantially more prevalent in the local universe, outnumbering the more energetic galaxies by a factor of $\sim$5 within a redshift of $z$~$\leq$~0.05. 

Employing CRPropa3 simulations, we estimate the mass-composition and energy spectra of UHECRs originating from FR0 galaxies for energies above 10$^{18.6}$~eV. This estimation fits data from the Pierre Auger Observatory (Auger) using three extensive air shower models; both constant and energy-dependent observed elemental fractions are considered. The simulation integrates an approximately isotropic distribution of FR0 galaxies, extrapolated from observed characteristics, with UHECR propagation in the intergalactic medium, incorporating various plausible configurations of extragalactic magnetic fields, both random and structured. We then compare the resulting emission spectral indices, rigidity cutoffs, and elemental fractions with recent Auger results. In total, 25 combined energy-spectrum and mass-composition fits are considered.

Beyond the cosmic-ray fluxes emitted by FR0 galaxies, this study predicts the secondary photon and neutrino fluxes from UHECR interactions with intergalactic cosmic photon backgrounds. The multimessenger approach, encompassing observational data and theoretical models, helps elucidate the contribution of low-luminosity FR0 radio galaxies to the total cosmic-ray energy density.
\end{abstract}


\end{frontmatter}
\newpage
\section{Introduction} 
\label{sec:intro}
Ultra-high-energy cosmic rays (UHECRs) are the most energetic particles observed in the Universe, with energies exceeding $10^{18}$ eV. Their origins are an open astrophysical question, with potential sources including active galactic nuclei (AGN), various radio galaxies, gamma-ray bursts, and starburst galaxies. The Pierre Auger Observatory (Auger), located in Argentina, was established as the largest ($\sim$3000 km$^2$) UHECR hybrid detector in the world to study these particles~\citep{PierreAuger:2015eyc}. Auger provides the most extensive statistics on the energy spectra, composition, and arrival directions of UHECR, offering essential information to explore their sources and propagation mechanisms.
  
Low-luminosity Fanaroff-Riley (FR0-class radio-loud jetted AGN) radio galaxies~\citep{Baldi2009} have emerged as a promising candidate class of sources of UHECRs detected on Earth. Recent surveys have highlighted that FR0s are approximately 5 times more prevalent in the local universe compared to their more energetic counterparts, Fanaroff-Riley Type 1 (FR1) and Type 2 (FR2) sources~\citep{Baldi:2017gao}, qualifying them as potentially substantial contributors to the UHECR energy density. Despite their similar core radio properties to FR1s~\citep{Croston2018}, FR0s display much weaker extended radio emission; however, their estimated jet power, which significantly exceeds the required cosmic-ray power per source~\citep{Heckman2014}, suggests they could substantially contribute to the UHECR flux measured on Earth. Studies utilizing the average FR0 spectral energy distribution suggest that these galaxies can accelerate UHECRs to the highest observed energies through hybrid acceleration, combining Fermi-I pre-acceleration with gradual shear acceleration~\citep{Merten:2021brk}. Therefore, they fulfill both the Hillas criterion~\citep{Hillas1984} and the energetics criteria~\citep{NaganoWatson,Bhattacharjee2000}. 

Multimessenger astrophysics, which involves the combined study of various cosmic messengers such as photons, neutrinos, and cosmic rays, offers an even more comprehensive understanding of astrophysical phenomena. This approach helps identify and characterize sources of UHECRs and refine our knowledge of the mechanisms that generate and propagate these particles. Regarding the multimessenger properties of FR0s, lep\-to-hadronic jet-disc modeling of core-emission from quiet low-luminosity AGN jets predicts that FR0s are expected to be relatively weak gam\-ma-ray and neutrino emitters~\citep{2023ApJ...955L..41B,Reimer:2024Lw}. This theoretical evaluation of weak multimessenger emission aligns with the results of searches for gam\-ma-ray-emitting FR0 galaxies. For example, stacking analyses were necessary to detect a very small unresolved FR0 population in \textit{Fermi}-LAT data and identify them as gam\-ma-ray emitters~\citep{2021ApJ...918L..39P, Khatiya:2023lkg}. A recent study~\citep{Partenheimer_2024} using various \textit{Fermi}-LAT catalogs as sources of cosmic rays in CRPropa simulations concluded that previously resolved (therefore overwhelmingly non-FR0) gamma-ray sources cannot account for the large-scale anisotropy of UHECR events above 8 EeV as established by the Pierre Auger Collaboration~\citep{Golup_ICRC2023}. This discrepancy led the authors of~\cite{Partenheimer_2024} to conclude that another population of UHECR sources, either lacking gamma-ray emission or unresolved by current-generation telescopes, must exist.

The effect on expected arrival distribution of diffuse UHECRs in the intergalactic magnetic field has been examined in several studies, e.g.,\ in~\cite{Eichmann2018, Wittkowski2018, Dundovic2019} that provide constraints on the source emission spectra and composition (e.g.,~\cite{Taylor2015, Das2019}). The Auger large angular scale arrival direction analysis points to the extragalactic origins of UHECR for energies E~$\geq$~8~EeV~\citep{Golup_ICRC2023}. A recent three-component combined-fit of the UHECR energy spectrum, extensive air shower (EAS) maxima distributions (X$_{\text{max}}$), and arrival directions supports an astrophysical model primarily comprised of homogeneous background sources and an adaptable addition from nearby source candidates~\citep{PierreAuger:2023htc}. For instance, if these additional nearby sources are a selection of 44 starburst galaxies, they could contribute up to a $\sim$20\% flux fraction at 40~EeV~\citep{PierreAuger:2023htc}---a sizable fraction of the remaining bulk of UHECRs may be from FR0s.

However, a considerable uncertainty in the strength and configuration of intergalactic magnetic fields confoun\-ds the extrapolation of UHECR measurements to their potential sources. Assuming a primordial field, theoretical modeling of the Universe's evolution results in structured field models with an average magnitude on the order of $\sim$10$^{-11}$ gauss~\citep{Dolag_2005, Hackstein:2017pex}. In contrast, fitting UHECR propagation simulations to data is a good fit with an average field magnitude of $\sim$1~nG~\citep{Lundquist:20233x, Mollerach:2013dza}. 

In this study, we leverage the strengths of multimessen\-ger observations to investigate the properties of FR0 radio galaxies as potential UHECR sources. We simulate isotropic FR0 UHECR emission, extrapolating from measured FR0 properties and propagating them to Earth through plausible intergalactic magnetic fields using CRPropa3~\citep{AlvesBatista:2022vem}. By a combined fit of these simulations to energy and composition using Auger data~\citep{Yushkov:2020nhr, Deligny:2020gzq}, we estimate the UHECR mass-composition and energy spectra emitted by FR0s. These combined energy-spectrum and composition fits are done using two nuclei emission models---a constant nuclei composition emission over the energy range and an evolving emission that depends on energy. Additionally, we present estimates of the cosmogenic secondary photon and neutrino fluxes resulting from UHECR interactions with cosmic photon backgrounds.

\section{Pierre Auger Observatory Data}
\label{sec:Data}
Auger performs high-precision and large-statistics analyses of energy spectra, mass-composition, and arrival directions of UHECR data which are essential for assessing theoretical models of cosmic-ray origins and propagation. This study utilizes the energy-spectrum and composition datasets to systematically extrapolate the observed characteristics of UHECRs to predictions of FR0 galaxy emission. The Auger data used in fitting simulations of FR0 radio galaxy UHECR emission is the publicly available energy spectrum from~\cite{Deligny:2020gzq} and composition from~\cite{Yushkov:2020nhr} (using the method of~\cite{2013JCAP...02..026P}) for energies E~$\geq$~10$^{18.6}$~eV. These results using 13 years of data are discussed further in Section~\ref{sec:results}. 

The composition is given in terms of mean log mass number \lnA, which is evaluated from fluorescence-detector-determined depth of shower maximum X$_{\text{max}}$ using three different extensive air-shower (EAS) models---\Sibyll~\citep{Riehn:2017mfm}, EPOS-LHC~\citep{Pierog:2013ria}, and \QGSJet{}~\citep{Ostapchenko:2010vb}. These are referred to in the following figures, tables, and in combination with magnetic field models as SIBYLL, EPOS, and QGS4 for brevity. UHECR data is an indirect experimental measurement of primary particle interaction, and their composition must be interpreted statistically through these EAS models that extrapolate the behavior of particle interactions in the atmosphere at much higher energies than available through direct experiments. The three most common EAS models are used in this study---the main two (EPOS and SIBYLL) are presented in figures for some clarity.



\section{Propagation Framework}
\label{sec:Propagation}

The open-source CRPropa3 (v3.2) framework~\citep{AlvesBatista:2022vem} is used to simulate UHECR primary nuclei propagation through intergalactic media. This software models interactions with cosmic backgrounds and deflections in magnetic fields, allowing us to trace the trajectories, composition, and energy changes of UHECRs from their sources to Earth. The five primary nuclei of hydrogen, helium, nitrogen, silicon, and iron are generated with an emission energy spectrum $\mathrm{d}N/\mathrm{d}E\propto$ E$_0^{-1}$. This $\gamma$ = 1 spectrum is generated for more statistics at the highest energies, and simulated events are reweighted for the combined fit to the data. 

Simulated UHECR interactions with the intergalactic cosmic microwave, UV, optical, IR (Gilmore12 model \citep{10.1111/j.1365-2966.2012.20841.x}), and radio (Protheroe96 model \citep{Protheroe:1996si}) photon backgrounds account for photopion production, nuclear photodisintegration, Bethe-Heitler process, gamma-gamma pair-production (single, double, and triplet), and inverse Compton scattering.  Additionally, universe-expansion adiabatic energy loss and unstable nuclear decays are considered. 

The CRPropa3 observer used is a 200~kpc sphere centered at the Milky Way center and intersecting nuclei are recorded for energies E~$\geq$~10$^{18.6}$~eV. Additional interactions and deflections in the last 200~kpc are expected to be negligible. To conserve computation time, particles that intersected with another sphere of 2 Gpc in radius or traveled more than 4 Gpc were discarded, as such distances are unachievable within the universe's lifetime. Secondary neutrinos with energies above 100~TeV that intersect the observer are recorded. Secondary photons with energies above 100~MeV are immediately recorded if a projected linear path intersects the observer and are then propagated using the \verb|DintElecaPropagation| module~\citep{Settimo:2013tua, Lee:1996fp}.

\subsection{Magnetic Fields}

\begin{figure*}[ht]
    \centering
    \subfloat[Subfigure 1][]{
    \includegraphics[width=.45\linewidth,trim={0cm 0cm 0cm 1cm},clip]{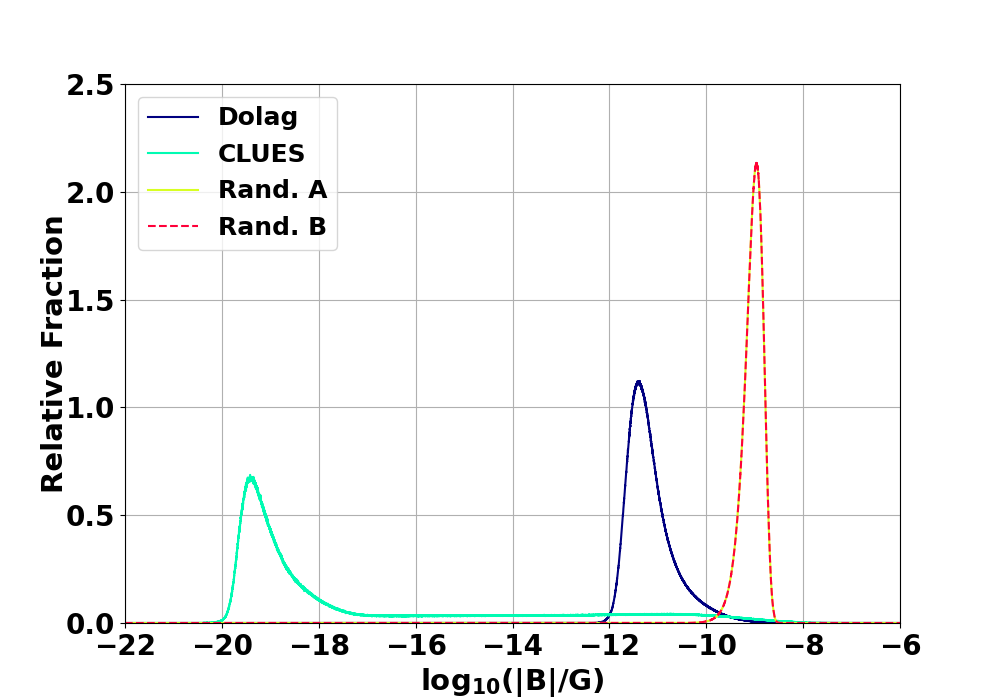}
    \label{fig:BmagHist}}
    \hspace{-0.5cm}
    \subfloat[Subfigure 2][]{
    \includegraphics[width=.45\linewidth,trim={0cm 0cm 0cm 1cm},clip]{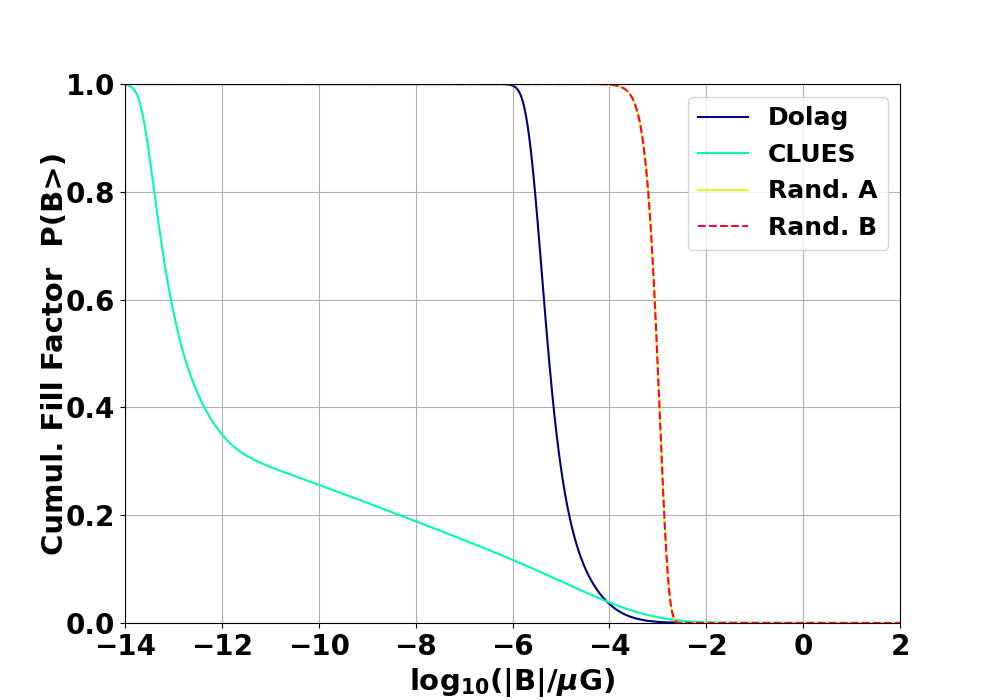}
    \label{fig:CumulativeFilling}}
    \caption{Magnetic field magnitude relative fraction (a) and inverse cumulative distribution (or filling factor) (b) of the simulated intergalactic magnetic field models used. The structured fields Dolag et al. ($\langle$B$\rangle$ = 0.047~nG, B$_{\mathrm{RMS}}$ = 11~nG)~\citep{Dolag_2005} and Hackstein et al.\ (CLUES) 'astrophysical1R' ($\langle$B$\rangle$ = 0.064~nG, B$_{\mathrm{RMS}}$~=~1.2~nG)~\citep{Hackstein:2017pex} and the two random $\langle$B$\rangle$~=~1~nG (B$_{\mathrm{RMS}}$ = 1.1~nG) fields with $l_{\mathrm{corr}}$ = 234~kpc (Rand.A) and $l_{\mathrm{corr}}$ = 647~kpc (Rand.B). The two random fields cannot be distinguished in this figure.}
    \label{fig:MagFields}
\end{figure*}

Charged nuclei deflections were simulated using four magnetic field models and a no-magnetic-field scenario. Two fields are turbulent and homogeneous with a mean strength \Bave = 1~nG (\Brms~= 1.1~nG) and length scales from 60~kpc to 1~Mpc (or 3~Mpc) and a Kolmogorov power spectrum for an average correlation length of $l_{\mathrm{corr}}$~=~234 kpc referred to as Rand.A (or $l_{\mathrm{corr}}$~=~647~kpc, Rand.B). The third model uses the Dolag et al.\ structured field~\citep{Dolag_2005}, while the fourth employs the Hackstein et al.\ (CLUES) 'astrophysical1R' structured field~\citep{Hackstein:2017pex}. Figure~\ref{fig:MagFields} shows the magnetic field strength distributions. This sampled range of magnetic fields are within theoretical expectations and were chosen based on their prevalence in recent UHECR studies and their expected impact on the propagation outcomes.

The Dolag et al.\ field~\citep{Dolag_2005} was created via a magnetohydrodynamic simulation of the primordial evolution of the local universe in a cube 340~Mpc on a side (repeated in simulation as needed) and has structures that are well corresponded with nearby galaxy clusters (Coma, Virgo, Centaurus, Hydra, Perseus, and A3627). The field also includes large-scale structures such as sheets and filaments. Though the average magnetic field is small (\Bave~= ~0.047~nG), the largest fields are on the order of $\sim$10~$\mu$G. Though these high-magnitude fields are a small proportion of the volume, this results in the large \Brms~=~11~nG. The reported expectation is that ultra-high-energy protons will, on average, have a smaller deflection than experimental sensitivities~\citep{Dolag_2005}.

The Hackstein et al.\ (CLUES) structured fields~\citep{Hackstein:2017pex} were also created via magnetohydrodynamic simulations within a smaller cube of 250~Mpc. The 'astrophysical1R' field (a nearly identical distribution to 'astrophysicalR') was chosen from the six CLUES fields. This field includes gas cooling and AGN feedback effects and also has structures corresponding with some nearby galaxy clusters (Coma, Centaurus, and Perseus) connected by strong filaments. Though the average magnetic field is larger than the one from Dolag et al. (\Bave~=~0.064~nG) due to a more substantial background, the largest fields are on the order of 100 times less ($\sim$0.1~$\mu$G), resulting in a smaller \Brms~=~1.2~nG. It is reported that a dipole amplitude is observed with a pure iron emission from isotropic sources~\citep{Hackstein:2017pex}.
 
\section{Simulating FR0s as UHECR Sources}
\label{sec:FR0Sources}

\begin{figure*}[ht]
    \vspace{-1em}
    \centering
    \includegraphics[width=0.9\textwidth,trim={0cm 1cm 0cm 1cm},clip]{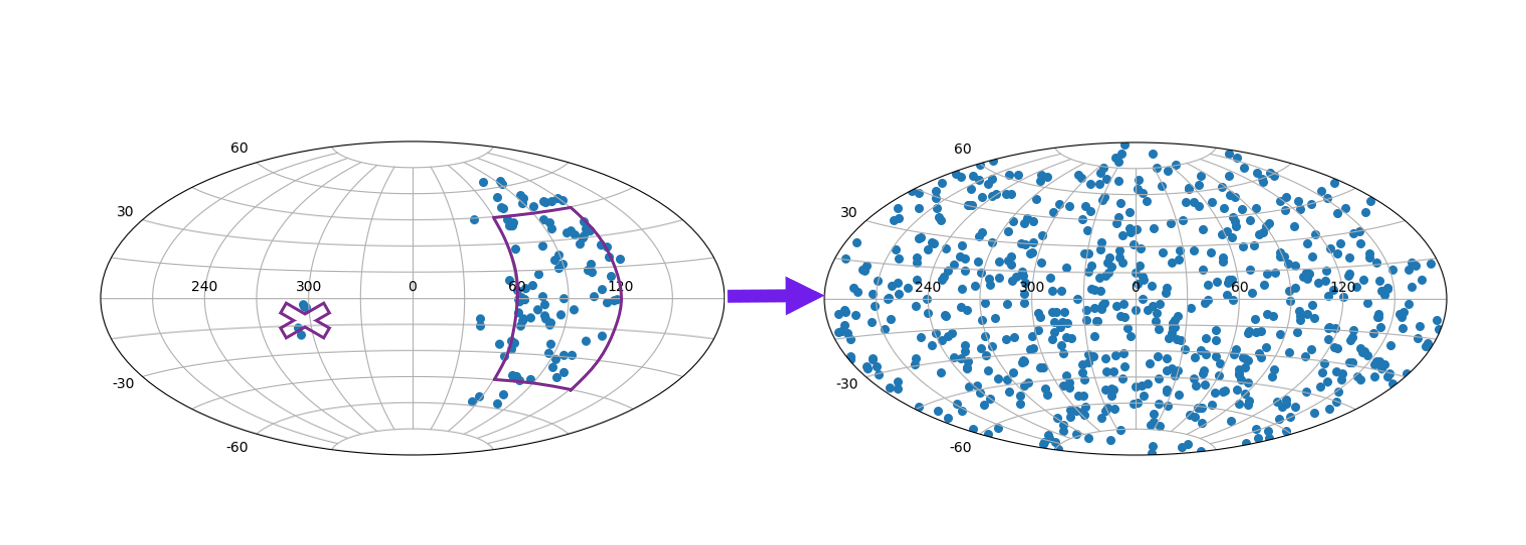}
    \caption{The FR0 catalog FR0CAT~\citep{Baldi:2017gao} in supergalactic coordinates (left) and a simulated FR0 distribution (right). Full-sky isotropic FR0 density is estimated from the well-sampled FR0CAT section. The crossed-out FR0 on the left figure were not used in the fits shown in Figure~\ref{fig:distributions}.}
    \label{fig:sky}
\end{figure*}

The simulated set of FR0s is upsampled from the well-sampled sky portion of the FR0CAT catalog~\citep{Baldi:2017gao} (limited to $z$~$\leq$~0.05) and extrapolated to a redshift $z$~$\leq$~0.2 distance. This area is the spherical rectangle in supergalactic coordinates of \hbox{-45}\Deg~to 45\Deg~SGB (latitude) and 60\Deg~to 120\Deg~SGL (longitude) (for a solid angle $\Omega=\pi/3\sqrt2$), as shown in Figure~\ref{fig:sky} on the left. Inside this 11.79\% sky coverage there are 76 cataloged FR0s within $z$~$\leq$~0.05 that appear isotropically distributed. Upsampling these data results in 645 FR0s, also with isotropically distributed pointing-directions within the same redshift limit (Figure~\ref{fig:sky}, right) and a total of $\sim$18,400 sources in an individual simulation within the extrapolated distance of $z$~$\leq$~0.2.

\begin{figure*}[hb]
    \centering
    \subfloat[Subfigure 1][]{
    \includegraphics[width=.33\linewidth]{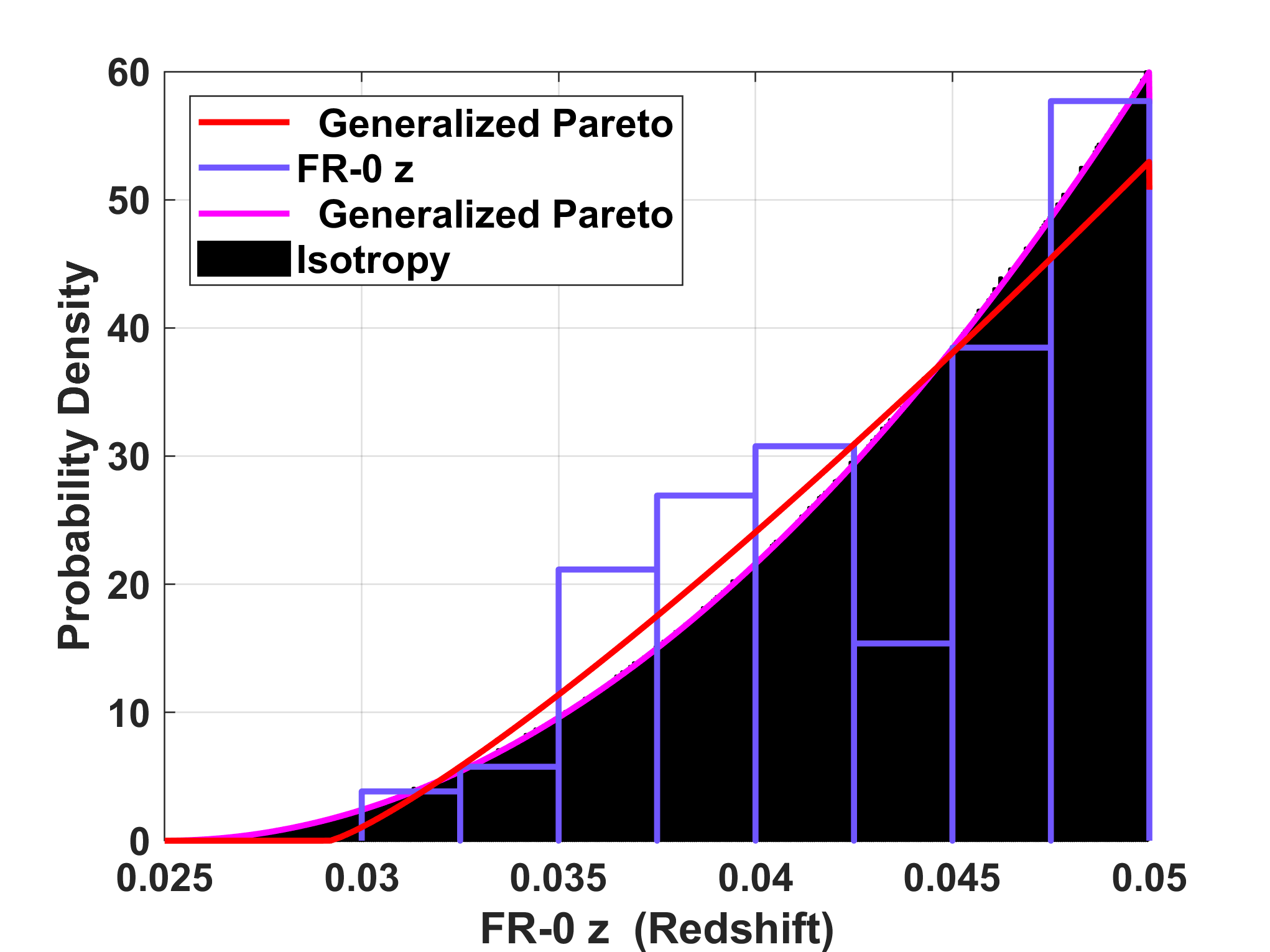}
    \label{fig:zhist}}
    \hspace{-0.5cm}
    \subfloat[Subfigure 2][]{
    \includegraphics[width=.33\linewidth]{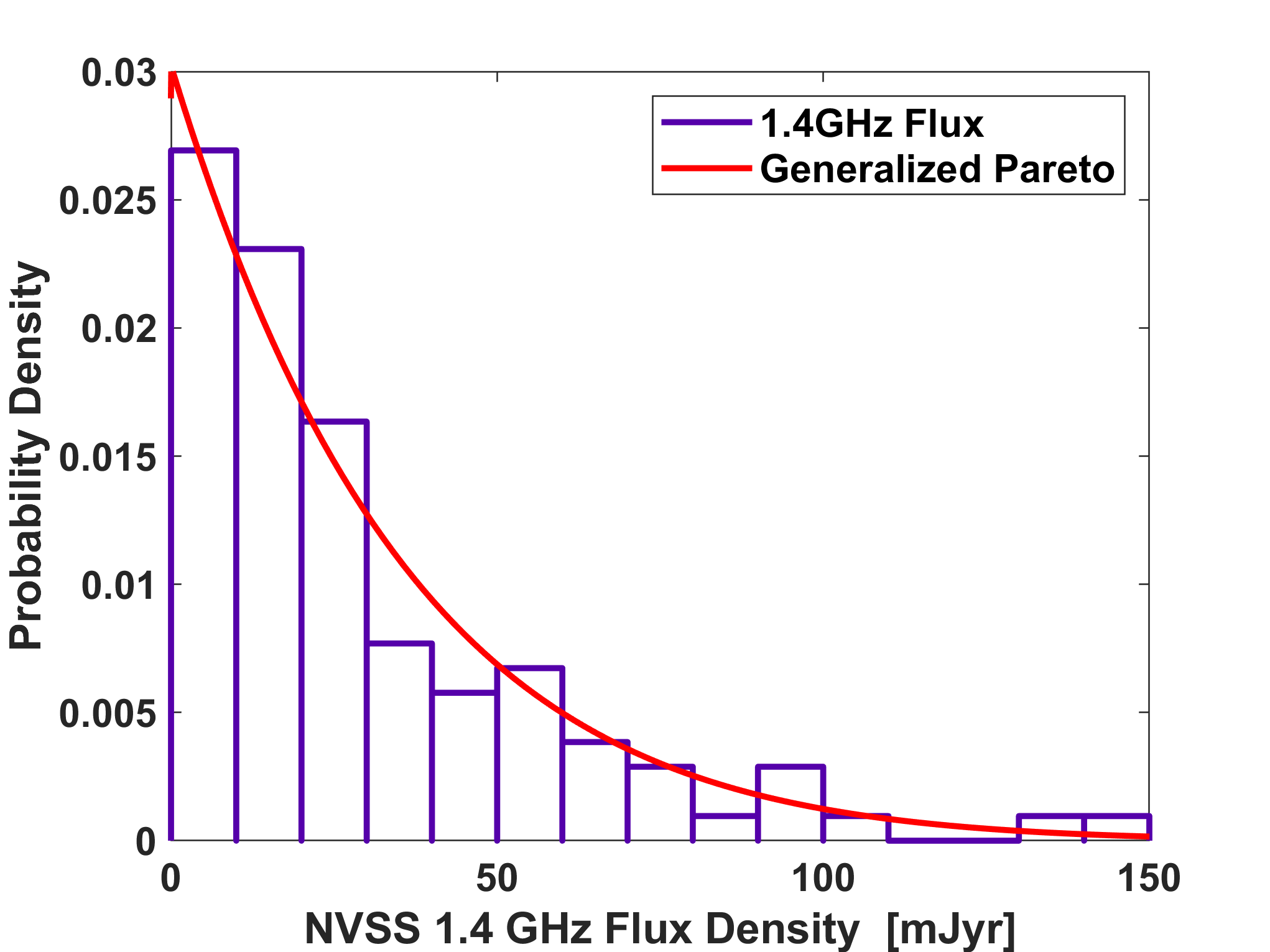}
    \label{fig:fluxhist}}
    \hspace{-0.69cm}
    \subfloat[Subfigure 3][]{
    \includegraphics[width=.33\linewidth]{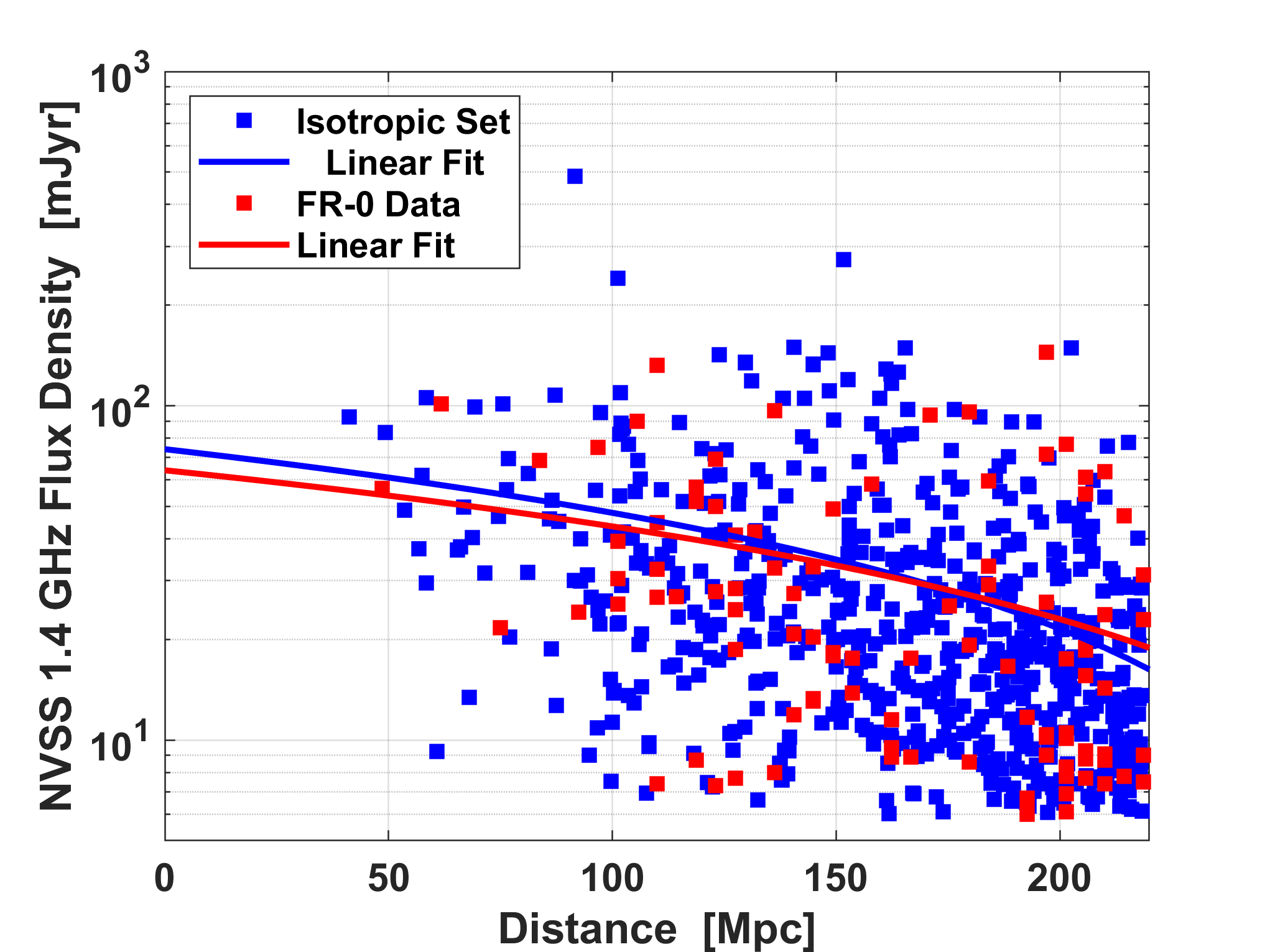}
    \label{fig:correlation}}
    \caption{(a) A Pareto distribution fit to the catalog data \citep{Baldi:2017gao} generates simulated FR0 redshifts. There is a $\sim$16\% probability that FR0s are volumetrically isotropic. (b) Relative simulated FR0 UHECR flux is proportional to radio output distribution. Generated by Pareto distribution fit to NVSS data ($p$-value deviation from Gaussian: 3.5$\times10^{-9}$)~\citep{Baldi:2017gao}. (c) Local universe source evolution modeled by preserving the redshift/flux correlation (Kendall’s correlation coefficient: -0.28, $p$-value:~4.6$\times10^{-5}$) from~\cite{Baldi:2017gao}.}
    \label{fig:distributions}
\end{figure*}

The simulated isotropic-pointing-direction FR0s match the redshift distance distribution of data shown in Figure~\ref{fig:zhist}, therefore the overall spatial distribution is not exactly volumetrically isotropic. The relative emission flux distribution is proportional to the radio output distribution (Figure~\ref{fig:fluxhist}). Furthermore, the simulated FR0s preserve the correlation between radio (a source jet power measure)/UHECR flux and redshift (Figure~\ref{fig:correlation}) to model the local universe source evolution. 

As this method is probabilistic based on observational FR0 data, it is more realistic than the usual functional source evolution form used for combined fits of (1+$z$)$^m$~\citep{PierreAuger:2016use, PierreAuger:2022atd, PierreAuger:2024hlp}. A fit to the emissivity versus redshift to this functional form results in $m = -31$. The cosmology used to convert FR0 catalog redshifts to distance was the Planck 2015 results~\citep{refId0}.

The UHECR flux of each simulated FR0 source is emitted isotropically. The isotropic emission model for FR0s is based on their morphological characteristics approximating that they appear to possess relatively slow bulk flows, which suggest a lack of strong directional biases in their emissions. Simulating directed emission is not expected to have a significant effect due to the large number of sources and the necessary assumption of isotropically distributed emission-cone pointing-directions. To refine our understanding of FR0 contributions, future work may however explore anisotropic emission scenarios, assessing their impacts on UHECR propagation and detection.

\subsection{Simulated Energy Spectrum}
\begin{figure}[htb]
    \centering
    \includegraphics[width=0.45\textwidth,trim={0cm 0cm 0cm 1cm},clip]{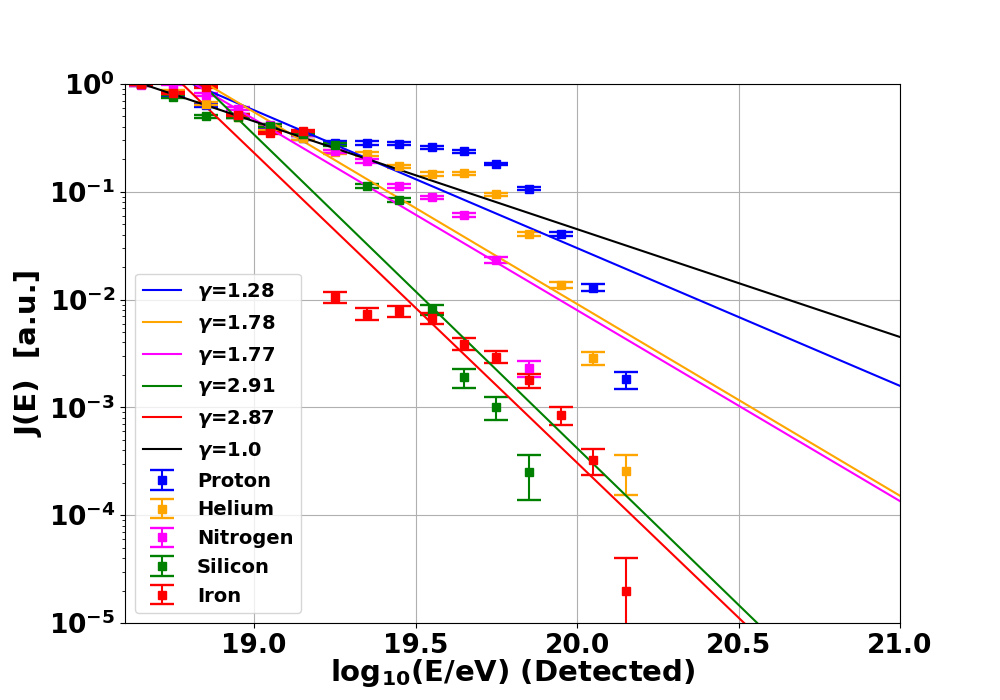}
    \caption{Relative energy spectra (maximum normalized to one) of simulations modified from the emitted E$_0^{-1}$ spectrum after intergalactic propagation. Shown are sets of nuclei that were initially proton, helium, nitrogen, silicon, and iron in the case with no intergalactic magnetic fields. Simple linear fits are shown to compare to the $\gamma=-1$ emission.}
    \label{fig:propspect}
\end{figure}

The simulated UHECR emission energies follow a uniform distribution with an E$_0^{-1}$ energy spectrum for energies 10$^{18.6}$~$\leq$~E$_0$/eV~$<$~10$^{20.2}$. This spectrum is modified by interactions in the intergalactic medium during propagation to the observer, as demonstrated for the no-magnetic-field case with an unmodified E$_0^{-1}$ emission spectrum in Figure~\ref{fig:propspect}. This propagated spectrum is modified when fitting to data by weighting the individual detected nuclei.

The upper bound of E$_0$~$<$~10$^{20.2}$ eV was chosen because it is within the FR0 source-modeled maximum limit for iron nuclei given in~\cite{Merten:2021brk}, while the highest energy event observed by the Pierre Auger Observatory is currently $\sim$10$^{20.2}$~eV~\citep{AbdulHalim_2023}. It is assumed that events with such extreme observed energies may originate from an exceptional non-FR0 source class.

The best-case scenario in~\cite{Merten:2021brk} (B-2) suggests a maximal mean rigidity of 10$^{19.2}$ V. This implies a maximal energy of 10$^{20.6}$ eV for iron, which is very close to 10$^{20.2}$ eV. The maximal energy in model B-2 is determined by the size and magnetic field strength of the emission region. It is likely that other reasonable combinations of parameter values could lead to even higher energies. Crucially, the maximal energies provided in the models of ~\cite{Merten:2021brk} are exponential cutoff energies, meaning particles can be further accelerated. This cutoff energy is modeled by the rigidity-dependent exponential cutoff fit parameter discussed in Section 5.1.

As shown, the results from the combined-fit simulation can help inform future source-modeling efforts.

\subsection{Emitted Nuclei}

Combined fits to the energy spectrum and estimated mean log mass number (\lnA) of data are done using sets of emitted proton, helium, nitrogen, silicon, and iron nuclei. These nuclei are commonly chosen in combined-fit analyses as each elemental species can be considered as a family of neighboring nuclei and this selection represents a comprehensive sampling of cosmic-ray nuclei observed in composition spectra. Studies suggest that heavier nuclei can originate from specific types of astrophysical environments, such as the dense, magnetized surroundings of AGN and elements heavier than iron should be rare enough to have a negligible impact on combined fits.

\begin{figure}[htb]
    \centering
    \includegraphics[width=0.45\textwidth,trim={0cm 0cm 0cm 1cm},clip]{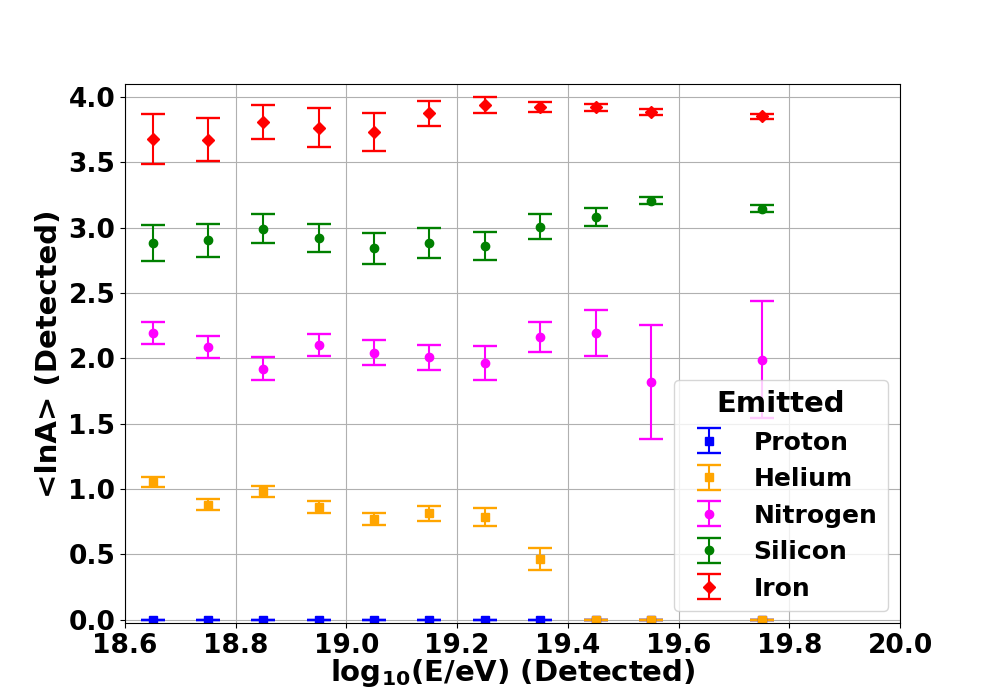}
    \caption{Emitted nuclei interact with the simulated intergalactic medium and are modified before observation. Shown are the emitted nuclei observed \lnA for the constant energy-independent nuclei fraction best-fit configuration CLUES-SIBYLL.}
    \label{fig:Sim_lnA}
\end{figure}

The emitted nuclei propagation is modeled according to their interaction cross-section with the photon fields, energy loss due to radiation emission, and deflection by cosmic magnetic fields as described in Section~\ref{sec:Propagation}, and are modified by the time they are observed. The resulting combined fit \lnA in bins of observed energy is shown in Figure~\ref{fig:Sim_lnA} for an example configuration and illustrates how different nuclei are modified from emission to detection. The observed \lnA slightly varies between the fitted configurations depending on the energy-spectrum fit parameters spectral index $\gamma$, rigidity-dependent exponential cutoff R$_{\text{cut}}$, and trajectory cutoff D$_{\text{cut}}$. Emitted proton is, of course, unmodified and has a \lnA$=0$. The emitted nuclei resulting observed \lnA for all evolving-fraction fits (see Section~\ref{evolvefit}) is shown in Appendix~\ref{appendix}.

\section{Combined Fit of Composition and Energy Spectrum}
\label{sec:fitting}
\subsection{Constant-Fraction Fits}
\label{subsec:constant}

For the constant energy-independent observed fraction fit, the simulated FR0 UHECR emission that best matches Auger data~\citep{Yushkov:2020nhr, Deligny:2020gzq} is found by minimizing the summed $\chi^2$ per degree of freedom for the composition and energy spectrum, considering eight input parameters. This fitting process utilizes four parameters representing the fractions of emitted protons, helium, nitrogen, silicon, and iron. These fractions, spanning the entire energy spectrum, are constrained to sum to 100\%. The remaining parameters include the emitted spectral index $\gamma$, rigidity-dependent exponential cutoff R$_{\text{cut}}$~\citep{PierreAuger:2016use, PierreAuger:2022atd, PierreAuger:2024hlp}, maximum particle-trajectory cutoff D$_{\text{cut}}$, and the energy-spectrum-only normalization $n$. 

The total cost function minimized is the sum of the $\chi^2$ per degree-of-freedom for composition and energy-spectrum comparisons between simulation and data:

\begin{equation}
\sum{\chi^2/N_{\mathrm{dof}}}=\sum{\chi_E^2/N^E_{\mathrm{dof}}}+\sum{\chi_C^2/N^C_{\mathrm{dof}}}.
\label{eq:chi2}
\end{equation}

The degrees of freedom for the energy spectrum are taken as $N^E_{\mathrm{dof}}= N^E_{\mathrm{bins}}-N^E_{\mathrm{params}} = 16 - 8 = 8$ and for the \lnA composition $N^C_{\mathrm{dof}}= N^C_{\mathrm{bins}}-N^C_{\mathrm{params}} = 11 - 7 = 4$.

For the sake of uniformity of our analysis, the accompanying variance of lnA (Var(lnA)) was not used in this work due to the unphysical negative values calculated for the QGS4 hadronic interaction model~\citep{Yushkov:2020nhr, 2013JCAP...02..026P}. A different method to find Var(lnA) was also reported in~\cite{Yushkov:2020nhr} using the correlation between X$_\text{max}$ and a variable proportional to EAS muon content (S$_{38}$) for the same three EAS models used in the present work~\citep{2016288}. The resulting variances are given for one energy bin (10$^{18.5}$~$\leq$ E/eV~$\leq$~10$^{19.0}$) as Var(lnA) $\approx$ 1.64 $\pm$ 0.92---this does not include uncertainty from the fact that EAS models do not adequately account for muons~\citep{TelescopeArray:2018eph, PhysRevD.109.102001}. The result is not a strong constraint and allows a wide range of UHECR nuclei composition mixes.

To minimize Equation~\ref{eq:chi2}, the SciPy optimization library~\citep{2020SciPy-NMeth} is employed. The process involves an initial global minimization using stochastic Differential Evolution~\citep{Storn1997}, followed by a refined deterministic local search to ensure precise fitting. Finally, a repeated deterministic nested fit is done, holding the spectrum parameters constant while varying the elemental content found in the previous step and vice versa.

This search method allows for a comprehensive exploration of the parameter space. The search limits include: spectral index $-4.5\leq\gamma\leq4.5$ (E$^{-\gamma}$), rigidity-dependent cutoff $10^{16} \leq R_{\mathrm{cut}}/\mathrm{V} \leq 10^{22}$, and maximum particle trajectory cutoff $841 \leq D_{\mathrm{cut}}/\mathrm{Mpc} \leq 2000$. The large range of rigidity-dependent exponential cutoff effectively results in two alternate models tested for---(dominant) exponential emission and (almost) pure power law.

Uncertainties on the fit parameters are calculated as bootstrapped 68.27\% confidence intervals~(1$\sigma$ Gaussian equivalent) around the best-fit values. The bootstrap samples include random sampling with replacement for the simulation outputs and random Gaussian perturbations to the data using the total systematic and statistical uncertainties. This way, uncertainties on the fitting method, simulation and data statistics, and systematic data uncertainties are considered. The result is conservative error bars on all fit parameters as bin-by-bin systematic correlations are unknown and not considered.

Generally, the data or bootstrap samples result in fits that do not reach the bounds of the parameters (except the Dolag-EPOS configuration bootstrap samples have some D$_{\mathrm{cut}}$ = 2000~Mpc). However, the Dolag-SIBYLL configuration yielded a best-fit spectral index of $\gamma$ = 0.5, leading to significantly suppressed photon and neutrino spectra. Given the variability in the Dolag-SIBYLL $\gamma$ distribution across bootstrap samples, with the highest peak at $\gamma$~=~2.29, the Dolag-SIBYLL spectral index is constrained to $\gamma \geq 1.5$ to ensure physically reasonable photon/neutrino spectra.

\subsection{Evolving-Fraction Fit}
\label{evolvefit}
For the evolving energy-dependent observed fraction fit, the simulated FR0 UHECR emission that best matches Auger data~\citep{Yushkov:2020nhr, Deligny:2020gzq} is found by minimizing the summed $\chi^2$ per bin for the composition and energy spectrum. This approach involves forty-eight input parameters, providing a detailed, bin-specific analysis. Forty-four parameters represent the fractions of emitted protons, helium, nitrogen, silicon, and iron, calculated for each of the eleven energy bins used in the Auger composition data, with each bin’s total fraction constrained to 100\%. The remaining parameters are those described in Section~\ref{subsec:constant}, namely the emitted spectral index $\gamma$, rigidity-dependent exponential cutoff R$_{\text{cut}}$~\citep{PierreAuger:2016use, PierreAuger:2022atd, PierreAuger:2024hlp}, maximum particle trajectory cutoff D$_{\text{cut}}$, and the energy-spectrum normalization $n$.

A constant fraction of emitted nuclei over the energy range has been assumed in previous UHECR combined-fit analyses such as discussed in Section~\ref{subsec:constant}. However, this assumption is unphysical, as the pre-acceleration composition of nuclei does not remain constant with energy. Additionally, acceleration processes, spallation products, and energy losses are all composition-dependent, further complicating the assumption of constancy across the energy spectrum. Though an evolving nuclei fraction dependent on energy is not a good model discriminator due to over-fitting, it is more capable of reflecting physical processes than a constant observed fraction model. This detailed modeling provides a deeper picture of how the best possible fit emission requirements vary with energy for each EAS model and magnetic field.

With essentially no degrees of freedom for composition across 44 nuclei fraction parameters, the cost function minimized is simply the sum of the chi-squared values per bin for the composition and energy-spectrum comparisons between simulation and data. The minimization begins using the constant-fraction fit as an initial search value. The fitting process then iteratively refines the model by splitting the elemental fractions into progressively more detailed energy ranges---first two ranges, then four, and finally eleven, corresponding to one fraction per energy bin. At each step, deterministic nested fitting is applied, holding the spectrum parameters constant while varying the elemental composition obtained from the previous step, and then vice versa. Additional refinement is achieved by performing a nested fit, varying one energy bin at a time while keeping the others constant, followed by a final fit of all 48 parameters simultaneously.

As before, this method facilitates an extensive global search of possible values of the parameters, though given the results of the constant-fraction fit $\gamma$ can be more constrained to $1\leq\gamma\leq3.5$ without a loss in bootstrap uncertainty calculation. Uncertainties on the fit parameters are calculated the same as Section~\ref{subsec:constant}.

\section{Results}
\label{sec:results}
\subsection{Constant-Fraction Fit}
\label{subsec:constant_results}

\begin{figure*}[t]
    \centering
    \subfloat[Subfigure 1][]{
    \includegraphics[width=.45\textwidth,trim={0cm 0cm 0cm 1cm},clip]{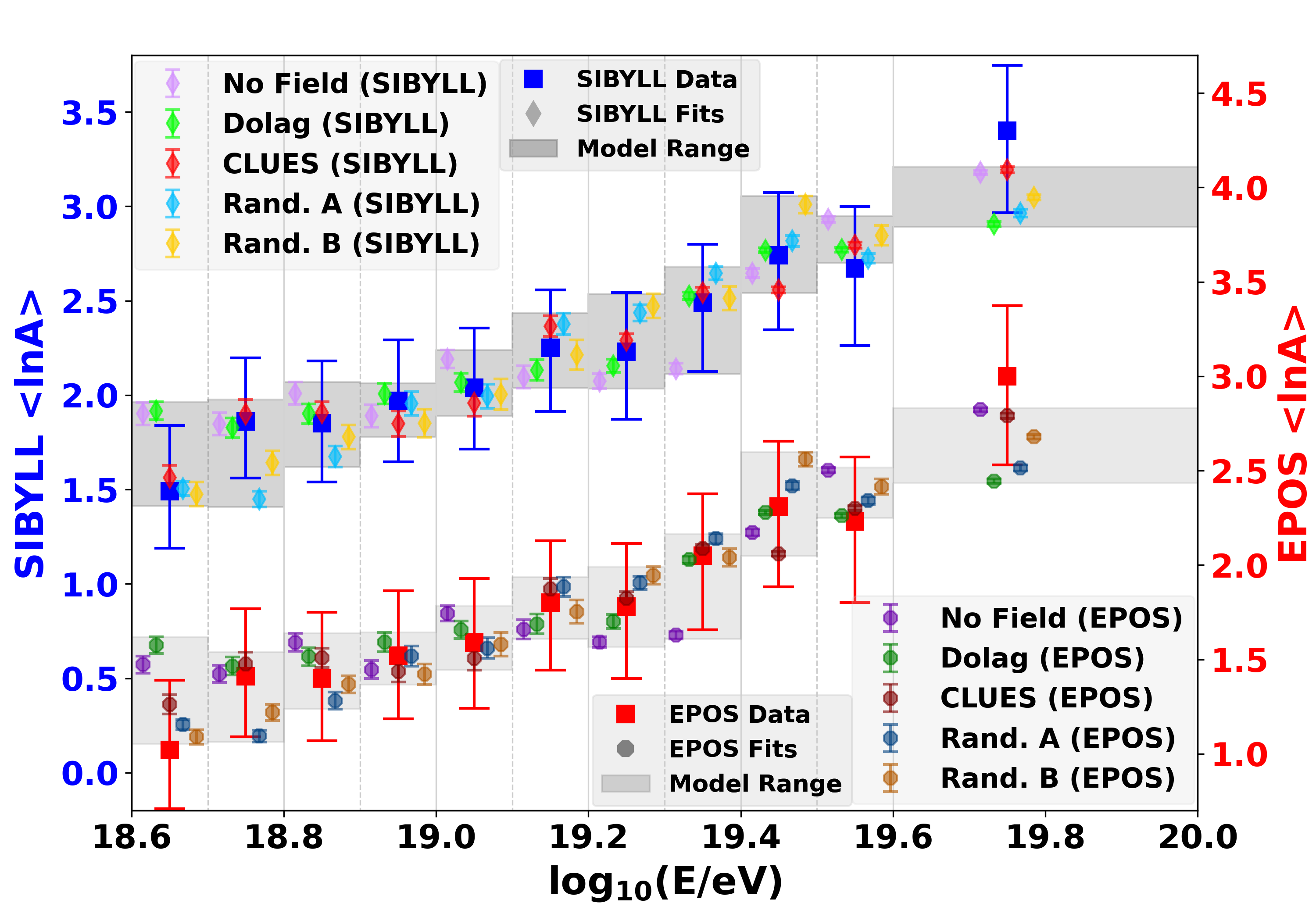}
    \label{fig:AmeanFit}}
    \subfloat[Subfigure 2][]{
    \includegraphics[width=.45\textwidth,trim={0cm 0cm 0cm 1cm},clip]{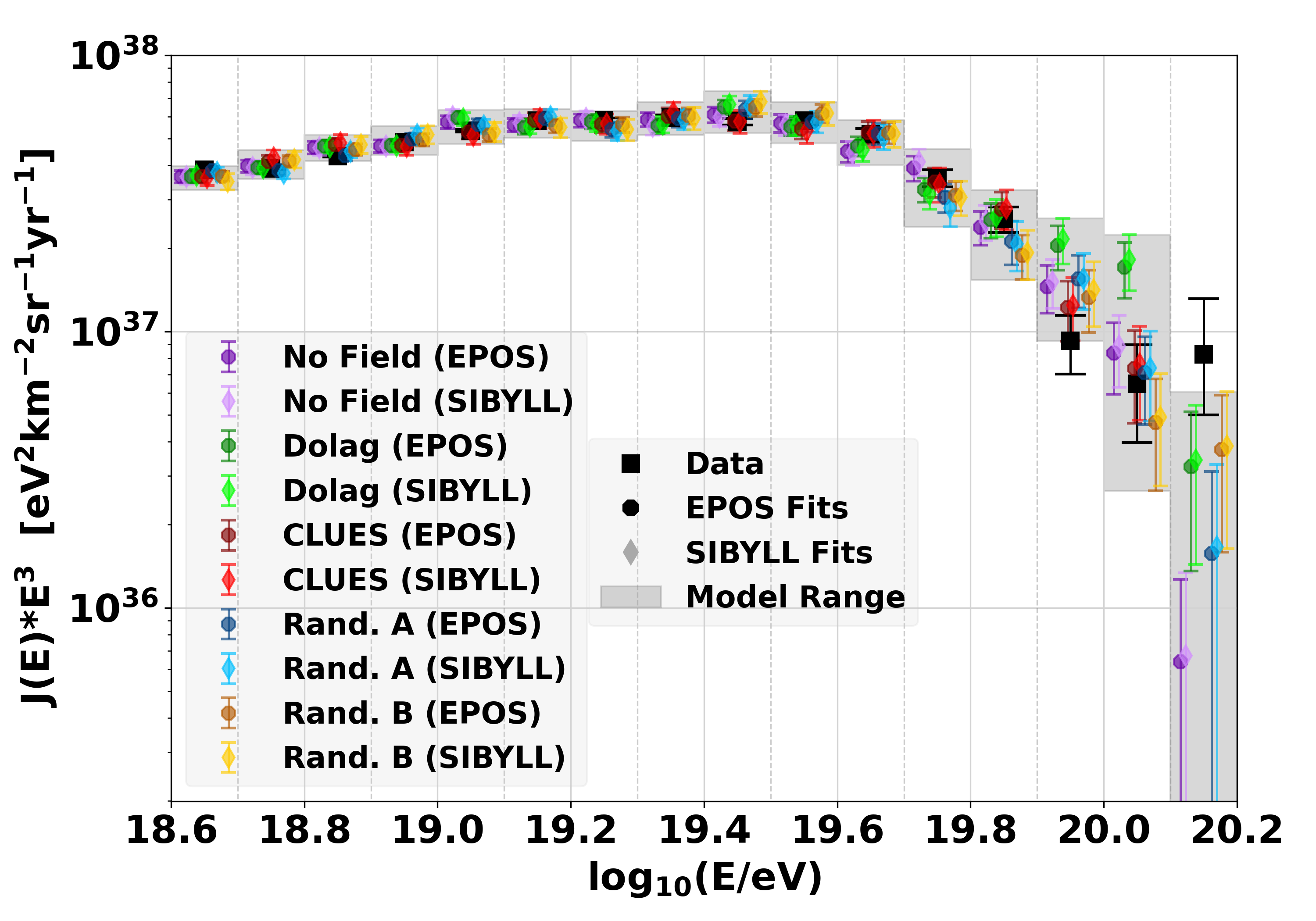}
    \label{fig:EspectFit}}
    \caption{Constant-fraction combined-fit composition and energy spectrum results for all simulated magnetic fields and two EAS models (EPOS-LHC (EPOS) and \Sibyll{} (SIBYLL)) compared to Auger data~\citep{Yushkov:2020nhr, Deligny:2020gzq}. The data error bars include statistical and systematic uncertainties. Offsets are applied to simulations within each bin on the x-axis for improved visibility. Grey areas display the $\pm1\sigma$~bounds of all the simulation configurations. Two legends are included---the first emphasizes figure element shapes (data: square, dark circle/light diamond: EPOS/SIBYLL hadronic model fits, grey box: range), while the second includes marker colors for all fits, where each base color represents a magnetic field model. (a) Mean log mass number \lnA. The blue left y-axis and blue square markers represent SIBYLL, while the red right y-axis and red square markers represent EPOS, each with adjusted limits for better visibility. (b) Energy spectra multiplied by E$^3$ for visibility.}
    \label{fig:Fits}
\end{figure*}

The constant observed fraction combined-fit results for all five simulated intergalactic media and two EAS models (EPOS-LHC and \Sibyll) compared to Auger data~\citep{Yushkov:2020nhr, Deligny:2020gzq} are shown in Figure~\ref{fig:Fits}. Results for the \QGSJet{} model are detailed in subsequent tables. For the observed mean log mass number \lnA (Figure~\ref{fig:AmeanFit}), the 1~nG random fields show a better fit for the lower energy, lighter composition, bins. In contrast, only the Dolag and Rand.B models sufficiently account for the data in the highest energy bin of the energy spectrum, as shown in Figure~\ref{fig:EspectFit}---though the Dolag model overshoots the previous two bins. The excellent fit of the longer correlation length 1~nG (Rand.B) field at the highest energies is somewhat unexpected, if FR0s do not contribute significantly at these extreme energies. The CLUES model, being on the order of J(E)*E$^3$~$\sim$~10$^{33}$ eV$^2$km$^{-2}$yr$^{-1}$str$^{-1}$, essentially does not contribute to the highest energy bin.

Table~\ref{tab:params} and Table~\ref{tab:nuclei} detail the best-fit parameters for each model configuration, along with the corresponding~1$\sigma$ equivalent confidence intervals that provide a measure of the uncertainty around these estimates. Using an 8-dimen\-sional multivariate (all fit parameters) kernel density estimation (KDE), Table~\ref{tab:paramsKDE} and Table~\ref{tab:nucleiKDE} present the most-probable values and 1$\sigma$ standard deviations derived from the bootstrap sample distributions.

Table~\ref{tab:params} and Table~\ref{tab:paramsKDE} show each EAS model and magnetic field's total $\Sigma \chi^2/\text{dof}$, emission spectral index~($\gamma$), exponential rigidity-dependent cutoff~(log$_{10}$(R$_{\text{cut}}$/V)), particle trajectory cutoff~(D$_{\text{cut}}$), and relative energy-spectrum normalization~($n$). Table~\ref{tab:nuclei} and Table~\ref{tab:nucleiKDE} show the fitted FR0 emission elemental fractions.

Table~\ref{tab:params} shows that the magnetic fields CLUES, Rand.A, and Rand.B generally exhibit the best fits. Specifically, the CLUES magnetic field yields the best fit for the EPOS-LHC and \Sibyll{} EAS models, while the Rand.A field is optimal for the \QGSJet{} model. The CLUES-SIBYLL configuration is the overall best fit, closely followed by CLUES-EPOS. Additionally, the EPOS-LHC model performs well paired with either random field model, demonstrating a robustness across different magnetic environments. When considering only EPOS and SIBYLL, the second-best magnetic field after CLUES is Rand.B.

Configurations with minimal magnetic fields, such as the 'no-field' or Dolag scenarios, are disfavored for all three EAS models as they have $\Sigma$$\chi$$^2$$>$3, indicating a significant influence of magnetic fields on fit accuracy. Notably, the \QGSJet{} EAS model typically shows a much poorer fit quality compared to the others, except when combined with the random fields, suggesting a sensitivity to specific intergalactic conditions.

\subsubsection{Energy-Spectrum Parameters}

\renewcommand{\arraystretch}{1.2}

\begin{table*}[hp]
\centering
\caption*{Constant-Fraction Energy-Spectrum Parameters}
\small
\rotatebox{0}{
\begin{tabular}{!{\vrule width 1.5pt} c | c | c | c | c | c | c !{\vrule width 1.5pt}}
\thickhline
\textbf{Field} & \textbf{Model} & \textbf{$\mathbf{\Sigma\chi^2/\text{dof}}$} &  \textbf{$\mathbf{\gamma}$} & \textbf{log$\mathbf{_{10}}$(R$\mathbf{_{cut}}$/V)} & \textbf{D$\mathbf{_{cut}}$/Mpc} & \textbf{$\mathbf{n}$}\\
\thickhline
\multirow{2}{*}{No Field} 
& SIBYLL & 3.21 & 2.51$^{+0.02}_{-0.67}$ & 19.36$^{+0.23}_{-0.31}$ & 843$^{+0}_{-1}$ & 1.337$^{+0.016}_{-0.003}$\\
\cline{2-7}
& EPOS & 3.15 & 2.50$^{+0.02}_{-0.16}$ & 19.40$^{+0.13}_{-0.06}$ & 843$^{+0}_{-0}$ & 1.337$^{+0.011}_{-0.004}$\\
\cline{2-7}
 & QGS4 & 3.47 & 2.47$^{+0.03}_{-0.08}$ & 19.43$^{+0.10}_{-0.03}$ & 843$^{+0}_{-0}$  & 1.338$^{+0.006}_{-0.006}$\\
\thickhline
\multirow{2}{*}{Dolag} 
& SIBYLL & 4.41 & 2.29$^{+0.06}_{-0.79}$ & 19.74$^{+0.00}_{-0.40}$ & 890$^{+320}_{-41}$ & 1.341$^{+0.008}_{-0.008}$\\
\cline{2-7}
& EPOS & 4.74 & 2.23$^{+0.11}_{-0.06}$ & 19.75$^{+0.03}_{-0.29}$ & 889$^{+230}_{-41}$ & 1.339$^{+0.007}_{-0.007}$\\
\cline{2-7}
 & QGS4 & 6.28 & 2.23$^{+0.08}_{-0.09}$ & 19.64$^{+0.10}_{-0.12}$ & 890$^{+47}_{-42}$ & 1.335$^{+0.005}_{-0.008}$\\
\thickhline
\multirow{2}{*}{CLUES} 
& SIBYLL & 1.76 & 2.54$^{+0.00}_{-0.19}$ & 19.45$^{+0.50}_{-0.12}$ & 842$^{+0}_{-0}$ & 1.354$^{+0.006}_{-0.015}$\\
\cline{2-7}
& EPOS & 1.87 & 2.43$^{+0.06}_{-0.13}$ & 19.51$^{+0.36}_{-0.07}$ & 842$^{+0}_{-1}$ & 1.347$^{+0.006}_{-0.011}$\\
\cline{2-7}
 & QGS4 & 3.10 & 2.32$^{+0.08}_{-0.05}$ & 19.56$^{+0.08}_{-0.07}$ & 841$^{+1}_{-0}$ & 1.334$^{+0.006}_{-0.007}$\\
\thickhline
\multirow{2}{*}{Rand.A}
& SIBYLL & 2.84 & 2.40$^{+0.07}_{-0.11}$ & 19.86$^{+0.12}_{-0.18}$ & 843$^{+80}_{-0}$ & 1.342$^{+0.007}_{-0.005}$\\
\cline{2-7}
& EPOS & 2.15 & 2.34$^{+0.08}_{-0.09}$ & 19.69$^{+0.19}_{-0.08}$ & 843$^{+76}_{-0}$ & 1.341$^{+0.006}_{-0.004}$\\
\cline{2-7}
 & QGS4 & 2.51 & 2.23$^{+0.07}_{-0.07}$ & 19.58$^{+0.07}_{-0.08}$ & 846$^{+69}_{-0}$ & 1.341$^{+0.005}_{-0.006}$\\
\thickhline
\multirow{2}{*}{Rand.B}
& SIBYLL & 2.57 & 2.47$^{+0.04}_{-0.16}$ & 19.71$^{+1.40}_{-0.08}$ & 843$^{+156}_{-0}$ & 1.346$^{+0.009}_{-0.006}$\\
\cline{2-7}
& EPOS & 2.29 & 2.33$^{+0.09}_{-0.15}$ & 19.60$^{+0.23}_{-0.09}$ & 843$^{+140}_{-0}$ & 1.346$^{+0.005}_{-0.006}$\\
\cline{2-7}
 & QGS4 & 2.60 & 1.97$^{+0.26}_{-0.05}$ & 19.52$^{+0.08}_{-0.07}$ & 854$^{+66}_{-11}$ & 1.343$^{+0.006}_{-0.004}$\\
\thickhline
\end{tabular}
}
\caption{The FR0 constant-fraction combined-fit results total sum chi-square per degree of freedom ($\Sigma\chi^2/\textrm{dof}$), spectral index ($\gamma$), exponential rigidity cutoff (log$_{10}$(R$_{\text{cut}}$/V)), trajectory cutoff (D$_{\text{cut}}$), and spectrum normalization (n) for all 15 configurations of magnetic field and EAS model. The three EAS models are EPOS-LHC (EPOS), \Sibyll{} (SIBYLL), and \QGSJet{} (QGS4).}
\label{tab:params}

\vspace{0.8cm} 

\centering
\caption*{Constant-Fraction Bootstrap Energy-Spectrum Parameters}
\small
\rotatebox{0}{
\begin{tabular}{!{\vrule width 1.5pt} c | c | c | c | c | c | c !{\vrule width 1.5pt}}
\thickhline
\textbf{Field} & \textbf{Model} & \textbf{$\mathbf{\Sigma\chi^2/\text{dof}}$} &  \textbf{$\mathbf{\gamma}$} & \textbf{log$\mathbf{_{10}}$(R$\mathbf{_{cut}/V}$)} & \textbf{D$\mathbf{_{cut}}$/Mpc} & \textbf{$\mathbf{n}$}\\
\thickhline
\multirow{2}{*}{No Field} 
& SIBYLL & 3.21 & 2.48 $^{+0.27}_{-0.27}$ & 19.38$^{+0.23}_{-0.23}$ & 843$^{+0}_{-0}$ & 1.341$^{+0.006}_{-0.006}$\\
\cline{2-7}
& EPOS & 3.15 & 2.48$^{+0.26}_{-0.26}$ & 19.40$^{+0.13}_{-0.13}$ & 843$^{+0}_{-0}$ & 1.338$^{+0.005}_{-0.005}$\\
\cline{2-7}
 & QGS4 & 3.47 & 2.45$^{+0.04}_{-0.04}$ & 19.43$^{+0.06}_{-0.06}$ & 843$^{+0}_{-0}$  & 1.335$^{+0.004}_{-0.004}$\\
\thickhline
\multirow{2}{*}{Dolag} 
& SIBYLL & 4.41 & 2.07$^{+0.23}_{-0.23}$ & 19.04$^{+0.57}_{-0.57}$ & 871$^{+126}_{-29}$ & 1.339$^{+0.005}_{-0.005}$\\
\cline{2-7}
& EPOS & 4.74 & 2.33$^{+0.38}_{-0.38}$ & 19.37$^{+0.24}_{-0.24}$ & 864$^{+180}_{-22}$ & 1.334$^{+0.004}_{-0.004}$\\
\cline{2-7}
 & QGS4 & 6.28 & 2.24$^{+0.37}_{-0.37}$ & 19.64$^{+0.07}_{-0.07}$ & 893$^{+86}_{-51}$ & 1.335$^{+0.004}_{-0.004}$\\
\thickhline
\multirow{2}{*}{CLUES} 
& SIBYLL & 1.76 & 2.54$^{+0.15}_{-0.15}$ & 19.41$^{+0.34}_{-0.34}$ & 842$^{+0}_{-0}$ & 1.352$^{+0.007}_{-0.007}$\\
\cline{2-7}
& EPOS & 1.87 & 2.41$^{+0.07}_{-0.07}$ & 19.51$^{+0.15}_{-0.15}$ & 842$^{+0}_{0}$ & 1.346$^{+0.005}_{-0.005}$\\
\cline{2-7}
 & QGS4 & 3.10 & 2.31$^{+0.05}_{-0.05}$ & 19.56$^{+0.06}_{-0.06}$ & 842$^{+0}_{-0}$ & 1.337$^{+0.004}_{-0.004}$\\
\thickhline
\multirow{2}{*}{Rand.A}
& SIBYLL & 2.84 & 2.36$^{+0.26}_{-0.26}$ & 19.87$^{+1.66}_{-1.66}$ & 864$^{+81}_{-22}$ & 1.343$^{+0.001}_{-0.001}$\\
\cline{2-7}
& EPOS & 2.15 & 2.31$^{+0.05}_{-0.05}$ & 19.67$^{+0.11}_{-0.11}$ & 853$^{+36}_{-11}$ & 1.343$^{+0.003}_{-0.003}$\\
\cline{2-7}
 & QGS4 & 2.51 & 2.25$^{+0.07}_{-0.07}$ & 19.59$^{+0.08}_{-0.08}$ & 854$^{+29}_{-12}$ & 1.342$^{+0.004}_{-0.004}$\\
\thickhline
\multirow{2}{*}{Rand.B}
& SIBYLL & 2.57 & 2.43$^{+0.06}_{-0.06}$ & 19.70$^{+10.20}_{-10.20}$ & 852$^{+53}_{-10}$ & 1.347$^{+0.005}_{-0.005}$\\
\cline{2-7}
& EPOS & 2.29 & 2.36$^{+0.10}_{-0.10}$ & 19.54$^{+7.2}_{-7.2}$ & 854$^{+64}_{-12}$ & 1.346$^{+0.000}_{-0.000}$\\
\cline{2-7}
 & QGS4 & 2.60 & 2.15$^{+0.14}_{-0.14}$ & 19.45$^{+0.62}_{-0.62}$ & 848$^{+48}_{-6}$ & 1.342$^{+0.003}_{-0.003}$\\
\thickhline
\end{tabular}
}
\caption{The FR0 constant-fraction combined-fit bootstrap distribution most-probable spectral index ($\gamma$), exponential rigidity cutoff (log$_{10}$(R$_{\text{cut}}$/V)), trajectory cutoff ($\text{D}_{\text{cut}}$), and spectrum normalization (n) for all 15 configurations of magnetic field and EAS model. The best fit results total sum chi-square per degree of freedom ($\Sigma\chi^2/\textrm{dof}$) from Table~\ref{tab:params} is also listed.}
\label{tab:paramsKDE}
\end{table*}

\begin{figure*}[hb]
    \centering
    \includegraphics[width=0.9\textwidth]{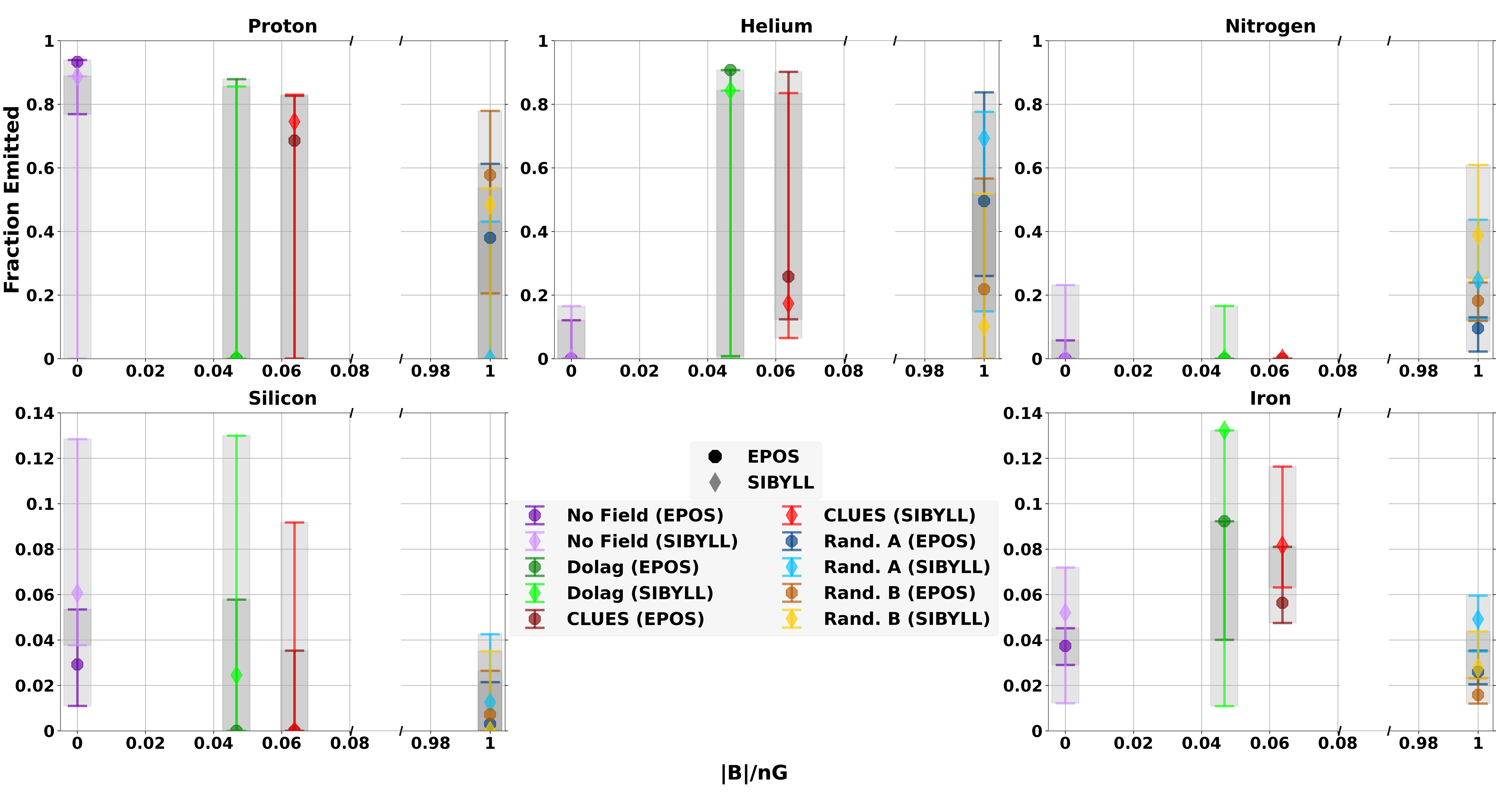}
    \caption{The constant observed fraction fit FR0 emitted nuclei fractions necessary to best fit the data for ten configurations based on the EPOS-LHC (EPOS) and \Sibyll{} (SIBYLL) hadronic interaction models versus magnetic field strength. Note the different y-axis limits for the top and bottom figures and the broken x-axis at $B=0.08$~nG. Grey bands extend from $\pm1\sigma$ for each configuration and is darkest in the ranges where the EAS models overlap. EPOS/SIBYLL fits are represented by circles/diamonds and darker/lighter colors. Each base color represents a magnetic field model.}
    \label{fig:nuclei}
\end{figure*}

Analysis of the constant-fraction energy-spectrum parameters of Table~\ref{tab:params} and Table~\ref{tab:paramsKDE} suggest that both the magnetic field strength \Bave and RMS \Brms influence the expected cosmic-ray source emission energy spectrum. However, the results are perhaps surprisingly stable given the wide range of magnetic field models and EAS model compositions. As \Bave or \Brms increases, the emission spectral index $\gamma$ does decrease---with generally harder emission spectra for \QGSJet. The fit values of the exponential rigidity-dependent cutoff, R$_{\text{cut}}$, generally increases with both \Bave and \Brms. For smaller magnetic fields, R$_{\text{cut}}$ decreases with a heavier elemental composition, while it increases with larger field strengths. The trajectory cutoff fit values, D$_{\text{cut}}$, remain fairly consistent across the fields tested, with the exception of the Dolag field, where a significantly higher \Brms of 11~nG introduces a noticeable deviation.

Also, for any given magnetic field, EAS models that result in a heavier composition \lnA (\Sibyll{} is heaviest \lnA as shown in Figure \ref{fig:AmeanFit} and \QGSJet~is the lightest) leads to a softer (increasing) spectral index E$^{-\gamma}$. An exception occurs with the bootstrap KDE most-probable $\gamma$ for Dolag-SIBYLL shown in Table~\ref{tab:paramsKDE}, as this configuration fit leads to multi-modal distributions (a 1-d KDE fit results in $\gamma$ = 2.30).

Compared to the previous FR0 combined-fit results, with sources limited to $z$~$\leq$~0.05 (and a slightly older CRPropa version), the emission spectra for the \QGSJet{} and EPOS-LHC EAS models are harder (a smaller $\gamma$). The exponential cutoff has also increased for all scenarios, with the exception of the Rand.B magnetic field~\citep{Lundquist:20233x}.

Excluding the disfavored magnetic fields (no-field and Dolag) and the disfavored CLUES-QGS4 configuration (th\-ose with $\Sigma$$\chi$$^2$$>$3), the bounds found for the best fit of these parameters from Table~\ref{tab:params} are:

\begin{itemize}
    \item 1.97 $\leq \gamma \leq$ 2.54
    \item 19.45 $\leq$ log$_{10}$(R$_{\text{cut}}$/V) $\leq$ 19.86
    \item 842 $\leq$ D$_{\text{cut}}$/Mpc $\leq$ 854
    \item 1.341 $\leq n \leq$ 1.354
\end{itemize}

For the bootstrap most-probable parameters from Table~\ref{tab:paramsKDE}, excluding the disfavored configurations, the ranges found are:

\begin{itemize}
    \item 2.15 $\leq \gamma \leq$ 2.54
    \item 19.41 $\leq$ log$_{10}$(R$_{\text{cut}}$/V) $\leq$ 19.87
    \item 842 $\leq$ D$_{\text{cut}}$/Mpc $\leq$ 864
    \item 1.342 $\leq n \leq$ 1.352
\end{itemize}

Compared to the recent Auger result of \cite{PierreAuger:2016use} these parameter values are similar to the reported second minima of their models and the low-$z$ source distance model ($z$~$<$~0.02).

\subsubsection{UHECR Composition Parameters}
\label{Constant_Comp}
\FloatBarrier
\renewcommand{\arraystretch}{1.2}
\begin{table*}[htbp]
\centering
\caption*{Constant-Fraction Composition Parameters}
\small
\rotatebox{0}{
\begin{tabular}{!{\vrule width 1.5pt} c | c | c | c | c | c | c | c !{\vrule width 1.5pt}}
\thickhline
\textbf{Field} & \textbf{Model} & \textbf{$\mathbf{f_H}$(\%)} + \textbf{$\mathbf{f_{He}}$(\%)} & \textbf{$\mathbf{f_H}$(\%)} & \textbf{$\mathbf{f_{He}}$(\%)} & \textbf{$\mathbf{f_N}$(\%)} & \textbf{$\mathbf{f_{Si}}$(\%)} & \textbf{$\mathbf{f_{Fe}}$(\%)} \\
\thickhline
\multirow{2}{*}{No Field}
& SIBYLL & 88.7$^{+0.2}_{-24.0}$ & 88.7$^{+0.1}_{-88.7}$ & 0.0$^{+16.5}_{-0.0}$ & 0.0$^{+23.1}_{-0.0}$ & 6.1$^{+6.8}_{-2.3}$ & 5.2$^{+2.0}_{-4.0}$ \\
\cline{2-8}
 & EPOS & 93.4$^{+0.7}_{-7.2}$ & 93.4$^{+0.6}_{-16.5}$ & 0.0$^{+12.1}_{-0.0}$ & 0.0$^{+5.8}_{-0.0}$ & 2.9$^{+2.4}_{-1.8}$ & 3.7$^{+0.8}_{-0.8}$ \\
\cline{2-8}
 & QGS4 & 97.6$^{+0.3}_{-1.1}$ & 97.6$^{+0.1}_{-8.8}$ & 0.0$^{+7.8}_{-0.0}$ & 0.0$^{+0.0}_{-0.0}$ & 0.2$^{+1.2}_{-0.2}$ & 2.2$^{+0.2}_{-0.5}$ \\
\thickhline
\multirow{2}{*}{Dolag} 
 & SIBYLL & 84.3$^{+5.2}_{-11.9}$ & 0.0$^{+85.6}_{-0.0}$ & 84.3$^{+0.0}_{-84.3}$ & 0.0$^{+16.6}_{-0.0}$ & 2.5$^{+10.5}_{-2.4}$ & 13.2$^{+0.0}_{-12.1}$ \\
\cline{2-8}
 & EPOS & 90.8$^{+3.7}_{-3.8}$ & 0.0$^{+87.9}_{-0.0}$ & 90.8$^{+0.0}_{-90.0}$ & 0.0$^{+0.0}_{-0.0}$ & 0.0$^{+5.8}_{-0.0}$ & 9.2$^{+0.0}_{-5.2}$ \\
\cline{2-8}
 & QGS4 & 97.1$^{+0.4}_{-1.0}$ & 54.4$^{+28.3}_{-35.4}$ & 42.7$^{+34.7}_{-28.4}$ & 0.0$^{+0.0}_{-0.0}$ & 0.0$^{+0.0}_{-0.0}$ & 2.9$^{+0.9}_{-0.5}$ \\
\thickhline
\multirow{2}{*}{CLUES}
& SIBYLL & 91.8$^{+0.3}_{-11.4}$ & 74.5$^{+8.5}_{-74.5}$ & 17.3$^{+66.2}_{-10.8}$ & 0.0$^{+0.0}_{-0.0}$ & 0.0$^{+9.2}_{-0.0}$ & 8.2$^{+3.5}_{-1.8}$ \\
\cline{2-8}
 & EPOS & 94.4$^{+0.4}_{-4.4}$ & 68.6$^{+14.1}_{-68.6}$ & 25.8$^{+64.4}_{-13.4}$ & 0.0$^{+0.0}_{-0.0}$ & 0.0$^{+3.5}_{-0.0}$ & 5.6$^{+2.5}_{-0.9}$ \\
\cline{2-8}
 & QGS4 & 97.4$^{+0.4}_{-0.5}$ & 68.8$^{+16.3}_{-15.8}$ & 28.6$^{+15.6}_{-16.1}$ & 0.0$^{+0.0}_{-0.0}$ & 0.0$^{+0.0}_{-0.0}$ & 2.6$^{+0.5}_{-0.4}$ \\
\thickhline
\multirow{2}{*}{Rand.A} 
 & SIBYLL & 69.3$^{+11.1}_{-19.2}$ & 0.0$^{+43.1}_{-0.0}$ & 69.3$^{+8.3}_{-54.4}$ & 24.5$^{+19.2}_{-12.3}$ & 1.3$^{+3.0}_{-1.3}$ & 4.9$^{+1.0}_{-1.4}$ \\
\cline{2-8}
 & EPOS & 87.6$^{+5.6}_{-4.2}$ & 38.0$^{+23.3}_{-38.0}$ & 49.6$^{+34.2}_{-23.5}$ & 9.5$^{+3.5}_{-7.3}$ & 0.3$^{+1.8}_{-0.3}$ & 2.6$^{+0.9}_{-0.5}$ \\
\cline{2-8}
 & QGS4 & 98.9$^{+0.0}_{-3.5}$ & 57.9$^{+18.4}_{-12.1}$ & 41.0$^{+10.0}_{-20.7}$ & 0.0$^{+3.5}_{-0.0}$ & 0.0$^{+0.1}_{-0.0}$ & 1.1$^{+0.2}_{-0.2}$ \\
\thickhline
\multirow{2}{*}{Rand.B} 
 & SIBYLL & 58.4$^{+10.5}_{-23.6}$ & 48.3$^{+5.3}_{-48.3}$ & 10.1$^{+41.8}_{-10.1}$ & 38.8$^{+22.2}_{-13.4}$ & 0.0$^{+3.5}_{-0.0}$ & 2.8$^{+1.6}_{-0.4}$ \\
\cline{2-8}
& EPOS & 79.5$^{+5.8}_{-6.5}$ & 57.7$^{+20.2}_{-37.2}$ & 21.8$^{+34.9}_{-21.8}$ & 18.2$^{+5.8}_{-6.3}$ & 0.7$^{+1.9}_{-0.7}$ & 1.6$^{+0.7}_{-0.4}$ \\
\cline{2-8}
 & QGS4 & 95.8$^{+1.7}_{-4.5}$ & 45.6$^{+41.0}_{-9.4}$ & 50.2$^{+9.0}_{-43.1}$ & 2.2$^{+5.1}_{-2.2}$ & 0.9$^{+0.6}_{-0.9}$ & 1.1$^{+0.2}_{-0.4}$ \\
\thickhline
\end{tabular}
}
\caption{The FR0 constant-fraction combined-fit nuclei emission percentages for proton, helium, nitrogen, silicon, and iron primaries for all 15 configurations of magnetic field and EAS model. Also, shown is the total fraction of the light-nuclei proton plus helium ($f_H+f_{He})$.}
\label{tab:nuclei}

\vspace{0.8cm} 

\caption*{Constant-Fraction Bootstrap Composition Parameters}
\small
\rotatebox{0}{
\begin{tabular}{!{\vrule width 1.5pt} c | c | c | c | c | c | c | c !{\vrule width 1.5pt}}
\thickhline
\textbf{Field} & \textbf{Model} & \textbf{$\mathbf{f_H}$(\%)} + \textbf{$\mathbf{f_{He}}$(\%)} & \textbf{$\mathbf{f_H}$(\%)} & \textbf{$\mathbf{f_{He}}$(\%)} & \textbf{$\mathbf{f_N}$(\%)} & \textbf{$\mathbf{f_{Si}}$(\%)} & \textbf{$\mathbf{f_{Fe}}$(\%)} \\
\thickhline
\multirow{2}{*}{No Field}
& SIBYLL & 86.3$^{+13.7}_{-15.6}$ & 85.6$^{+14.4}_{-21.7}$ & 0.6$^{+15.2}_{-0.6}$ & 2.8$^{+15.8}_{-2.8}$ & 5.6$^{+3.1}_{-3.1}$ & 5.4$^{+2.3}_{-2.3}$ \\
\cline{2-8}
 & EPOS & 92.8$^{+5.3}_{-5.3}$ & 91.4$^{+8.6}_{-13.2}$ & 1.4$^{+10.7}_{-1.4}$ & 0.6$^{+5.3}_{-0.6}$ & 2.9$^{+1.5}_{-1.5}$ & 3.7$^{+0.8}_{-0.8}$ \\
\cline{2-8}
 & QGS4 & 97.6$^{+0.8}_{-0.8}$ & 96.4$^{+3.6}_{-7.6}$ & 1.2$^{+7.4}_{-1.2}$ & 0.0$^{+0.8}_{-0.0}$ & 0.2$^{+0.5}_{-0.2}$ & 2.2$^{+0.2}_{-0.2}$ \\
\thickhline
\multirow{2}{*}{Dolag} 
 & SIBYLL & 85.2$^{+8.0}_{-8.0}$ & 84.4 $^{+15.6}_{-22.6}$ & 0.9$^{+22.1}_{-0.9}$ & 0.0$^{+7.6}_{-0.0}$ & 13.0$^{+3.9}_{-3.9}$ & 1.8$^{+3.1}_{-1.8}$ \\
\cline{2-8}
 & EPOS & 94.8$^{+4.9}_{-4.9}$ & 84.5$^{+15.5}_{-23.6}$ & 10.3$^{+22.4}_{-10.3}$ & 0.1$^{+4.7}_{-0.1}$ & 0.2$^{+2.2}_{-0.2}$ & 4.9$^{+1.7}_{-1.7}$ \\
\cline{2-8}
 & QGS4 & 97.0$^{+0.6}_{-0.6}$ & 54.7$^{+18.0}_{-18.0}$ & 42.3$^{+17.7}_{-17.7}$ & 0.0$^{+0.5}_{-0.0}$ & 0.0$^{+0.3}_{-0.0}$ & 2.9$^{+0.5}_{-0.5}$ \\
\thickhline
\multirow{2}{*}{CLUES}
& SIBYLL & 91.6$^{+3.6}_{-3.6}$ & 81.2$^{+18.8}_{-23.3}$ & 10.6$^{+20.7}_{-10.6}$ & 0.0$^{+2.2}_{-0.0}$ & 0.7$^{+3.4}_{-0.7}$ & 7.5$^{+1.9}_{-1.9}$ \\
\cline{2-8}
 & EPOS & 94.2$^{+1.8}_{-1.8}$ & 66.7$^{+21.8}_{-21.8}$ & 27.6$^{+20.5}_{-20.5}$ & 0.0$^{+0.5}_{-0.0}$ & 0.2$^{+1.6}_{-0.2}$ & 5.5$^{+1.0}_{-1.0}$ \\
\cline{2-8}
 & QGS4 & 97.5$^{+0.4}_{-0.4}$ & 67.6$^{+12.3}_{-12.3}$ & 29.9$^{+12.1}_{-12.1}$ & 0.0$^{+0.0}_{-0.0}$ & 0.0$^{+0.2}_{-0.0}$ & 2.5$^{+0.3}_{-0.3}$ \\
\thickhline
\multirow{2}{*}{Rand.A} 
 & SIBYLL & 70.0$^{+11.4}_{-11.4}$ & 5.3$^{+14.7}_{-5.3}$ & 63.9$^{+17.5}_{-17.5}$ & 25.1$^{+11.8}_{-11.8}$ & 0.7$^{+1.5}_{-0.7}$ & 5.0$^{+0.9}_{-0.9}$ \\
\cline{2-8}
 & EPOS & 87.6$^{+4.6}_{-4.6}$ & 36.2$^{+15.9}_{-15.9}$ & 51.5$^{+16.0}_{-16.0}$ & 9.5$^{+4.9}_{-4.9}$ & 0.2$^{+0.8}_{-0.2}$ & 2.7$^{+0.5}_{-0.5}$ \\
\cline{2-8}
 & QGS4 & 98.4$^{+1.1}_{-1.1}$ & 60.6$^{+10.1}_{-10.1}$ & 37.7$^{+10.4}_{-10.4}$ & 0.5$^{+1.2}_{-0.5}$ & 0.0$^{+0.2}_{-0.0}$ & 1.1$^{+0.2}_{-0.2}$ \\
\thickhline
\multirow{2}{*}{Rand.B} 
 & SIBYLL & 59.3$^{+11.1}_{-11.1}$ & 45.7$^{+15.3}_{-15.3}$ & 15.2$^{+14.4}_{-14.4}$ & 35.8$^{+11.5}_{-11.5}$ & 0.6$^{+1.3}_{-0.6}$ & 2.7$^{+0.6}_{-0.6}$ \\
\cline{2-8}
& EPOS & 82.0$^{+4.5}_{-4.5}$ & 73.3$^{+16.3}_{-16.3}$ & 9.5$^{+14.8}_{-9.5}$ & 15.6$^{+4.7}_{-4.7}$ & 0.2$^{+1.0}_{-0.2}$ & 1.4$^{+0.4}_{-0.4}$ \\
\cline{2-8}
 & QGS4 & 93.6$^{+1.8}_{-1.8}$ & 80.4$^{+14.2}_{-14.2}$ & 13.2$^{+14.7}_{-13.2}$ & 5.6$^{+2.1}_{-2.1}$ & 0.0$^{+0.4}_{-0.0}$ & 0.7$^{+0.2}_{-0.2}$ \\
\thickhline
\end{tabular}
}
\caption{The FR0 constant-fraction combined-fit bootstrap distribution most-probable nuclei emission percentages for proton, helium, nitrogen, silicon, and iron primaries for all 15 configurations.}
\label{tab:nucleiKDE}
\end{table*}
The FR0 emission fractions for constant (observed) fraction fits of emitted proton, helium, nitrogen, silicon, and iron nuclei are shown in Figure~\ref{fig:nuclei}, Table~\ref{tab:nuclei}, and Table~\ref{tab:nucleiKDE}. Due to significant propagation effects, where a large percentage of emitted helium transforms into observed protons (as detailed in Figure~\ref{fig:Sim_lnA}), the best-fit emission fractions for protons and helium exhibit some degeneracy. This results in instability in determining whether protons or helium dominate for a given magnetic field. The combined light component fraction ($f_\mathrm{H} + f_{\mathrm{He}}$) and the corresponding KDE most-probable values of Table~\ref{tab:nucleiKDE} show more stable results.

These results indicate that magnetic field strength can significantly impact the relative nuclei abundances of the source-emitted cosmic rays. As the magnetic field strength increases, the emission fractions of protons and helium remains relatively stable. In contrast, the fractions of nitrogen nuclei tend to increase, whereas heavier nuclei such as silicon and iron show a decrease in their relative fractions.

Compared to the previous FR0 combined-fit results, with sources limited to $z$~$\leq$~0.05, the helium fractions have significantly increased for the \QGSJet{} and EPOS-LHC EAS models (except for the no-field scenario)~\citep{Lundquist:20233x}. The amount of proton has commensurately decreased for these configurations (except for the no-field scenario). Also, comparatively, there is now more nitrogen for the Rand.A-EPOS and Rand.B-EPOS configurations and less nitrogen for the Rand.A-QGS4 and Rand.B-QGS4 configurations. The fraction of silicon has decreased for all configurations except for a small increase for Rand.B-QGS4. Finally, the iron fraction has increased for all magnetic fields except for the no-field scenario~\citep{Lundquist:20233x}.

Excluding the disfavored configurations, the ranges fou\-nd for the FR0 emission best-fit elemental fractions (in~\%) from Table~\ref{tab:nuclei} are:

\begin{itemize}
    \item 58.4 $\leq f_\mathrm{H} + f_{\mathrm{He}} \leq$ 98.9
    \item 0.0 $\leq f_\mathrm{H} \leq$ 74.5
    \item 10.1 $\leq f_{\mathrm{He}} \leq$ 69.3
    \item 0.0 $\leq f_\mathrm{N} \leq$ 38.8
    \item 0.0 $\leq f_\mathrm{Si} \leq$ 1.3
    \item 1.1 $\leq f_\mathrm{Fe} \leq$ 8.2
\end{itemize}

The analysis shows a broad range of elemental fractions, with the most notable being a substantial variation in the combined proton and helium fraction, ranging from 58.4\% to 98.9\%, and the fraction of nitrogen ranging from zero to 38.8\%. These ranges reflect the effect of interactions between magnetic fields and nuclei during cosmic-ray propagation.

For the bootstrap most-probable parameters from Table~\ref{tab:nucleiKDE}, excluding the disfavored configurations, the ranges found for the FR0 emission elemental fractions (\%) are:

\begin{itemize}
    \item 59.3 $\leq f_\mathrm{H} + f_{\mathrm{He}} \leq$ 98.4
    \item 5.3 $\leq f_\mathrm{H} \leq$ 81.2
    \item 9.5 $\leq f_{\mathrm{He}} \leq$ 63.9
    \item 0 $\leq f_\mathrm{N} \leq$ 35.8
    \item 0 $\leq f_\mathrm{Si} \leq$ 0.7
    \item 0.7 $\leq f_\mathrm{Fe} \leq$ 7.5
\end{itemize}

It can be seen that the amount of silicon required is negligible for all configurations and its absence is not likely to affect the results significantly.

What may be expected from the constant-fraction combined fits using the older Dominguez11 IR model is an increase in emitted helium and a decrease in proton with a hard emission spectrum---though this result could simply be due to fitting uncertainties~\citep{PierreAuger:2016use, PierreAuger:2022atd}. A more thorough comparison done in~\cite{AlvesBatista:2015jem} with soft injection iron nuclei shows a smaller variation on the energy spectrum and a maximum 10\% difference on \lnA at an energy of E = 10$^{19.6}$~eV.

\subsubsection{Multimessenger Photons and Neutrinos}

\begin{figure*}[t]
    \centering
    \subfloat[Subfigure 1][]{
    \includegraphics[width=.45\textwidth]{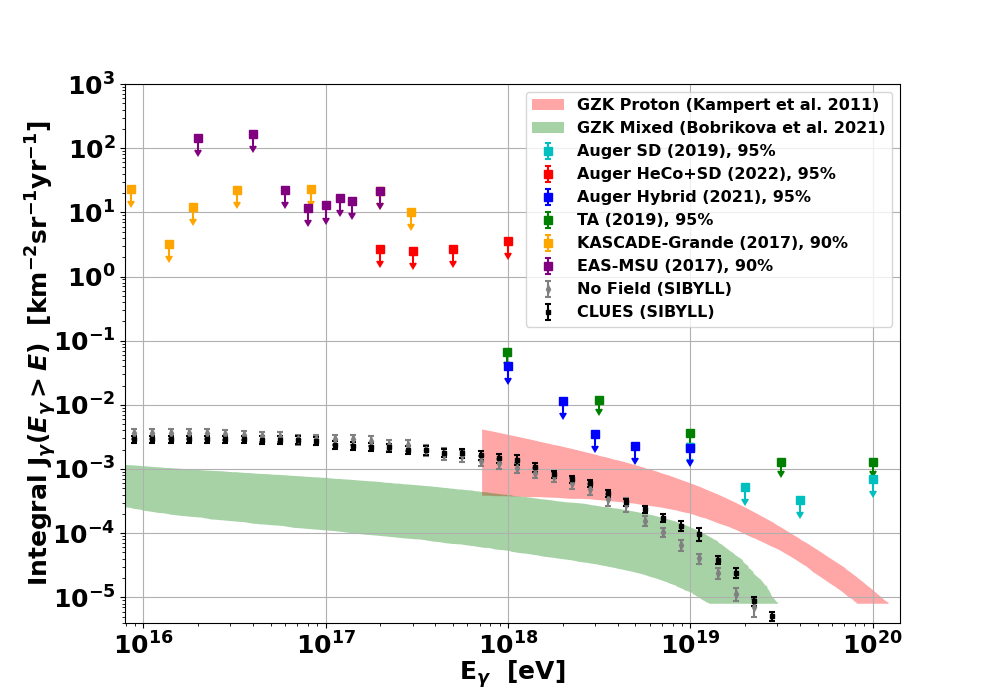}
    \label{fig:photon}}
    \subfloat[Subfigure 2][]{
    \includegraphics[width=.45\textwidth]{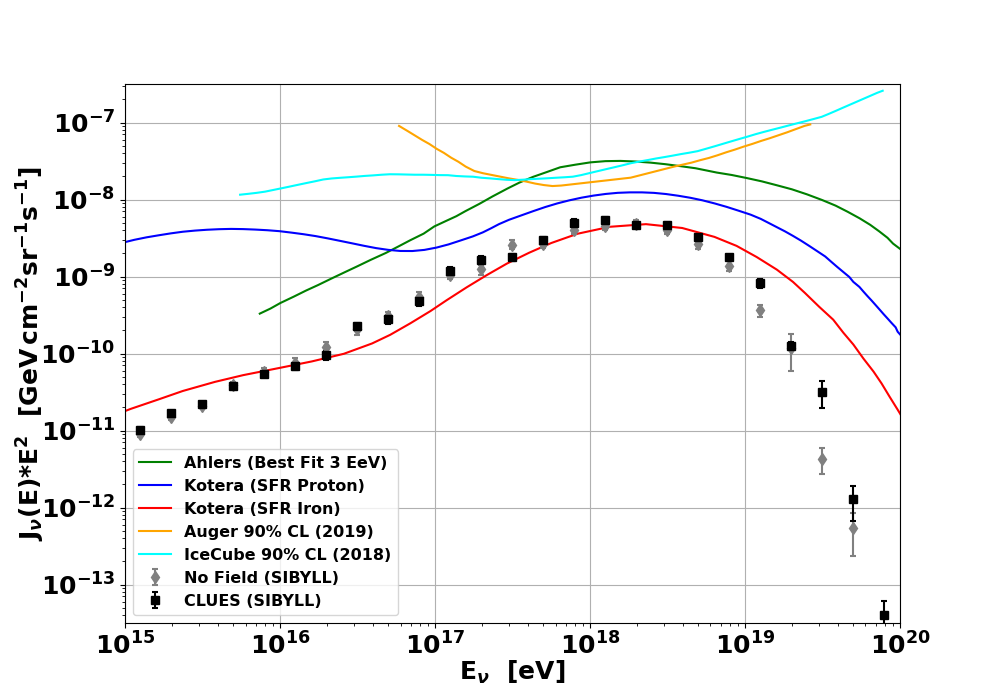}
    \label{fig:neutrino}}
    \caption{Constant-fraction fit cosmogenic photon and neutrino spectra for the best-fit CLUES-SIBYLL configuration (in black) compared to the no-field case (in grey). (a) Integral cosmogenic photon spectrum compared with two theoretical models and experimental upper limits as in~\cite{PierreAuger:2022uwd}. (b) All-flavor neutrino spectrum compared with three theoretical models and experimental upper limits~\citep{IceCube:2018fhm, PierreAuger:2019ens, 2010JCAP...10..013K}.}
    \label{fig:multimessenger}
\end{figure*}

The constant-fraction fit cosmogenic integral photon and all-flavor neutrino spectra are shown in Figures~\ref{fig:photon} and~\ref{fig:neutrino}, respectively, for the best-fit configuration (compared to the no-field scenario)---the CLUES structured magnetic field and \Sibyll-based mass-composition. Although the \Sibyll~hadronic interaction model results in a heavier reconstructed mass for data compared to the other models, it predicts a 92\% emission of light nuclei (protons and helium) from FR0 sources propagated through the CLUES magnetic field. At lower energies, the integral photon flux aligns closely with a theoretical prediction for a pure proton cosmogenic source, while at higher energies, it transitions into a range predicted for a mixed composition.

\begin{figure*}[t]
    \centering
    \subfloat[Subfigure 1][]{
    \includegraphics[width=.45\textwidth]{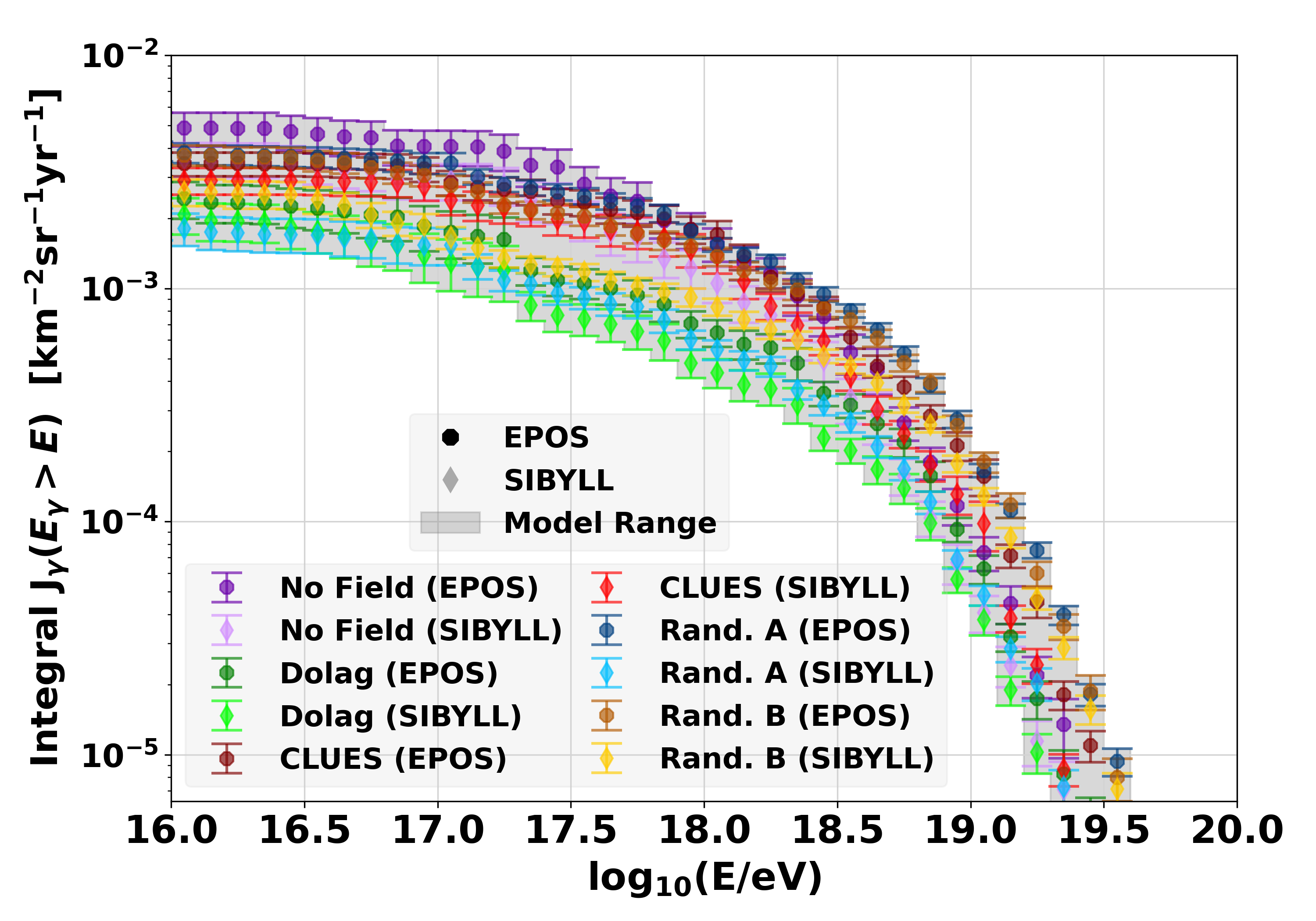}
    \label{fig:allphoton}}
    \subfloat[Subfigure 2][]{
    \includegraphics[width=.45\textwidth]{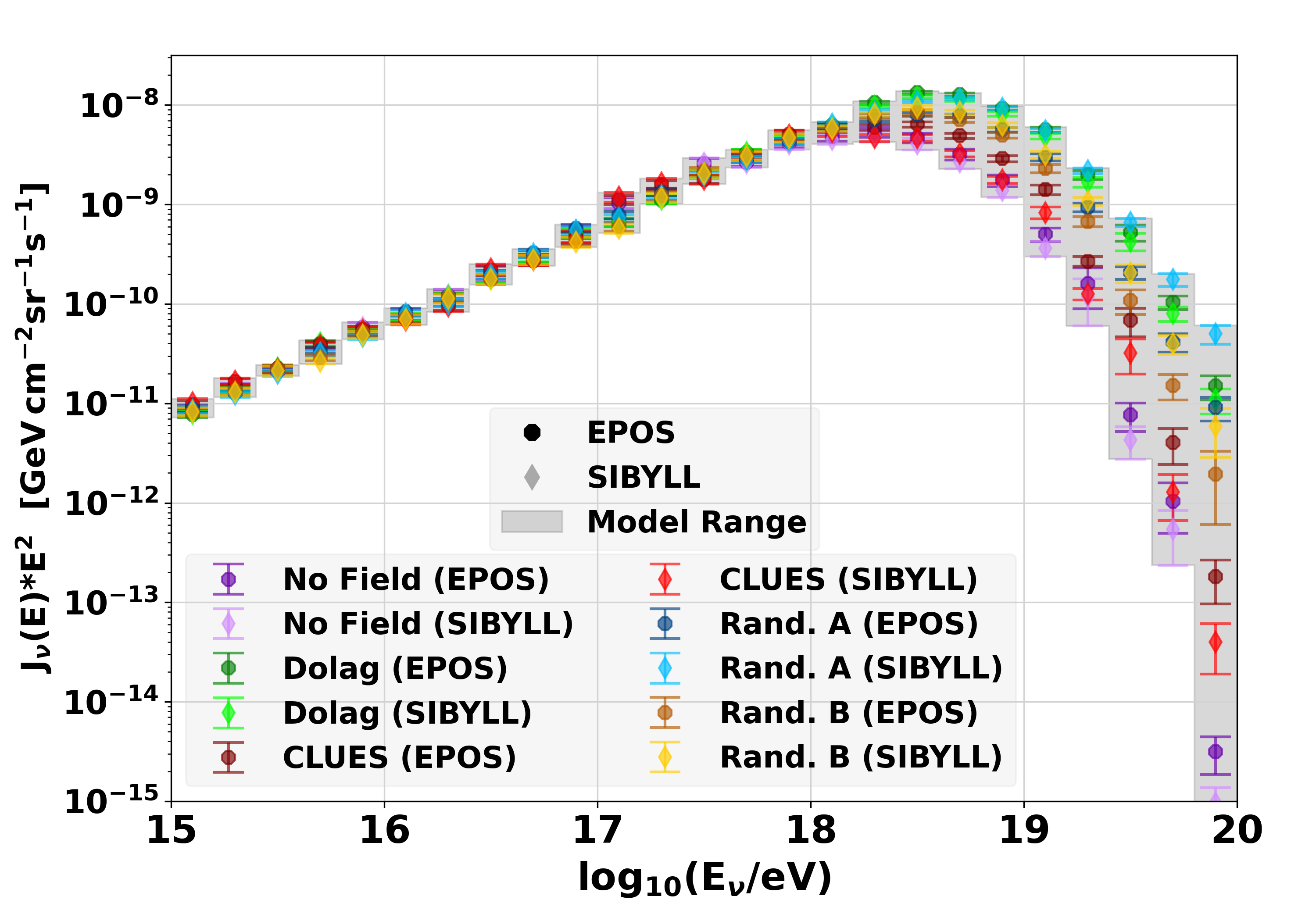}
    \label{fig:allneutrino}}
    \caption{Constant-fraction fit cosmogenic photon and neutrino spectra for the ten configurations based on the EPOS-LHC (EPOS) and \Sibyll{} (SIBYLL) hadronic interaction models and all five tested magnetic fields. Grey areas display the $\pm1\sigma$~bounds of all the simulation configurations. EPOS/SIBYLL fits are represented by circles/diamonds and darker/lighter colors. Each base color represents a magnetic field model. (a) Integral cosmogenic photon spectra using the axes of Figure~\ref{fig:photon}. (b) All-flavor neutrino spectra multiplied by E$^2$ using the axes of Figure~\ref{fig:neutrino}.}
    \label{fig:multi}
\end{figure*}

Notably, the neutrino flux resembles predictions for a pure iron source, a result possibly influenced by the simulation's constraint to relatively nearby FR0 sources ($z$~$\leq$~0.2). However, although the previous FR0 combined-fit results, limited to sources with $z$~$\leq$~0.05, show fewer low-energy neutrinos and more high-energy neutrinos, the differences are not significant~\citep{Lundquist:20233x}. Considering the broad range of theoretical predictions and current experimental upper limits, the simulated FR0 fluxes falls within a reasonable range. Overall, in both photon and neutrino spectra, the presence of a magnetic field generally leads to increased fluxes, with this effect being most pronounced at the highest energies, as shown in Figure~\ref{fig:multi}.

The total number of cosmogenic photons largely depends on the fraction of emitted protons. Looking at the total number of photons with energies E$>$10$^{16}$ eV, shown by the first bin of Figure~\ref{fig:allphoton}, the configuration with the highest photon count is the No-Field-EPOS which has the highest proton emission at 93.5\%. The lowest photons count is from Rand.A-SIBYLL, which has the lowest proton emission at 0\% (the next two lowest photon counts come from the Dolag configurations with 0\% proton emission). Without a magnetic field, the number of detected photons is approximately 55 times lower for iron than for protons, as the energy is divided among the 56 nucleons, reducing the energy available for photopion production. Additionally, given the same emission, an intergalactic magnetic field increases the number of detected photons. A proton emission with $\gamma$ = 1, without exponential cutoff or trajectory cutoff (just the raw simulation before fitting), results in approximately three times more photons in the Rand.B 1ng-3Mpc magnetic field than in the no-field case.

As with photons, the number of detected neutrinos increases with the magnetic field. Comparing no-field to the Rand.B 1ng-3Mpc case, protons produce 1.03 times more neutrinos, while iron produces 1.6 times more with the magnetic field. However, inversely from photons, the number of detected neutrinos increases with mass number: the no-field iron raw simulation results in approximately 1.3 times more neutrinos than the proton case.

In summary, all things being equal for a given source emission, turning on an intergalactic magnetic field increases the number of cosmogenic photons and neutrinos. Since we are fitting to data, each configuration's emission is different, leading to differences in composition and energy spectra. The energy spectrum is also important, as can most easily be seen in the no-field scenario, which has the softest emission spectrum and strongest exponential cutoff, resulting in the lowest photon and neutrino counts at the highest energies.

\subsection{Evolving-Fraction Fits}
\label{subsec:evolving_results}
\begin{figure*}[htb]
    \centering
    \subfloat[Subfigure 1][]{
    \includegraphics[width=.45\textwidth]{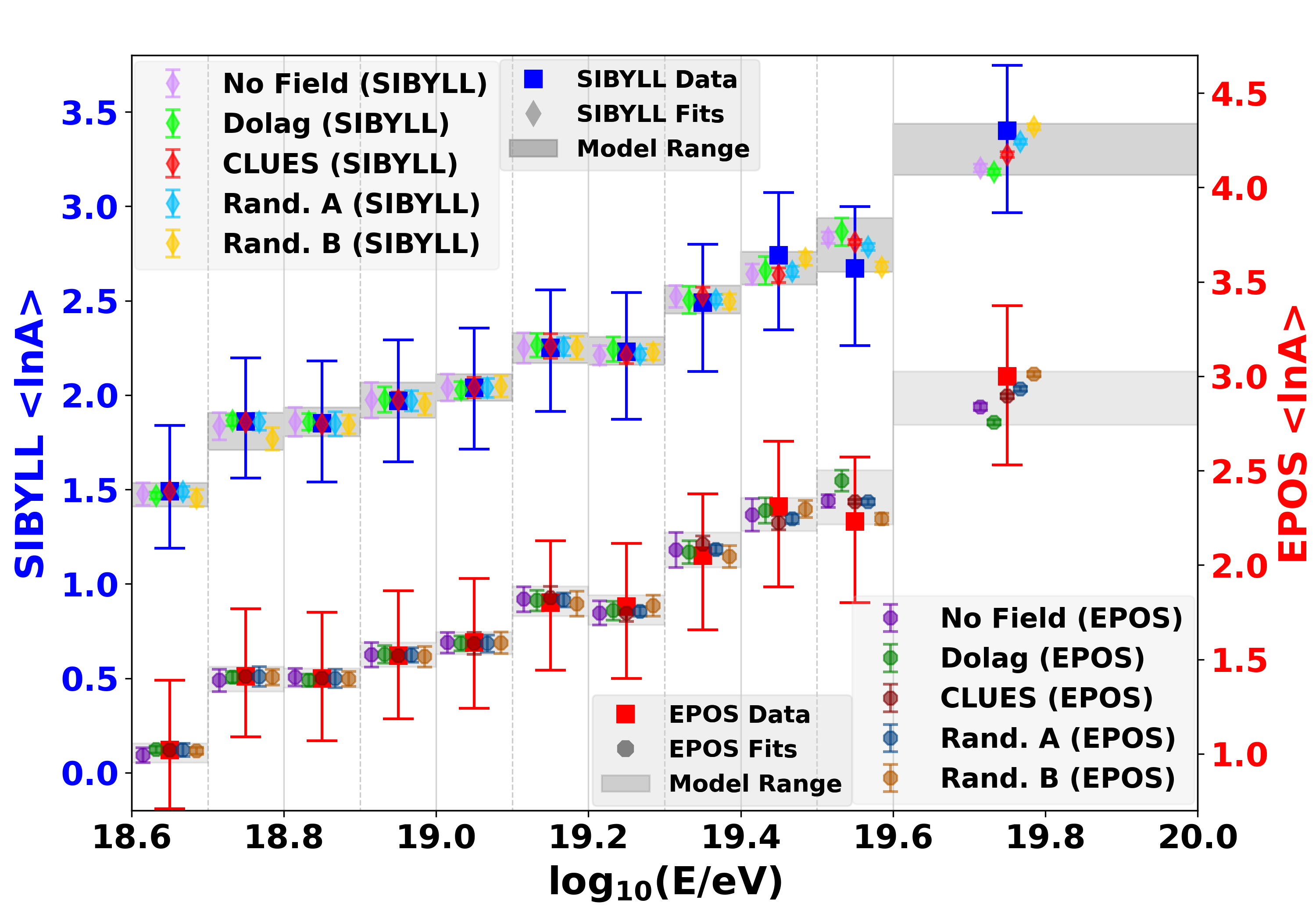}
    \label{fig:AmeanFit_evolved}}
    \subfloat[Subfigure 2][]{
    \includegraphics[width=.45\textwidth]{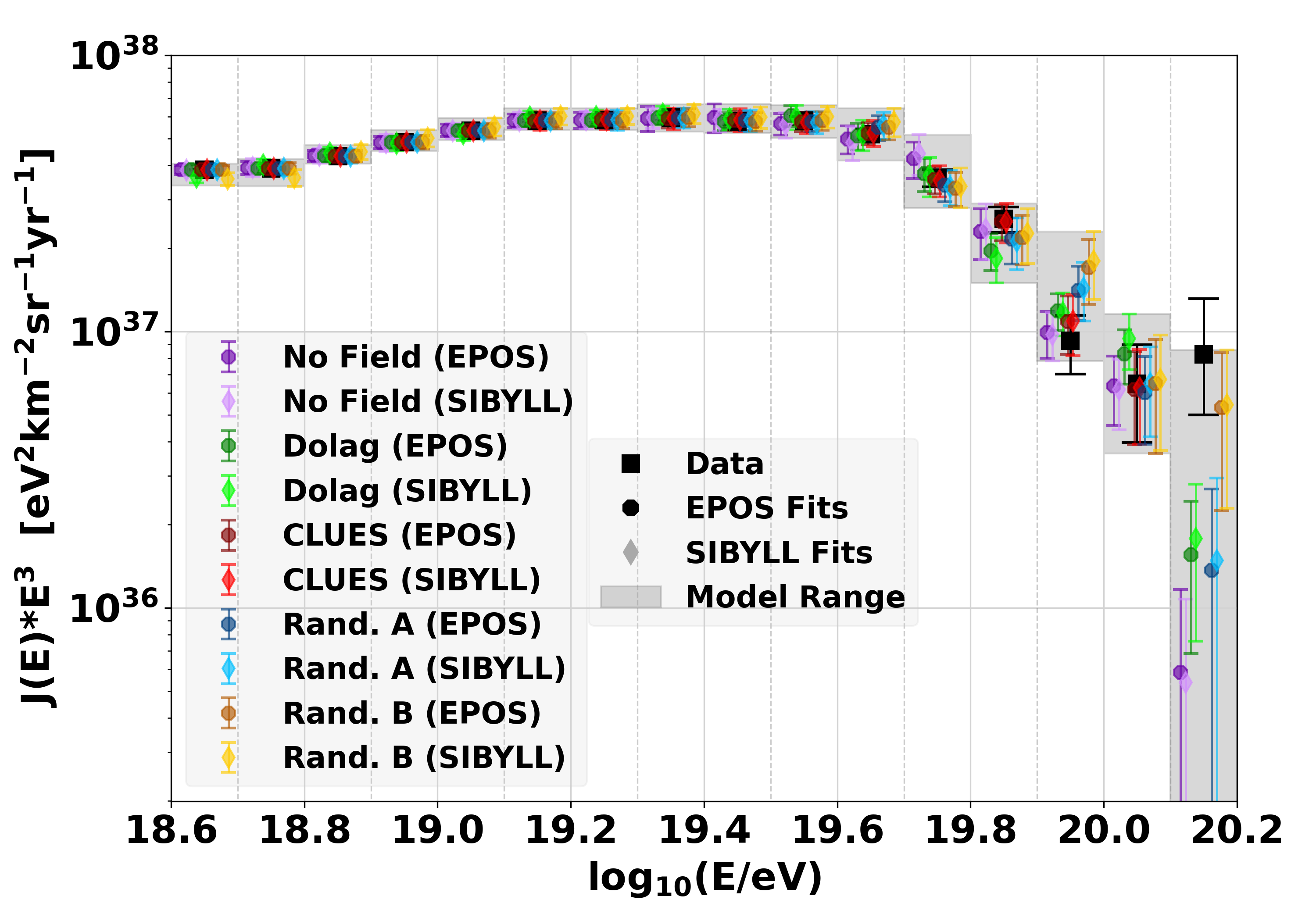}
    \label{fig:EspectFit_evolved}}
    \caption{The evolving-fraction combined-fit composition and energy-spectrum results for all simulated magnetic fields and two EAS models (EPOS-LHC (EPOS) and \Sibyll{} (SIBYLL)) compared to Auger data~\citep{Yushkov:2020nhr, Deligny:2020gzq}. Offsets are applied to simulations within each bin on the x-axis for improved visibility. Grey areas display the $\pm1\sigma$~bounds of all the simulation configurations. Two legends are included---the first emphasizes figure element shapes (data: square, dark circle/light diamond: EPOS/SIBYLL hadronic model fits, grey box: range), while the second includes marker colors for all fits, where each base color represents a magnetic field model. (a) Mean log mass number \lnA. The blue left y-axis and blue square markers represent SIBYLL, while the red right y-axis and red square markers represent EPOS, each with adjusted limits for better visibility. (b) Energy spectra multiplied by E$^3$ for visibility.}
    \label{fig:Fits_evolved}
\end{figure*}

The evolving-fraction combined-fit results for all five simulated intergalactic media and two EAS models \\(EPOS-LHC and \Sibyll) compared to Auger data~\citep{Yushkov:2020nhr, Deligny:2020gzq} are shown in Figure~\ref{fig:Fits_evolved}. The \QGSJet{} model was excluded from the evolving-fraction analysis. The mean log mass number \lnA shown in Figure~\ref{fig:AmeanFit_evolved} illustrates that the inclusion of 44 nuclei fractions leads to a nearly perfect fit across all configurations due to the large number of parameters.

However, similar to the constant-fraction fits 1~nG random fields results having a better fit to the \lnA lower energy bins, these fields are the best fit to the two highest-energy heavier bins with the evolving fractions. As with the constant-fraction fits only the Rand.B configurations sufficiently account for the highest-energy bin of the energy spectra as shown in Figure~\ref{fig:EspectFit_evolved}. Continuing the trend observed in constant-fraction fits, the superior performance of the longer correlation length 1~nG (Rand.B) field at the highest energies remains surprising, challenging the assumption that FR0s are minor contributors at extreme energies. And again, the CLUES model essentially does not contribute to the highest energy bin.

The best-fit parameters are presented in Table~\ref{tab:params_evolved} and Figure~\ref{fig:nuclei_evolved}, along with corresponding~1$\sigma$ equivalent confidence intervals around the best fit. Table~\ref{tab:paramsKDE_evolved} shows the most-probable value of the bootstrap sample distributions using an four-dimensional multivariate kernel density estimation (KDE) and the 1$\sigma$ standard deviations of the KDEs. A 48-dimensional KDE estimation has not been done due to time and computing constraints and the 4D KDE should be a reasonable approximation.

For each EAS model and magnetic field, Table~\ref{tab:params_evolved} and Table~\ref{tab:paramsKDE_evolved} show the total $\Sigma \chi^2/\text{bin}$, emission spectral index~($\gamma$), exponential rigidity-dependent cutoff~(log$_{10}$(R$_{\text{cut}}$/V)), particle trajectory cutoff~(D$_{\text{cut}}$), and relative energy-spectrum normalization~($n$). Figure~\ref{fig:nuclei_evolved} shows the fitted FR0 emission elemental fractions as a function of energy for all 10 configurations.

As indicated in Table~\ref{tab:params_evolved}, similar to findings from constant-fraction fits, the CLUES, Rand.A, and Rand.B magnetic fields consistently deliver the best fit results. The Rand.B field, in particular, does stand out as the optimal fit for both the EPOS-LHC and \Sibyll{} models, with the Rand.-B-SIBYLL configuration achieving the overall best fit (Ra\-nd.B-EPOS is a very close second). EPOS-LHC is the best fit for all magnetic field models except Rand.B-EPOS but the difference from SIBYLL is very small. The second-best magnetic field after Rand.B is CLUES. 

\subsubsection{Energy-Spectrum Parameters}

Overall, the evolving-fraction fit energy-spectrum parameters of Table~\ref{tab:params_evolved} and Table~\ref{tab:paramsKDE_evolved} again indicate that both magnetic field strength \Bave, and RMS \Brms, have an impact on the expected cosmic-ray source emission energy spectrum. Generally, the emission spectral index, E$^{-\gamma}$, increases with the magnetic field strength \Bave. However, the exception is the Rand.B field model where this trend does not apply. The exponential rigidity-dependent cutoff, R$_{\text{cut}}$, generally rises with an increase in \Brms, except in the case of the Rand.A field model. Furthermore, for smaller magnetic fields, R$_{\text{cut}}$ decreases with a heavier composition, while for larger fields, this trend is reversed, demonstrating the complex interaction between magnetic field strength and particle composition. The trajectory cutoff, D$_{\text{cut}}$, is fairly stable across most fields, with the notable exception of the Dolag field which has a significantly higher \Brms of 11~nG.

\renewcommand{\arraystretch}{1.2}
\begin{table*}[htbp]
\centering
\caption*{Evolving-Fraction Energy-Spectrum Parameters}
\small
\rotatebox{0}{
\begin{tabular}{!{\vrule width 1.5pt} c | c | c | c | c | c | c !{\vrule width 1.5pt}}
\thickhline
\textbf{Field} & \textbf{Model} & \textbf{$\mathbf{\Sigma\chi^2/\text{bin}}$} &  \textbf{$\mathbf{\gamma}$} & \textbf{log$\mathbf{_{10}}$(R$\mathbf{_{cut}}$/V)} & \textbf{D$\mathbf{_{cut}}$/Mpc} & \textbf{$\mathbf{n}$}\\
\thickhline
\multirow{2}{*}{No Field} 
& SIBYLL & 0.370 & 2.10$^{+0.47}_{-0.14}$ & 19.19$^{+0.84}_{-0.20}$ & 844$^{+0}_{-1}$ & 1.344$^{+0.014}_{-0.005}$\\
\cline{2-7}
& EPOS & 0.326 & 1.94$^{+0.60}_{-0}$ & 19.22$^{+0.75}_{-0.01}$ & 844$^{+0}_{-1}$ & 1.343$^{+0.006}_{-0.005}$\\
\thickhline
\multirow{2}{*}{Dolag} 
& SIBYLL & 0.469 & 2.28$^{+0.14}_{-0.84}$ & 19.89$^{+0.31}_{-0.98}$ & 907$^{+450}_{-50}$ & 1.328$^{+0.023}_{-0.000}$\\
\cline{2-7}
& EPOS & 0.446 & 2.31$^{+0.12}_{-0.20}$ & 19.89$^{+0.40}_{-0.38}$ & 889$^{+146}_{-39}$ & 1.342$^{+0.006}_{-0.011}$\\
\thickhline
\multirow{2}{*}{CLUES} 
& SIBYLL & 0.307 & 2.65$^{+0.00}_{-0.29}$ & 19.58$^{+0.27}_{-0.25}$ & 841$^{+1}_{-0}$ & 1.342$^{+0.013}_{-0.007}$\\
\cline{2-7}
& EPOS & 0.295 & 2.52$^{+0.09}_{-0.22}$ & 19.58$^{+0.51}_{-0.20}$ & 842$^{+0}_{-1}$ & 1.342$^{+0.006}_{-0.004}$\\
\thickhline
\multirow{2}{*}{Rand.A}
& SIBYLL & 0.344 & 2.64$^{+0.20}_{-0.25}$ & 19.95$^{+0.50}_{-0.56}$ & 855$^{+93}_{-7}$ & 1.343$^{+0.008}_{-0.005}$\\
\cline{2-7}
& EPOS & 0.341 & 2.65$^{+0.05}_{-0.30}$ & 19.89$^{+0.94}_{-0.32}$ & 844$^{+113}_{-0}$ & 1.342$^{+0.006}_{-0.006}$\\
\thickhline
\multirow{2}{*}{Rand.B}
& SIBYLL & 0.236 & 2.35$^{+0.34}_{-0.30}$ & 19.40$^{+1.64}_{-0.32}$ & 843$^{+171}_{-0}$ & 1.346$^{+0.009}_{-0.006}$\\
\cline{2-7}
& EPOS & 0.237 & 2.21$^{+0.34}_{-0.45}$ & 19.51$^{+1.20}_{-0.38}$ & 843$^{+174}_{-0}$ & 1.342$^{+0.004}_{-0.007}$\\
\thickhline
\end{tabular}
}
\caption{The FR0 evolving-fraction combined-fit results total sum chi-square per bin ($\Sigma\chi^2/\textrm{bin}$), spectral index ($\gamma$), exponential rigidity cutoff (log$_{10}$(R$_{\text{cut}}$/V)), trajectory cutoff (D$_{\text{cut}}$), and spectrum normalization (n) for all 10 configurations used in the fit, including the EPOS-LHC (EPOS) and \Sibyll{} (SIBYLL) hadronic interaction models and all five magnetic fields.}
\label{tab:params_evolved}
\end{table*}

Despite the no-field and Dolag configurations consistently showing poorer fits, similar to the constant-fraction scenario, the inclusion of a large number of parameters in the evolving-fraction fits leads to excellent overall fit quality for all configurations. Therefore, the bounds found for the best fit of these parameters from Table~\ref{tab:params}, using all configurations, are:

\begin{itemize}
    \item 1.94 $\leq \gamma \leq$ 2.65
    \item 19.19 $\leq$ log$_{10}$(R$_{\text{cut}}$/V) $\leq$ 19.95
    \item 841 $\leq$ D$_{\text{cut}}$/Mpc $\leq$ 907
    \item 1.328 $\leq n \leq$ 1.346
\end{itemize}

Note, that these bounds are larger than the constant-fraction fit bounds due the inclusion of the no-field and Dolag configurations.

\renewcommand{\arraystretch}{1.2}
\begin{table*}[htb]
\centering
\caption*{Evolving-Fraction Bootstrap Energy-Spectrum Parameters}
\small
\rotatebox{0}{
\begin{tabular}{!{\vrule width 1.5pt} c | c | c | c | c | c | c !{\vrule width 1.5pt}}
\thickhline
\textbf{Field} & \textbf{Model} & \textbf{$\mathbf{\Sigma\chi^2/\text{bin}}$} &  \textbf{$\mathbf{\gamma}$} & \textbf{log$\mathbf{_{10}}$(R$\mathbf{_{cut}/V}$)} & \textbf{D$\mathbf{_{cut}}$/Mpc} & \textbf{$\mathbf{n}$}\\
\thickhline
\multirow{2}{*}{No Field} 
& SIBYLL & 0.370 & 2.44$^{+0.23}_{-0.23}$ & 19.34$^{+0.25}_{-0.25}$ & 843$^{+0}_{-0}$ & 1.345$^{+0.006}_{-0.006}$\\
\cline{2-7}
& EPOS & 0.326 & 2.47$^{+0.12}_{-0.12}$ & 19.40$^{+0.10}_{-0.10}$ & 843$^{+0}_{0}$ & 1.340$^{+0.004}_{-0.004}$\\
\thickhline
\multirow{2}{*}{Dolag} 
& SIBYLL & 0.469 & 2.26$^{+0.28}_{-0.28}$ & 19.23$^{+0.25}_{-0.25}$ & 867$^{+193}_{-193}$ & 1.335$^{+0.004}_{-0.004}$\\
\cline{2-7}
& EPOS & 0.446 & 2.26$^{+0.19}_{-0.19}$ & 19.74$^{+0.18}_{-0.18}$ & 890$^{+107}_{-107}$ & 1.341$^{+0.003}_{-0.003}$\\
\thickhline
\multirow{2}{*}{CLUES} 
& SIBYLL & 0.307 & 2.40$^{+0.07}_{-0.07}$ & 19.71$^{+0.12}_{-0.12}$ & 842$^{+0}_{-0}$ & 1.347$^{+0.013}_{-0.007}$\\
\cline{2-7}
& EPOS & 0.295 & 2.34$^{+0.06}_{-0.06}$ & 19.68$^{+0.10}_{-0.10}$ & 842$^{+0}_{-0}$ & 1.341$^{+0.005}_{-0.005}$\\
\thickhline
\multirow{2}{*}{Rand.A}
& SIBYLL & 0.344 & 2.37$^{+0.05}_{-0.05}$ & 19.83$^{+0.11}_{-0.11}$ & 853$^{+35}_{-35}$ & 1.342$^{+0.003}_{-0.003}$\\
\cline{2-7}
& EPOS & 0.341 & 2.33$^{+0.05}_{-0.05}$ & 19.68$^{+0.08}_{-0.08}$ & 851$^{+30}_{-30}$ & 1.342$^{+0.003}_{-0.003}$\\
\thickhline
\multirow{2}{*}{Rand.B}
& SIBYLL & 0.236 & 2.41$^{+0.10}_{-0.10}$ & 19.74$^{+0.41}_{-0.41}$ & 850$^{+43}_{-43}$ & 1.348$^{+0.004}_{-0.004}$\\
\cline{2-7}
& EPOS & 0.237 & 2.31$^{+0.11}_{-0.11}$ & 19.59$^{+0.20}_{-0.20}$ & 853$^{+49}_{-49}$ & 1.345$^{+0.004}_{-0.004}$\\
\thickhline
\end{tabular}
}
\caption{The FR0 evolving-fraction combined-fit bootstrap distribution most-probable spectral index ($\gamma$), exponential rigidity cutoff (log$_{10}$(R$_{\text{cut}}$/V)), trajectory cutoff (D$_{\text{cut}}$), and spectrum normalization (n) for all 10 configurations used in the fit, including the EPOS-LHC (EPOS) and \Sibyll{} (SIBYLL) hadronic interaction models and all five magnetic fields. The best-fit results total sum chi-square per bin ($\Sigma\chi^2/\textrm{bin}$) from Table~\ref{tab:params_evolved} is also listed.}
\label{tab:paramsKDE_evolved}
\end{table*}

For the bootstrap KDE fit most-probable parameters from Table~\ref{tab:paramsKDE_evolved} the ranges found are:

\begin{itemize}
    \item 2.26 $\leq \gamma \leq$ 2.47
    \item 19.34 $\leq$ log$_{10}$(R$_{\text{cut}}$/V) $\leq$ 19.83
    \item 842 $\leq$ D$_{\text{cut}}$/Mpc $\leq$ 890
    \item 1.335 $\leq n \leq$ 1.348
\end{itemize}

Compared to the recent Auger result of~\cite{PierreAuger:2016use} these parameter values are similar to the reported second minima of their models and the low-$z$ source distance model ($z$~$<$~0.02).

\subsubsection{UHECR Composition Parameters}
\begin{figure*}[hp]
    \centering
    \subfloat[Subfigure 1][]{
    \includegraphics[width=0.45\textwidth]{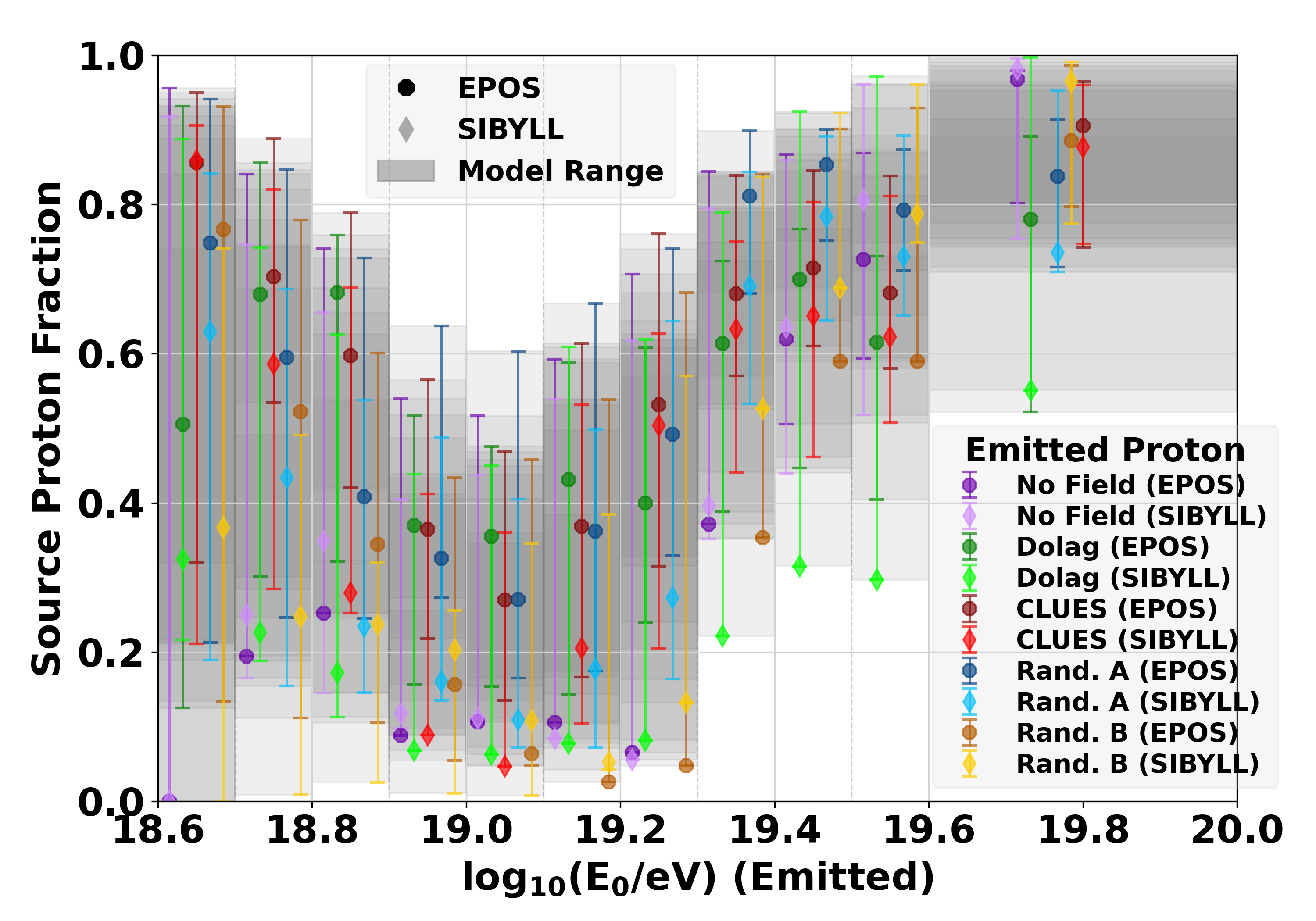}
    \label{fig:H_evolved}}
    \subfloat[Subfigure 2][]{
    \includegraphics[width=0.45\textwidth]{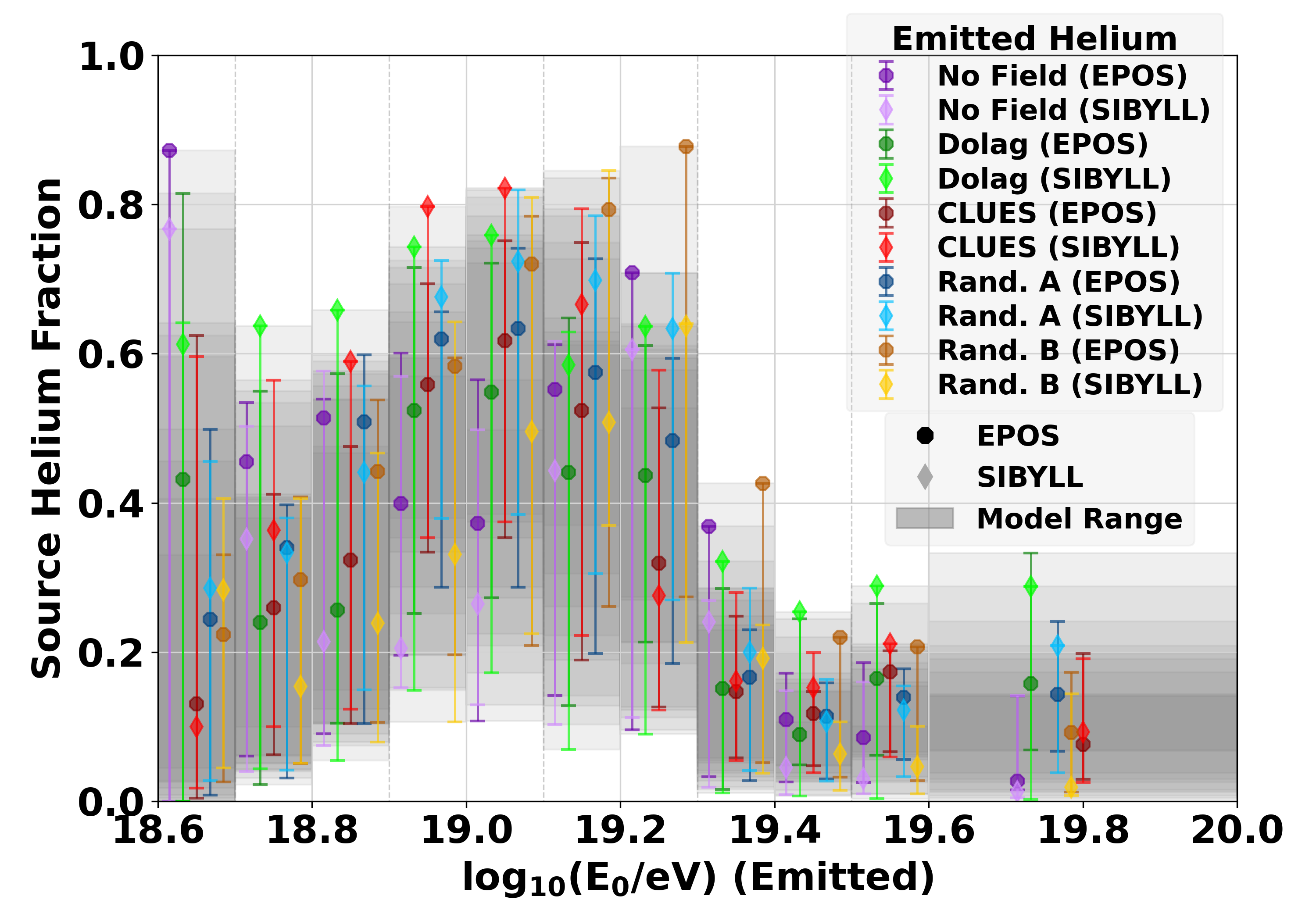}
    \label{fig:He_evolved}}\\
    \subfloat[Subfigure 3][]{
    \includegraphics[width=0.45\textwidth]{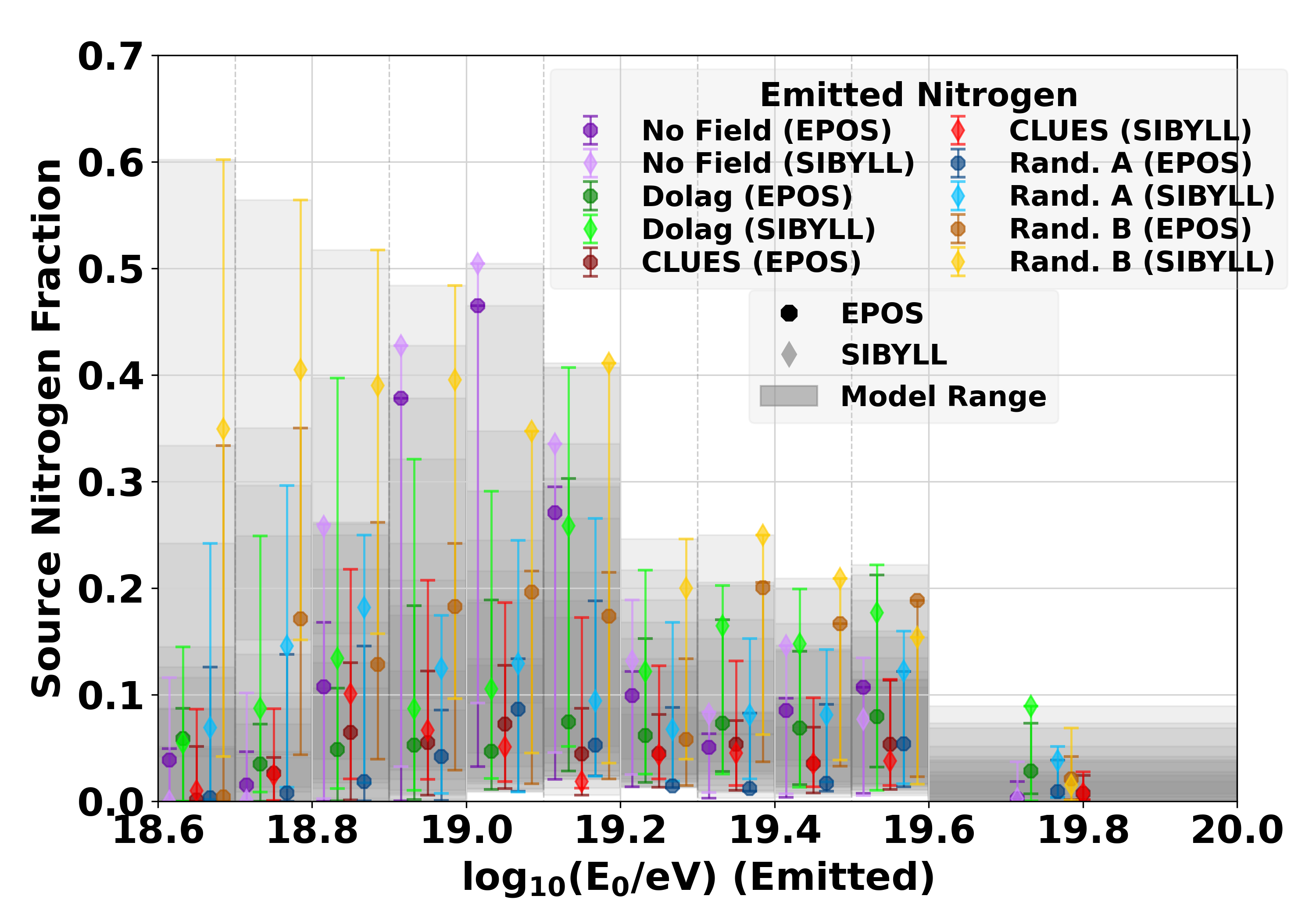}
    \label{fig:N_evolved}}
    \subfloat[Subfigure 4][]{
    \includegraphics[width=0.45\textwidth]{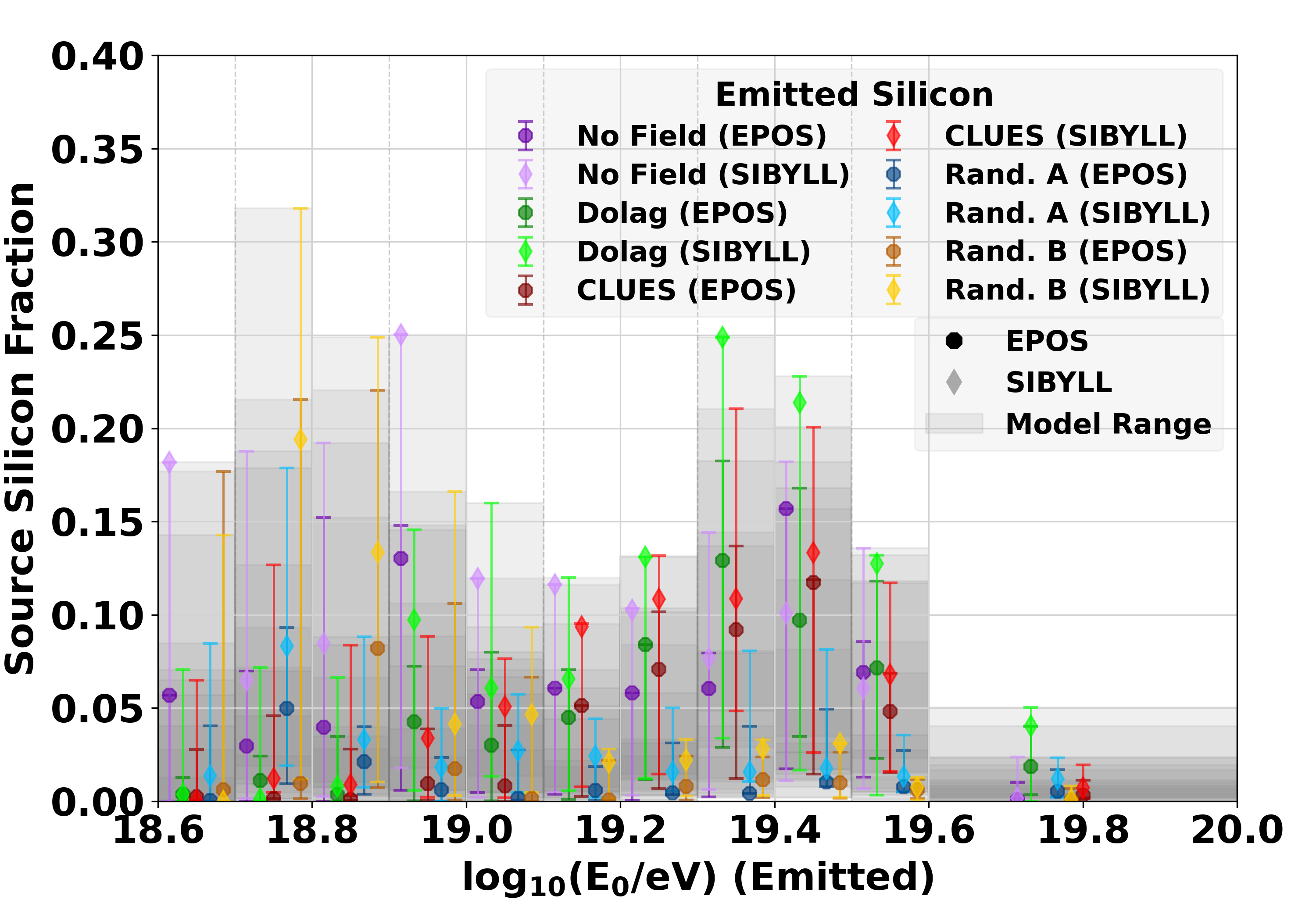}
    \label{fig:Si_evolved}}\\
    \subfloat[Subfigure 5][]{
    \includegraphics[width=0.45\textwidth]{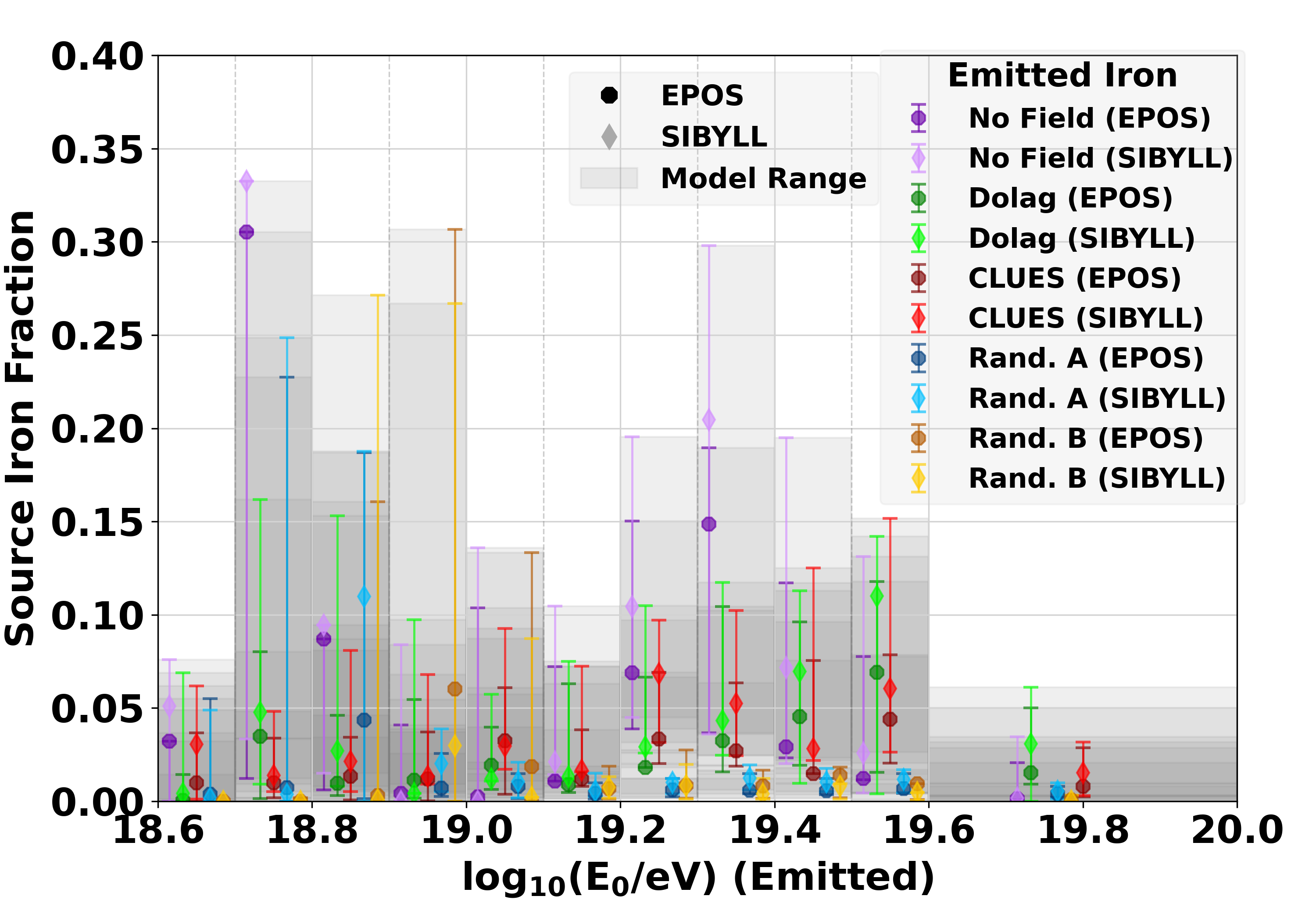}
    \label{fig:Fe_evolved}}
    \subfloat[Subfigure 6][]{
    \includegraphics[width=0.45\textwidth]{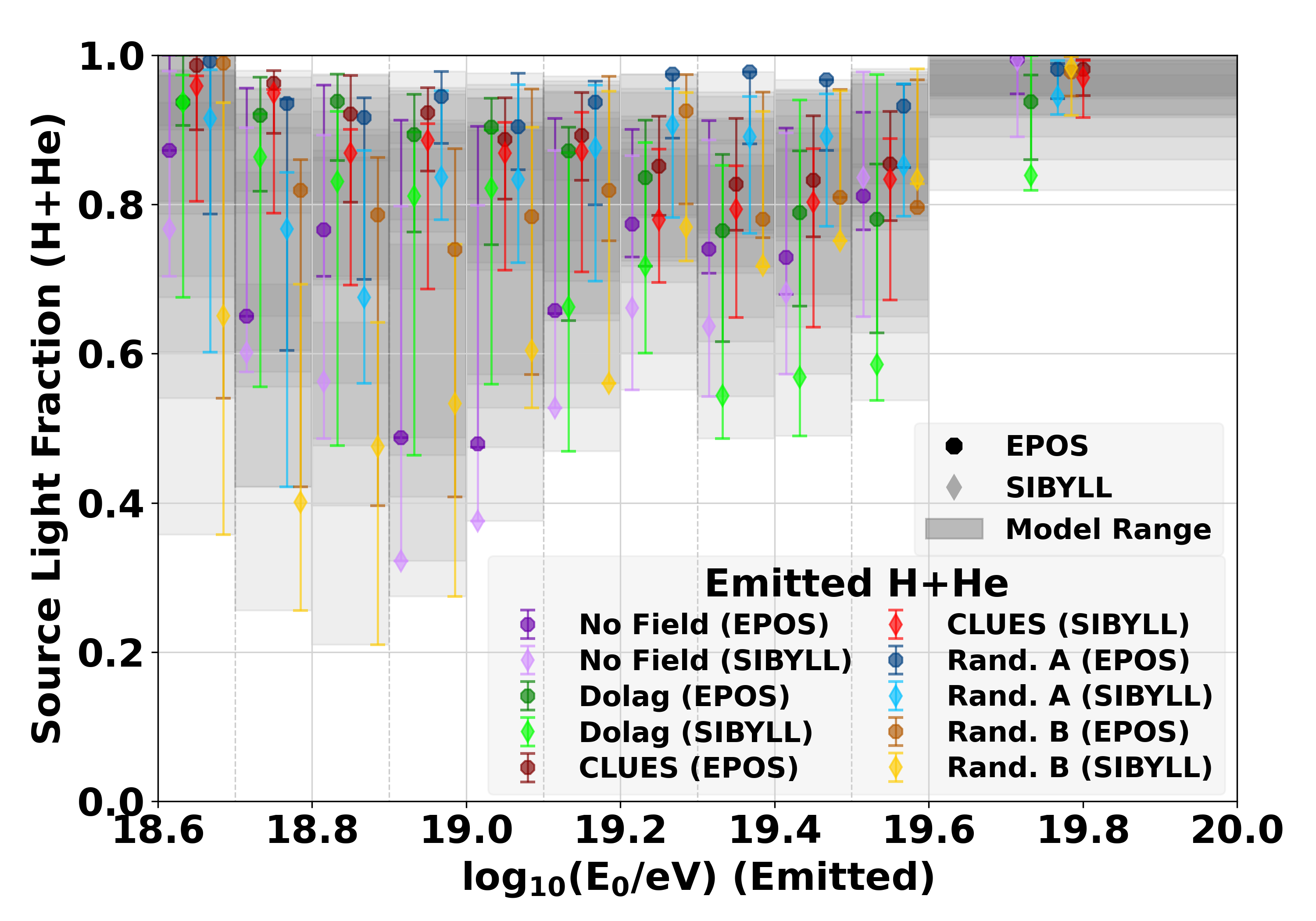}
    \label{fig:Light_evolved}}
    \caption{The evolving-fraction fit FR0 emitted nuclei fractions versus emitted energy necessary to best fit the data for the ten configurations based on the EPOS-LHC (EPOS) and \Sibyll{} (SIBYLL) hadronic interaction models and all five magnetic fields. Additionally, the lightest components---proton (hydrogen) plus helium are shown in (f). Offsets are applied to simulations within each bin on the x-axis for improved visibility. Grey bands extend from $\pm1\sigma$ for each configuration and is darkest where the ranges overlap. EPOS/SIBYLL fits are represented by circles/diamonds and darker/lighter colors. Each base color represents a magnetic field model.}
    \label{fig:nuclei_evolved}
\end{figure*}

The FR0 emission fractions for the evolving energy-dependent (observed) fraction fit of proton, helium, nitrogen, silicon, and iron emitted nuclei are shown in Figure~\ref{fig:nuclei_evolved}. Similar to the constant-fraction fits, the emission fractions of protons and helium exhibit instability, with dominance of each varying due to propagation effects, as shown in Figure~\ref{fig:Sim_lnA}. The combined fraction of light components ($f_\mathrm{H} + f_{He}$) demonstrates greater stability across different configurations as shown in Figure~\ref{fig:Light_evolved}. Interestingly, for all configurations the fraction of emitted protons increases with energy above 10$^{19}$ eV as shown in Figure~\ref{fig:H_evolved}.

These results suggest that the magnetic field strength can significantly impact the expected composition of the source-emitted cosmic rays, influencing the relative fractions of each element. With increasing magnetic field strength, the average emission percentages of protons and helium remain relatively stable. In contrast, the average fraction of nitrogen nuclei tends to increase, while heavier nuclei such as silicon and iron tend to decrease.

The ranges found for the energy-bin average FR0-emiss\-ion best-fit elemental fractions (in~\%) from Figure~\ref{fig:nuclei_evolved} are:

\begin{itemize}
    \item 63.3 $\leq f_\mathrm{H} + f_{\mathrm{He}} \leq$ 95.0
    \item 21.8 $\leq f_\mathrm{H} \leq$ 60.7
    \item 27.0 $\leq f_{\mathrm{He}} \leq$ 52.6
    \item 2.9 $\leq f_\mathrm{N} \leq$ 28.4
    \item 1.1 $\leq f_\mathrm{Si} \leq$ 10.5
    \item 0.6 $\leq f_\mathrm{Fe} \leq$ 6.4
\end{itemize}

The observed ranges for average FR0 emission fractions across energy bins indicate that lighter elements like hydrogen and helium dominate, whereas it can be seen that the amount of heavier elements like iron contribute minimally for all configurations, and their absence is not likely to affect the results significantly.

As with the constant-fraction fits Section~\ref{Constant_Comp}, using the older Dominguez11 IR model may result in a small increase in emitted helium and a decrease in proton~\citep{PierreAuger:2016use, AlvesBatista:2015jem, PierreAuger:2022atd}.

\subsubsection{Multimessenger Photons and Neutrinos}
\begin{figure*}[hb]
    \centering
    \subfloat[Subfigure 1][]{
    \includegraphics[width=.45\textwidth]{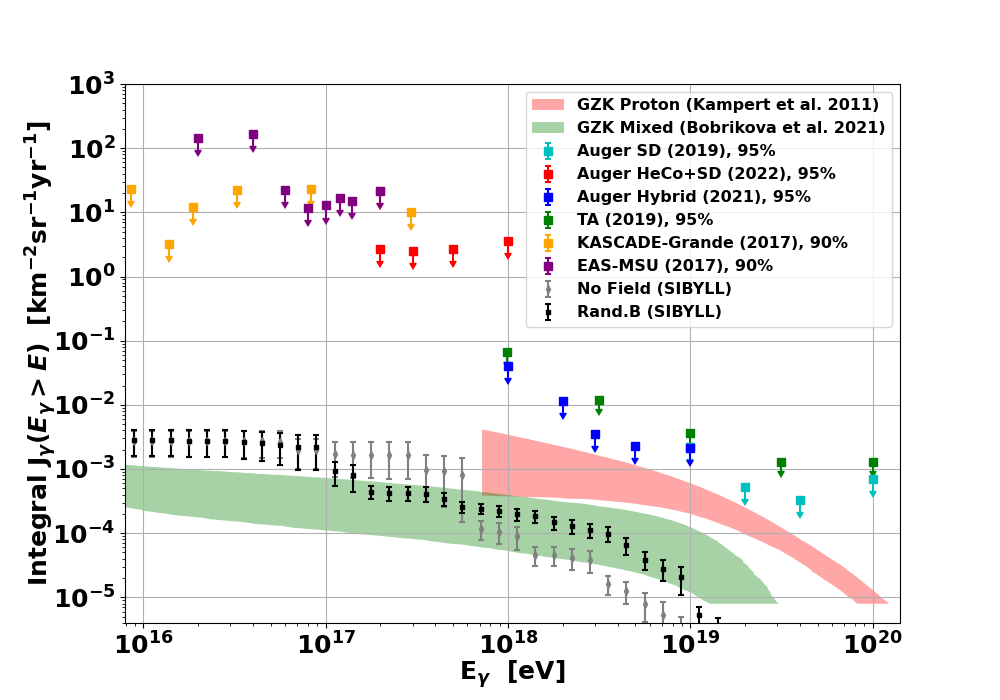}
    \label{fig:photon_evolved}}
    \subfloat[Subfigure 2][]{
    \includegraphics[width=.45\textwidth]{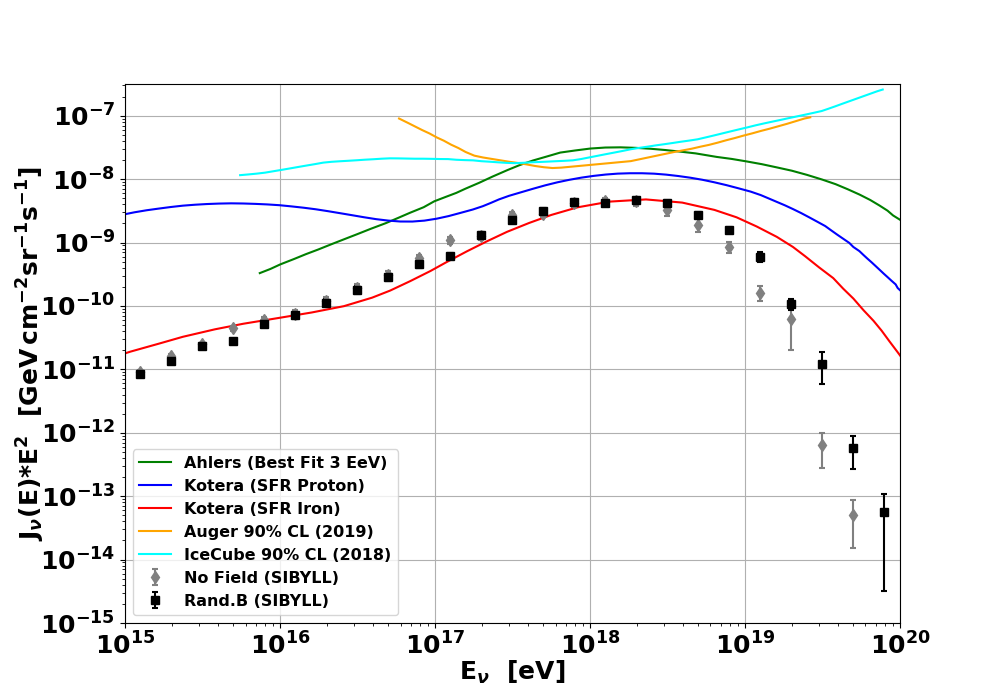}
    \label{fig:neutrino_evolved}}
    \caption{Evolving-fraction fit cosmogenic photon and neutrino spectra for the best-fit Rand.B-SIBYLL configuration (in black) compared to the no-field case (in grey). (a) Integral cosmogenic photon spectrum compared with two theoretical models and experimental upper limits as in~\cite{PierreAuger:2022uwd}. (b) All-flavor neutrino spectrum compared with three theoretical models and experimental upper limits~\citep{IceCube:2018fhm, PierreAuger:2019ens, 2010JCAP...10..013K}.}
    \label{fig:multimessenger_evolved}
\end{figure*}
The evolving-fraction fit cosmogenic integral photon and all-flavor neutrino spectra are shown in Figures~\ref{fig:photon_evolved} and~\ref{fig:neutrino_evolved}. These results are for the best-fit configuration using the Rand.B 1 nG random magnetic field and \Sibyll-based mass-composition, compared to the no-field scenario. Although the \Sibyll~hadronic interaction model reconstructs a heavier mass for data relative to other models, the evolving-fraction fit still predicts that approximately 66\% of the FR0 emission consists of light nuclei. 

\begin{figure*}[ht]
    \centering
    \subfloat[Subfigure 1][]{
    \includegraphics[width=.45\textwidth]{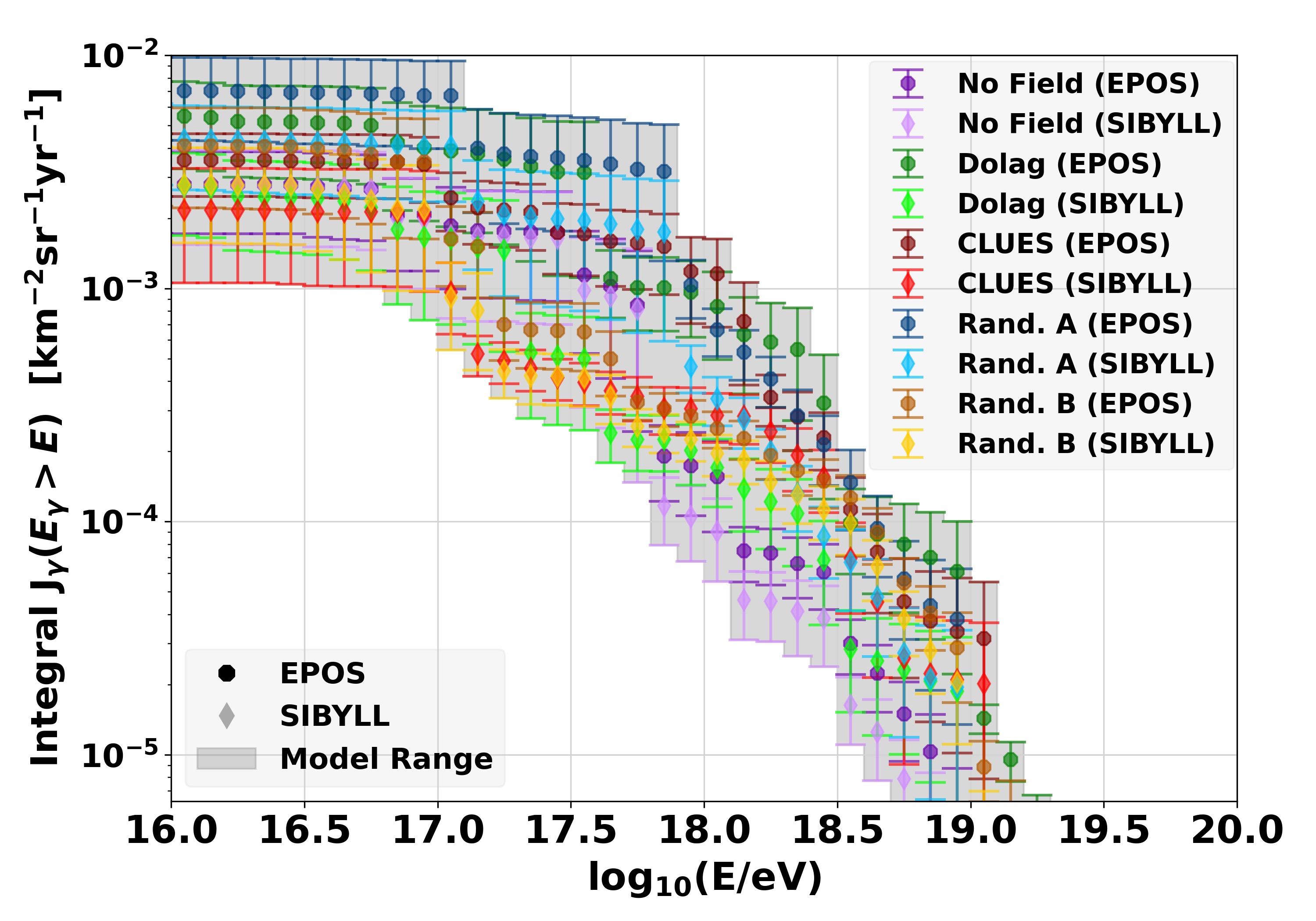}
    \label{fig:allphoton_evolved}}
    \subfloat[Subfigure 2][]{
    \includegraphics[width=.45\textwidth]{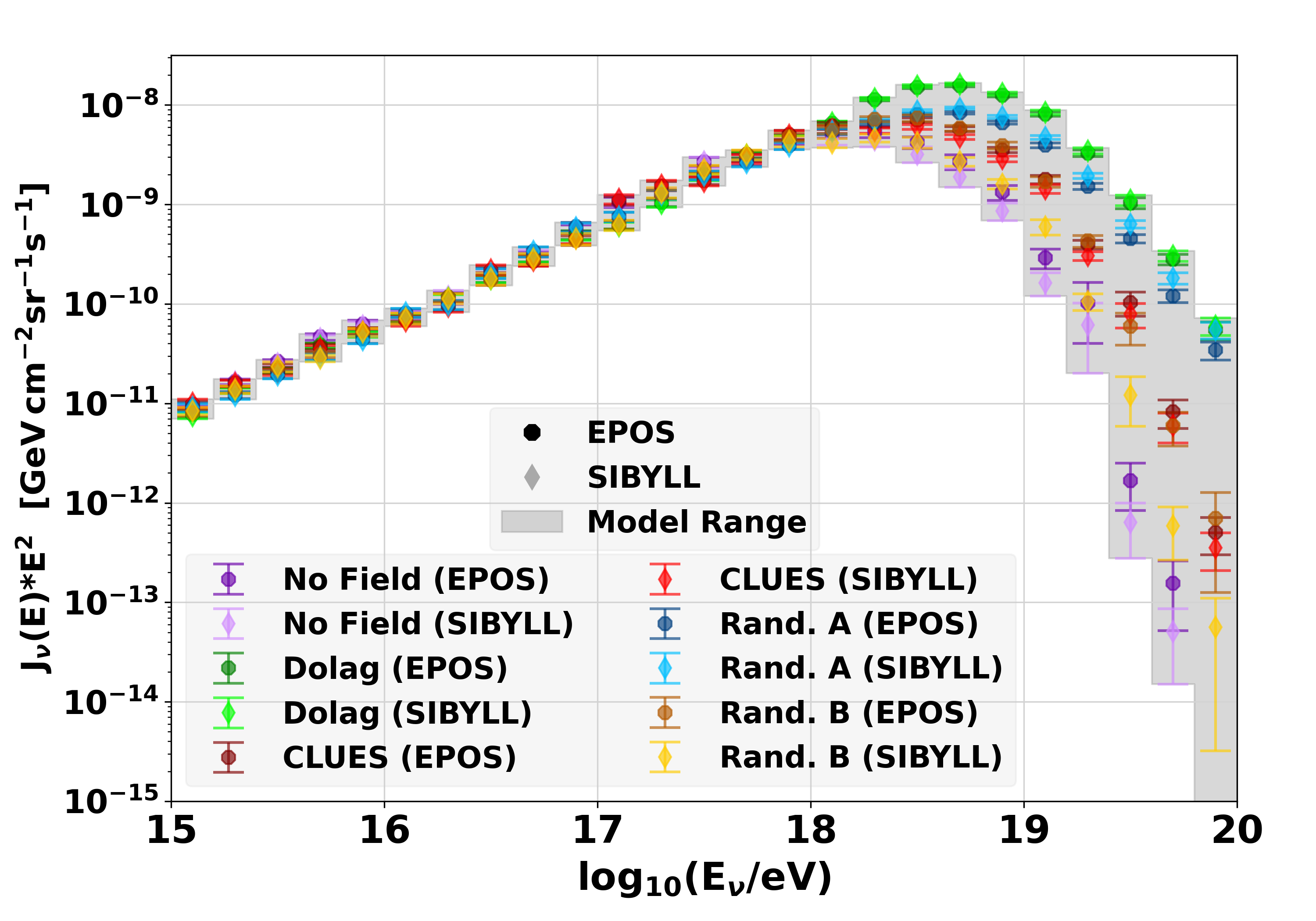}
    \label{fig:allneutrino_evolved}}
    \caption{Evolving-fraction fit cosmogenic photon and neutrino spectra for the ten configurations based on the EPOS-LHC (EPOS) and \Sibyll{} (SIBYLL) hadronic interaction models and all five tested magnetic fields. Grey areas display the $\pm1\sigma$~bounds of all the simulation configurations. EPOS/SIBYLL fits are represented by circles/diamonds and darker/lighter colors. Each base color represents a magnetic field model. (a) Integral cosmogenic photon spectra using the axes of Figure~\ref{fig:photon_evolved}. (b) All-flavor neutrino spectra multiplied by E$^2$ using the axes of Figure~\ref{fig:neutrino_evolved}.}
    \label{fig:multi_evolved}
\end{figure*}

At lower energies (up to $E_\gamma~\leq~10^{17}$~eV), the integral photon flux is compatible with pure proton cosmogenic predictions, shifting towards a theoretical mixed composition at higher energies. Interestingly, the neutrino flux resembles predictions for a pure iron prediction, which may be partially influenced by the simulation constraint of relatively close FR0 sources ($z$~$\leq$~0.2). Given the wide range of theoretical predictions and current experimental upper limits the FR0 simulated flux is reasonable. Overall, for both photon and neutrino spectra, it can be seen that a magnetic field results in a larger flux, particularly at higher energies, as shown in Figure~\ref{fig:multi_evolved}.

\section{Summary \& Conclusions}
\label{sec:summary}

This work presents an in-depth study of FR0 radio galaxies, a prevalent but less luminous class of jetted active galaxies, and their potential emission features as ultra-high-energy cosmic-ray (UHECR) sources. Although individually less luminous than more powerful radio galaxies, the abundance of FR0 galaxies may render them significant contributors to the UHECR energy density observed on Earth.

To investigate this hypothesis, extensive CRPropa3~\citep{AlvesBatista:2022vem} simulations were conducted, modeling isotropic cosmic-ray emissions extrapolated from measured FR0 properties (de\-nsity and luminosity distributions and luminosity-redshift source evolution). These simulations were then fit to the mass-composition (expressed in mean log mass number \lnA) and energy spectra of UHECR data from the Pierre Auger Observatory~\citep{Yushkov:2020nhr, Deligny:2020gzq}. 

The results from these combined fits, utilizing two fit models, three EAS models, and five intergalactic magnetic field configurations (both random and structured), indicate that emissions from FR0 galaxies can effectively account for the observed UHECR energy spectrum and composition. The overall compatibility, in addition to the results of our previous study of the capability of FR0 radio galaxies to accelerate cosmic rays up to ultra-high energies~\citep{Merten:2021brk}, supports the viability of FR0s as significant contributors to the observed UHECR flux. Furthermore, the secondary photon and neutrino fluxes predicted from UHECR interactions are well within current observational limits, emphasizing the potential of future multimessen\-ger observations to further refine our understanding of FR0 galaxies as UHECR sources.

For the most extreme energy bin, the configurations tested are generally below the expected flux (only Dolag and \textit{longer} correlation length 1~nG Rand.B magnetic field models have \textit{overlapping} error bars, as shown in Figure~\ref{fig:EspectFit}). Additionally, over all configurations, there is a small underfitting of the last energy bin of \lnA (only Dolag and the \textit{smaller} correlation length 1~nG Rand.A magnetic field models are \textit{outside} the error bars as shown in Figure~\ref{fig:AmeanFit}). These two results may allow the possibility of a different extreme-energy heavy nuclei source type to become the dominant UHECR flux contributor.

These findings underscore previous results that show the importance of considering the effects of intergalactic magnetic fields on UHECR propagation~(such as~\cite{PierreAuger:2024hlp}). Results across the different magnetic field models tested indicate their significant influence on the arrival energy spectrum and composition of cosmic rays from FR0 galaxies, as well as up to orders-of-magnitude variations on the secondary photon and neutrino spectra. Larger magnetic fields result in better fits and a need for a greater emission of nuclei heavier than helium - additionally, generally, slightly harder emission spectra ($\langle$$\gamma$$\rangle$ $\approx$ 2.4 compatible with Fermi shock acceleration), higher energy exponential cutoffs ($\langle$log$_{10}$(R$_{\text{cut}}$/V)$\rangle$ $\approx$ 19.6), and an increase in the highest energy secondary photons and neutrinos. The impact of magnetic field strength on light-nuclei emission is the greatest for the more modern \Sibyll{} and EPOS-LHC hadronic interaction models.

The Hackstein et al. (CLUES) 'astrophysical1R' mo\-del~\citep{Hackstein:2017pex}, characterized by $\langle$B$\rangle$ = 0.064~nG and B$_{\mathrm{RMS}}$ = 1.2~nG, emerged as the best-fit magnetic field model for the constant nuclei fraction fits. This is compatible with a previous estimate in \cite{Abbasi_2020} of extragalactic magnetic fields using detected cosmic-ray multiplets. Using the given $\langle$Z*B*D$\rangle$ = 30~nG*Mpc, the magnetic field of $\langle$B$\rangle$ = 0.064~nG, and the CLUES-SIBYLL best-fit composition ($\langle$Z$\rangle$ $\approx$ 2.7) results in an average particle trajectory length of about 1.7$\times$10$^{2}$~Mpc. Using the mean energy cut E~$\geq$ 27.7~EeV from \cite{Abbasi_2020} and the CLUES-SIBYLL best-fit composition, the mean trajectory length of simulated particles is \\$\sim$2.3$\times$10$^{2}$~Mpc.

In conclusion, our comprehensive study provides strong evidence supporting the significant contribution of FR0 radio galaxies to the UHECR energy density. This work not only enhances our understanding of potential UHECR sources but also opens new avenues for multimessenger astrophysics investigations.

\section*{Acknowledgments}
Financial support was received from the Austrian Science Fund (FWF) under grant agreement number I 4144-N27 and the Slovenian Research Agency---ARIS (project no. N1-0111). LM acknowledges support from the DFG within the Collaborative Research Center SFB1491 "Cosmic Interacting Matters---From Source to Signal". MB has for this project received funding from the European Union’s Horizon 2020 research and innovation program under the Marie Sklodowska-Curie grant agreement No 847476. GB acknowledges financial support from the State Agency for Research of the Spanish MCIU through the "Center of Excellence Severo Ochoa" award to the Instituto de Astrofísica de Andalucía (SEV-2017-0709) and from the Spanish "Ministerio de Ciencia e Innovaci\'on” (MICINN)
 through grant PID2019-107847RB-C44. The views and opinions expressed herein do not necessarily reflect those of the European Commission. The Slovenian National Supercomputing Network provided extensive computing support.
\bibliographystyle{elsarticle-num}
\bibliography{main}{}

\begin{thebibliography}{10}
\expandafter\ifx\csname url\endcsname\relax
  \def\url#1{\texttt{#1}}\fi
\expandafter\ifx\csname urlprefix\endcsname\relax\def\urlprefix{URL }\fi
\expandafter\ifx\csname href\endcsname\relax
  \def\href#1#2{#2} \def\path#1{#1}\fi

\bibitem{PierreAuger:2015eyc}
Aab, A., et~al., {The Pierre Auger Cosmic Ray Observatory}, Nucl. Instrum. Meth. A 798 (2015) 172--213.
\newblock \href {http://arxiv.org/abs/1502.01323} {\path{arXiv:1502.01323}}, \href {http://dx.doi.org/10.1016/j.nima.2015.06.058} {\path{doi:10.1016/j.nima.2015.06.058}}.

\bibitem{Baldi2009}
{Baldi}, R.D., {Capetti}, A., {Radio and spectroscopic properties of miniature radio galaxies: revealing the bulk of the radio-loud AGN population}, Astron. and Astrophys. 508~(2) (2009) 603--614.
\newblock \href {http://arxiv.org/abs/0910.4261} {\path{arXiv:0910.4261}}, \href {http://dx.doi.org/10.1051/0004-6361/200913021} {\path{doi:10.1051/0004-6361/200913021}}.

\bibitem{Baldi:2017gao}
Baldi, R.D., {Capetti, A.}, {Massaro, F.}, {FR0CAT: a FIRST catalog of FR 0 radio galaxies}, Astron. Astrophys. 609 (2018) A1.
\newblock \href {http://arxiv.org/abs/1709.00015} {\path{arXiv:1709.00015}}, \href {http://dx.doi.org/10.1051/0004-6361/201731333} {\path{doi:10.1051/0004-6361/201731333}}.

\bibitem{Croston2018}
{Croston}, J.H., {Ineson}, J., {Hardcastle}, M.J., {Particle content, radio-galaxy morphology, and jet power: all radio-loud AGN are not equal}, Mon. Not. R. Astron. Soc. 476~(2) (2018) 1614--1623.
\newblock \href {http://arxiv.org/abs/1801.10172} {\path{arXiv:1801.10172}}, \href {http://dx.doi.org/10.1093/mnras/sty274} {\path{doi:10.1093/mnras/sty274}}.

\bibitem{Heckman2014}
{Heckman}, T.M., {Best}, P.N., {The Coevolution of Galaxies and Supermassive Black Holes: Insights from Surveys of the Contemporary Universe}, Annu. Rev. Astron. Astrophys. 52 (2014) 589--660.
\newblock \href {http://arxiv.org/abs/1403.4620} {\path{arXiv:1403.4620}}, \href {http://dx.doi.org/10.1146/annurev-astro-081913-035722} {\path{doi:10.1146/annurev-astro-081913-035722}}.

\bibitem{Merten:2021brk}
Merten, L., et~al., {Scrutinizing FR0 radio galaxies as ultra-high-energy cosmic ray source candidates}, Astropart. Phys. 128 (2021) 102564.
\newblock \href {http://arxiv.org/abs/2102.01087} {\path{arXiv:2102.01087}}, \href {http://dx.doi.org/10.1016/j.astropartphys.2021.102564} {\path{doi:10.1016/j.astropartphys.2021.102564}}.

\bibitem{Hillas1984}
{Hillas}, A.M., {The Origin of Ultra-High-Energy Cosmic Rays}, Annu. Rev. Astron. Astrophys. 22 (1984) 425--444.
\newblock \href {http://dx.doi.org/10.1146/annurev.aa.22.090184.002233} {\path{doi:10.1146/annurev.aa.22.090184.002233}}.

\bibitem{NaganoWatson}
{Nagano}, M., {Watson}, A.A., {Observations and implications of the ultrahigh-energy cosmic rays}, Rev. Mod. Phys. 72~(3) (2000) 689--732.
\newblock \href {http://dx.doi.org/10.1103/RevModPhys.72.689} {\path{doi:10.1103/RevModPhys.72.689}}.

\bibitem{Bhattacharjee2000}
{Bhattacharjee}, P., {Sigl}, G., {Origin and propagation of extremely high energy cosmic rays}, \physrep{} 327 (2000) 109--247.
\newblock \href {http://arxiv.org/abs/astro-ph/9811011} {\path{arXiv:astro-ph/9811011}}, \href {http://dx.doi.org/10.1016/S0370-1573(99)00101-5} {\path{doi:10.1016/S0370-1573(99)00101-5}}.

\bibitem{2023ApJ...955L..41B}
{Boughelilba}, M., {Reimer}, A., {On the Subparsec-scale Core Composition of FR 0 Radio Galaxies}, Astrophys. J. Lett. 955~(2) (2023) L41.
\newblock \href {http://arxiv.org/abs/2310.06398} {\path{arXiv:2310.06398}}, \href {http://dx.doi.org/10.3847/2041-8213/acf83c} {\path{doi:10.3847/2041-8213/acf83c}}.

\bibitem{Reimer:2024Lw}
Reimer, A., Boughelilba, M., Merten, L., Da~Vela, P., {Low-luminosity jetted AGN as particle multi-messenger sources}, PoS TAUP2023 (2024) 119.
\newblock \href {http://dx.doi.org/10.22323/1.441.0119} {\path{doi:10.22323/1.441.0119}}.

\bibitem{2021ApJ...918L..39P}
{Paliya}, V.S., {A New Gamma-Ray-emitting Population of FR0 Radio Galaxies}, Astrophys. J. Lett. 918~(2) (2021) L39.
\newblock \href {http://arxiv.org/abs/2108.11701} {\path{arXiv:2108.11701}}, \href {http://dx.doi.org/10.3847/2041-8213/ac2143} {\path{doi:10.3847/2041-8213/ac2143}}.

\bibitem{Khatiya:2023lkg}
{Khatiya}, N.S., et~al., {Characterizing the $\gamma$-ray Emission from FR0 Radio Galaxies}, arXiv e-prints\href {http://arxiv.org/abs/2310.19888} {\path{arXiv:2310.19888}}, \href {http://dx.doi.org/10.48550/arXiv.2310.19888} {\path{doi:10.48550/arXiv.2310.19888}}.

\bibitem{Partenheimer_2024}
{Partenheimer}, A., {Fang}, K., {Alves Batista}, R., {de Almeida}, R.M., {Ultra-high-energy Cosmic-Ray Sources Can Be Gamma-Ray Dim}, Astrophys. J. Lett. 967~(1) (2024) L15.
\newblock \href {http://arxiv.org/abs/2404.17631} {\path{arXiv:2404.17631}}, \href {http://dx.doi.org/10.3847/2041-8213/ad4359} {\path{doi:10.3847/2041-8213/ad4359}}.

\bibitem{Golup_ICRC2023}
Abdul~Halim, A., et~al., {An update on the arrival direction studies made with data from the Pierre Auger Observatory}, PoS ICRC2023 (2023) 252.
\newblock \href {http://dx.doi.org/10.22323/1.444.0252} {\path{doi:10.22323/1.444.0252}}.

\bibitem{Eichmann2018}
{Eichmann}, B., et~al., {Ultra-high-energy cosmic rays from radio galaxies}, J. Cosmol. Astropart. Phys. 2018~(2) (2018) 036.
\newblock \href {http://arxiv.org/abs/1701.06792} {\path{arXiv:1701.06792}}, \href {http://dx.doi.org/10.1088/1475-7516/2018/02/036} {\path{doi:10.1088/1475-7516/2018/02/036}}.

\bibitem{Wittkowski2018}
{Wittkowski}, D., {Kampert}, K.H., {On the Anisotropy in the Arrival Directions of Ultra-high-energy Cosmic Rays}, Astrophys. J. Lett. 854~(1) (2018) L3.
\newblock \href {http://arxiv.org/abs/1710.05617} {\path{arXiv:1710.05617}}, \href {http://dx.doi.org/10.3847/2041-8213/aaa2f9} {\path{doi:10.3847/2041-8213/aaa2f9}}.

\bibitem{Dundovic2019}
{Dundovi{\'c}}, A., {Sigl}, G., {Anisotropies of ultra-high energy cosmic rays dominated by a single source in the presence of deflections}, J. Cosmol. Astropart. Phys. 2019~(1) (2019) 018.
\newblock \href {http://arxiv.org/abs/1710.05517} {\path{arXiv:1710.05517}}, \href {http://dx.doi.org/10.1088/1475-7516/2019/01/018} {\path{doi:10.1088/1475-7516/2019/01/018}}.

\bibitem{Taylor2015}
{Taylor}, A.M., {Ahlers}, M., {Hooper}, D., {Indications of negative evolution for the sources of the highest energy cosmic rays}, \prd{} 92~(6) (2015) 063011.
\newblock \href {http://arxiv.org/abs/1505.06090} {\path{arXiv:1505.06090}}, \href {http://dx.doi.org/10.1103/PhysRevD.92.063011} {\path{doi:10.1103/PhysRevD.92.063011}}.

\bibitem{Das2019}
{Das}, S., {Razzaque}, S., {Gupta}, N., {Ultrahigh energy cosmic rays and neutrinos from light nuclei composition}, \prd{} 99~(8) (2019) 083015.
\newblock \href {http://arxiv.org/abs/1809.05321} {\path{arXiv:1809.05321}}, \href {http://dx.doi.org/10.1103/PhysRevD.99.083015} {\path{doi:10.1103/PhysRevD.99.083015}}.

\bibitem{PierreAuger:2023htc}
Abdul~Halim, A., et~al., {Constraining models for the origin of ultra-high-energy cosmic rays with a novel combined analysis of arrival directions, spectrum, and composition data measured at the Pierre Auger Observatory}, J. Cosmol. Astropart. Phys. 01 (2024) 022.
\newblock \href {http://arxiv.org/abs/2305.16693} {\path{arXiv:2305.16693}}, \href {http://dx.doi.org/10.1088/1475-7516/2024/01/022} {\path{doi:10.1088/1475-7516/2024/01/022}}.

\bibitem{Dolag_2005}
Dolag, K., Grasso, D., Springel, V., Tkachev, I., {Constrained simulations of the magnetic field in the local Universe and the propagation of ultrahigh energy cosmic rays}, J. Cosmol. Astropart. Phys. 01 (2005) 009.
\newblock \href {http://arxiv.org/abs/astro-ph/0410419} {\path{arXiv:astro-ph/0410419}}, \href {http://dx.doi.org/10.1088/1475-7516/2005/01/009} {\path{doi:10.1088/1475-7516/2005/01/009}}.

\bibitem{Hackstein:2017pex}
Hackstein, S., et~al., {Simulations of ultra-high energy cosmic rays in the local Universe and the origin of cosmic magnetic fields}, Mon. Not. Roy. Astron. Soc. 475~(2) (2018) 2519--2529.
\newblock \href {http://arxiv.org/abs/1710.01353} {\path{arXiv:1710.01353}}, \href {http://dx.doi.org/10.1093/mnras/stx3354} {\path{doi:10.1093/mnras/stx3354}}.

\bibitem{Lundquist:20233x}
Lundquist, J.P., et~al., {The UHECR-FR0 Radio Galaxy Connection: A Multi-Messenger Study of Energy Spectra/Composition Emission and Extragalactic Magnetic Field Propagation}, PoS ICRC2023 (2023) 1512.
\newblock \href {http://arxiv.org/abs/2308.10803} {\path{arXiv:2308.10803}}, \href {http://dx.doi.org/10.22323/1.444.1512} {\path{doi:10.22323/1.444.1512}}.

\bibitem{Mollerach:2013dza}
Mollerach, S., Roulet, E., {Magnetic diffusion effects on the ultra-high energy cosmic ray spectrum and composition}, J. Cosmol. Astropart. Phys. 10 (2013) 013.
\newblock \href {http://arxiv.org/abs/1305.6519} {\path{arXiv:1305.6519}}, \href {http://dx.doi.org/10.1088/1475-7516/2013/10/013} {\path{doi:10.1088/1475-7516/2013/10/013}}.

\bibitem{AlvesBatista:2022vem}
Alves~Batista, R., et~al., {CRPropa 3.2 \textemdash{} an advanced framework for high-energy particle propagation in extragalactic and galactic spaces}, J. Cosmol. Astropart. Phys. 09 (2022) 035.
\newblock \href {http://arxiv.org/abs/2208.00107} {\path{arXiv:2208.00107}}, \href {http://dx.doi.org/10.1088/1475-7516/2022/09/035} {\path{doi:10.1088/1475-7516/2022/09/035}}.

\bibitem{Yushkov:2020nhr}
Yushkov, A., {Mass Composition of Cosmic Rays with Energies above 10$^{17.2}$ eV from the Hybrid Data of the Pierre Auger Observatory}, PoS ICRC2019 (2020) 482.
\newblock \href {http://dx.doi.org/10.22323/1.358.0482} {\path{doi:10.22323/1.358.0482}}.

\bibitem{Deligny:2020gzq}
Deligny, O., {The energy spectrum of ultra-high energy cosmic rays measured at the Pierre Auger Observatory and at the Telescope Array}, PoS ICRC2019 (2020) 234.
\newblock \href {http://arxiv.org/abs/2001.08811} {\path{arXiv:2001.08811}}, \href {http://dx.doi.org/10.22323/1.358.0234} {\path{doi:10.22323/1.358.0234}}.

\bibitem{2013JCAP...02..026P}
Abreu, P., et~al., {Interpretation of the depths of maximum of extensive air showers measured by the Pierre Auger Observatory}, J. Cosmol. Astropart. Phys. 02 (2013) 026.
\newblock \href {http://arxiv.org/abs/1301.6637} {\path{arXiv:1301.6637}}, \href {http://dx.doi.org/10.1088/1475-7516/2013/02/026} {\path{doi:10.1088/1475-7516/2013/02/026}}.

\bibitem{Riehn:2017mfm}
Riehn, F., et~al., {The hadronic interaction model SIBYLL 2.3c and Feynman scaling}, PoS ICRC2017 (2018) 301.
\newblock \href {http://arxiv.org/abs/1709.07227} {\path{arXiv:1709.07227}}, \href {http://dx.doi.org/10.22323/1.301.0301} {\path{doi:10.22323/1.301.0301}}.

\bibitem{Pierog:2013ria}
Pierog, T., et~al., {EPOS LHC: Test of collective hadronization with data measured at the CERN Large Hadron Collider}, Phys. Rev. C 92~(3) (2015) 034906.
\newblock \href {http://arxiv.org/abs/1306.0121} {\path{arXiv:1306.0121}}, \href {http://dx.doi.org/10.1103/PhysRevC.92.034906} {\path{doi:10.1103/PhysRevC.92.034906}}.

\bibitem{Ostapchenko:2010vb}
Ostapchenko, S., {Monte Carlo treatment of hadronic interactions in enhanced Pomeron scheme: QGSJET-II model}, Phys. Rev. D 83 (2011) 014018.
\newblock \href {http://arxiv.org/abs/1010.1869} {\path{arXiv:1010.1869}}, \href {http://dx.doi.org/10.1103/PhysRevD.83.014018} {\path{doi:10.1103/PhysRevD.83.014018}}.

\bibitem{10.1111/j.1365-2966.2012.20841.x}
{Gilmore, R.C.}, {Somerville, R.S.}, {Primack, J.R.}, {Domínguez, A.}, {Semi-analytic modelling of the extragalactic background light and consequences for extragalactic gamma-ray spectra}, Mon. Not. R. Astron. Soc. 422~(4) (2012) 3189--3207.
\newblock \href {http://arxiv.org/abs/1104.0671} {\path{arXiv:1104.0671}}, \href {http://dx.doi.org/10.1111/j.1365-2966.2012.20841.x} {\path{doi:10.1111/j.1365-2966.2012.20841.x}}.

\bibitem{Protheroe:1996si}
{Protheroe, R.J. and Biermann, P.L.}, {A new estimate of the extragalactic radio background and implications for ultra-high-energy $\gamma$-ray propagation}, Astropart. Phys. 6 (1996) 45--54, [Erratum: Astropart. Phys. 7, 181 (1997)].
\newblock \href {http://arxiv.org/abs/astro-ph/9605119} {\path{arXiv:astro-ph/9605119}}, \href {http://dx.doi.org/10.1016/S0927-6505(96)00041-2} {\path{doi:10.1016/S0927-6505(96)00041-2}}.

\bibitem{Settimo:2013tua}
{Settimo, M. and De Domenico, M.}, {Propagation of extragalactic photons at ultra-high energy with the $\mathit{EleCa}$ code}, Astropart. Phys. 62 (2015) 92--99.
\newblock \href {http://arxiv.org/abs/1311.6140} {\path{arXiv:1311.6140}}, \href {http://dx.doi.org/10.1016/j.astropartphys.2014.07.011} {\path{doi:10.1016/j.astropartphys.2014.07.011}}.

\bibitem{Lee:1996fp}
{Lee, S.}, {On the propagation of extragalactic high energy cosmic and $\gamma$ rays}, Phys. Rev. D 58 (1998) 043004.
\newblock \href {http://arxiv.org/abs/astro-ph/9604098} {\path{arXiv:astro-ph/9604098}}, \href {http://dx.doi.org/10.1103/PhysRevD.58.043004} {\path{doi:10.1103/PhysRevD.58.043004}}.

\bibitem{PierreAuger:2016use}
{Aab, A.}, et~al., {Combined fit of spectrum and composition data as measured by the Pierre Auger Observatory}, J. Cosmol. Astropart. Phys. 04 (2017) 038, [Erratum: JCAP 03, E02 (2018)].
\newblock \href {http://arxiv.org/abs/1612.07155} {\path{arXiv:1612.07155}}, \href {http://dx.doi.org/10.1088/1475-7516/2017/04/038} {\path{doi:10.1088/1475-7516/2017/04/038}}.

\bibitem{PierreAuger:2022atd}
{Abdul Halim}, A., et~al., {Constraining the sources of ultra-high-energy cosmic rays across and above the ankle with the spectrum and composition data measured at the Pierre Auger Observatory}, J. Cosmol. Astropart. Phys. 05 (2023) 024.
\newblock \href {http://arxiv.org/abs/2211.02857} {\path{arXiv:2211.02857}}, \href {http://dx.doi.org/10.1088/1475-7516/2023/05/024} {\path{doi:10.1088/1475-7516/2023/05/024}}.

\bibitem{PierreAuger:2024hlp}
Abdul~Halim, A., Abreu, P., Aglietta, M., et~al., {Impact of the Magnetic Horizon on the Interpretation of the {Pierre Auger Observatory} Spectrum and Composition Data}, JCAP 07 (2024) 094.
\newblock \href {http://arxiv.org/abs/2404.03533} {\path{arXiv:2404.03533}}, \href {http://dx.doi.org/10.1088/1475-7516/2024/07/094} {\path{doi:10.1088/1475-7516/2024/07/094}}.

\bibitem{refId0}
{Adam, R.}, et~al., \emph{Planck} 2015 results - {I. O}verview of products and scientific results, Astron. Astrophys. 594 (2016) A1.
\newblock \href {http://arxiv.org/abs/1502.01582} {\path{arXiv:1502.01582}}, \href {http://dx.doi.org/10.1051/0004-6361/201527101} {\path{doi:10.1051/0004-6361/201527101}}.

\bibitem{AbdulHalim_2023}
Halim, A.A., et~al., \href{https://dx.doi.org/10.3847/1538-4365/aca537}{A catalog of the highest-energy cosmic rays recorded during phase i of operation of the pierre auger observatory}, The Astrophysical Journal Supplement Series 264~(2) (2023) 50.
\newblock \href {http://dx.doi.org/10.3847/1538-4365/aca537} {\path{doi:10.3847/1538-4365/aca537}}.
\newline\urlprefix\url{https://dx.doi.org/10.3847/1538-4365/aca537}

\bibitem{2016288}
Aab, A., et~al., Evidence for a mixed mass composition at the ‘ankle’ in the cosmic-ray spectrum, Phys. Let. B 762 (2016) 288--295.
\newblock \href {http://arxiv.org/abs/1609.08567} {\path{arXiv:1609.08567}}, \href {http://dx.doi.org/10.1016/j.physletb.2016.09.039} {\path{doi:10.1016/j.physletb.2016.09.039}}.

\bibitem{TelescopeArray:2018eph}
Abbasi, R.U., et~al., {Study of muons from ultrahigh energy cosmic ray air showers measured with the Telescope Array experiment}, Phys. Rev. D 98~(2) (2018) 022002.
\newblock \href {http://arxiv.org/abs/1804.03877} {\path{arXiv:1804.03877}}, \href {http://dx.doi.org/10.1103/PhysRevD.98.022002} {\path{doi:10.1103/PhysRevD.98.022002}}.

\bibitem{PhysRevD.109.102001}
Abdul~Halim, A., et~al., Testing hadronic-model predictions of depth of maximum of air-shower profiles and ground-particle signals using hybrid data of the {Pierre Auger Observatory}, Phys. Rev. D 109 (2024) 102001.
\newblock \href {http://arxiv.org/abs/2401.10740} {\path{arXiv:2401.10740}}, \href {http://dx.doi.org/10.1103/PhysRevD.109.102001} {\path{doi:10.1103/PhysRevD.109.102001}}.

\bibitem{2020SciPy-NMeth}
Virtanen, P., et~al., {{SciPy} 1.0: fundamental algorithms for scientific computing in Python}, Nat. Methods 17 (2020) 261--272.
\newblock \href {http://arxiv.org/abs/1907.10121} {\path{arXiv:1907.10121}}, \href {http://dx.doi.org/10.1038/s41592-019-0686-2} {\path{doi:10.1038/s41592-019-0686-2}}.

\bibitem{Storn1997}
Storn, R., Price, K., {Differential Evolution -- A Simple and Efficient Heuristic for Global Optimization over Continuous Spaces}, J. Glob. Optim. 11~(4) (1997) 341--359.
\newblock \href {http://dx.doi.org/10.1023/A:1008202821328} {\path{doi:10.1023/A:1008202821328}}.

\bibitem{AlvesBatista:2015jem}
Alves~Batista, R., et~al., {Effects of uncertainties in simulations of extragalactic UHECR propagation, using CRPropa and SimProp}, J. Cosmol. Astropart. Phys. 10 (2015) 063.
\newblock \href {http://arxiv.org/abs/1508.01824} {\path{arXiv:1508.01824}}, \href {http://dx.doi.org/10.1088/1475-7516/2015/10/063} {\path{doi:10.1088/1475-7516/2015/10/063}}.

\bibitem{PierreAuger:2022uwd}
Abreu, P., et~al., {A Search for Photons with Energies Above 2x10$^{17}$ eV Using Hybrid Data from the Low-Energy Extensions of the Pierre Auger Observatory}, Astrophys. J. 933~(2) (2022) 125.
\newblock \href {http://arxiv.org/abs/2205.14864} {\path{arXiv:2205.14864}}, \href {http://dx.doi.org/10.3847/1538-4357/ac7393} {\path{doi:10.3847/1538-4357/ac7393}}.

\bibitem{IceCube:2018fhm}
Aartsen, M.G., et~al., {Differential limit on the extremely-high-energy cosmic neutrino flux in the presence of astrophysical background from nine years of IceCube data}, Phys. Rev. D 98~(6) (2018) 062003.
\newblock \href {http://arxiv.org/abs/1807.01820} {\path{arXiv:1807.01820}}, \href {http://dx.doi.org/10.1103/PhysRevD.98.062003} {\path{doi:10.1103/PhysRevD.98.062003}}.

\bibitem{PierreAuger:2019ens}
Aab, A., et~al., {Probing the origin of ultra-high-energy cosmic rays with neutrinos in the EeV energy range using the Pierre Auger Observatory}, J. Cosmol. Astropart. Phys. 10 (2019) 022.
\newblock \href {http://arxiv.org/abs/1906.07422} {\path{arXiv:1906.07422}}, \href {http://dx.doi.org/10.1088/1475-7516/2019/10/022} {\path{doi:10.1088/1475-7516/2019/10/022}}.

\bibitem{2010JCAP...10..013K}
{Kotera}, K., {Allard}, D., {Olinto}, A.V., {Cosmogenic neutrinos: parameter space and detectabilty from PeV to ZeV}, J. Cosmol. Astropart. Phys. 2010~(10) (2010) 013.
\newblock \href {http://arxiv.org/abs/1009.1382} {\path{arXiv:1009.1382}}, \href {http://dx.doi.org/10.1088/1475-7516/2010/10/013} {\path{doi:10.1088/1475-7516/2010/10/013}}.

\bibitem{Abbasi_2020}
Abbasi, R.U., et~al., {Evidence for a Supergalactic Structure of Magnetic Deflection Multiplets of Ultra-high-energy Cosmic Rays}, Astrophys. J. 899~(1) (2020) 86.
\newblock \href {http://arxiv.org/abs/2005.07312} {\path{arXiv:2005.07312}}, \href {http://dx.doi.org/10.3847/1538-4357/aba26c} {\path{doi:10.3847/1538-4357/aba26c}}.

\end{thebibliography}

\newpage
\clearpage
\appendix
\onecolumn
\section{Evolving Fraction Fit} \label{appendix}

\begin{figure*}[!htbp]
    \centering
    \subfloat[Subfigure 1][]{
    \includegraphics[width=0.45\textwidth]{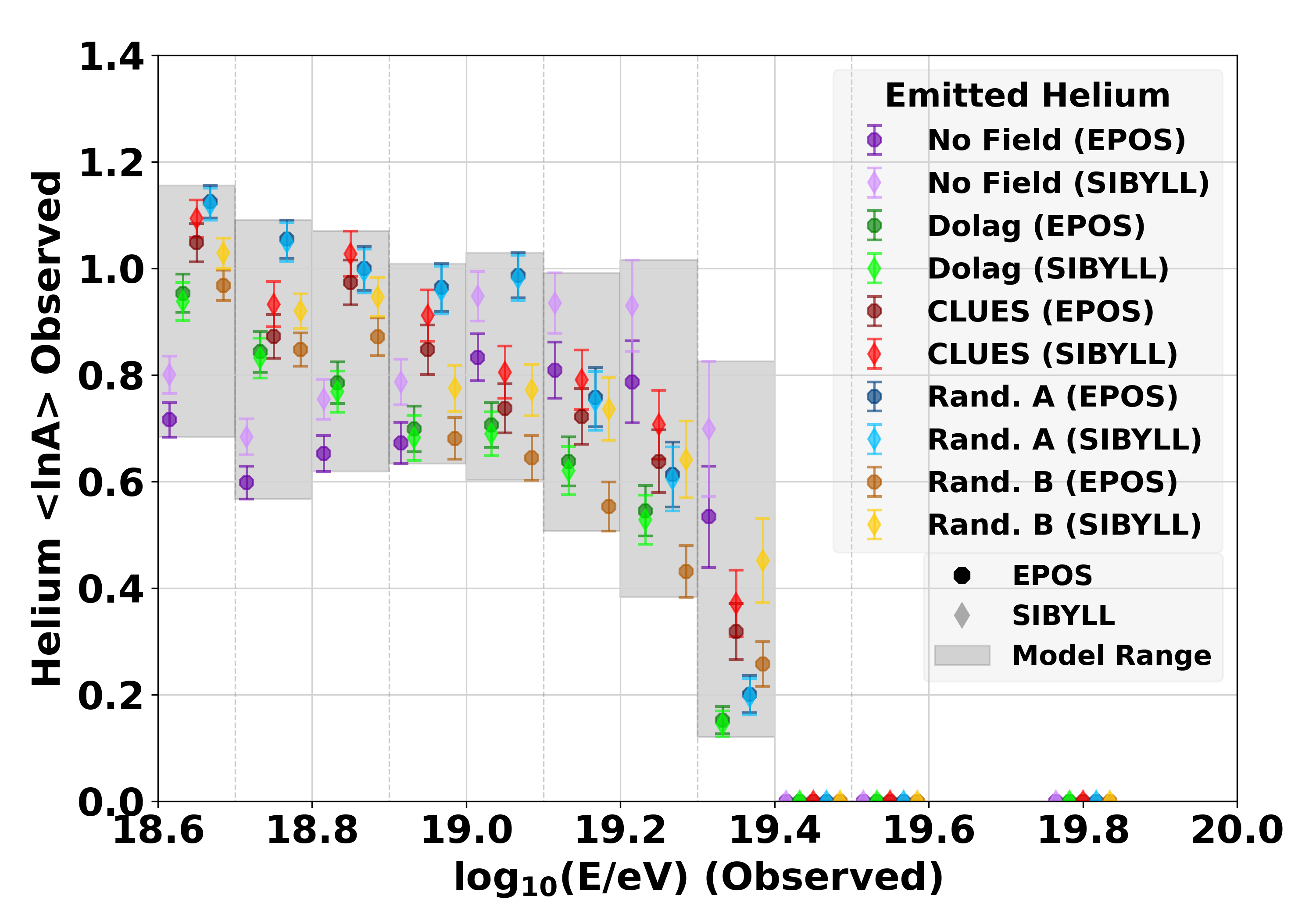}
    \label{fig:lnAHe_evolved}}
    \subfloat[Subfigure 2][]{
    \includegraphics[width=0.45\textwidth]{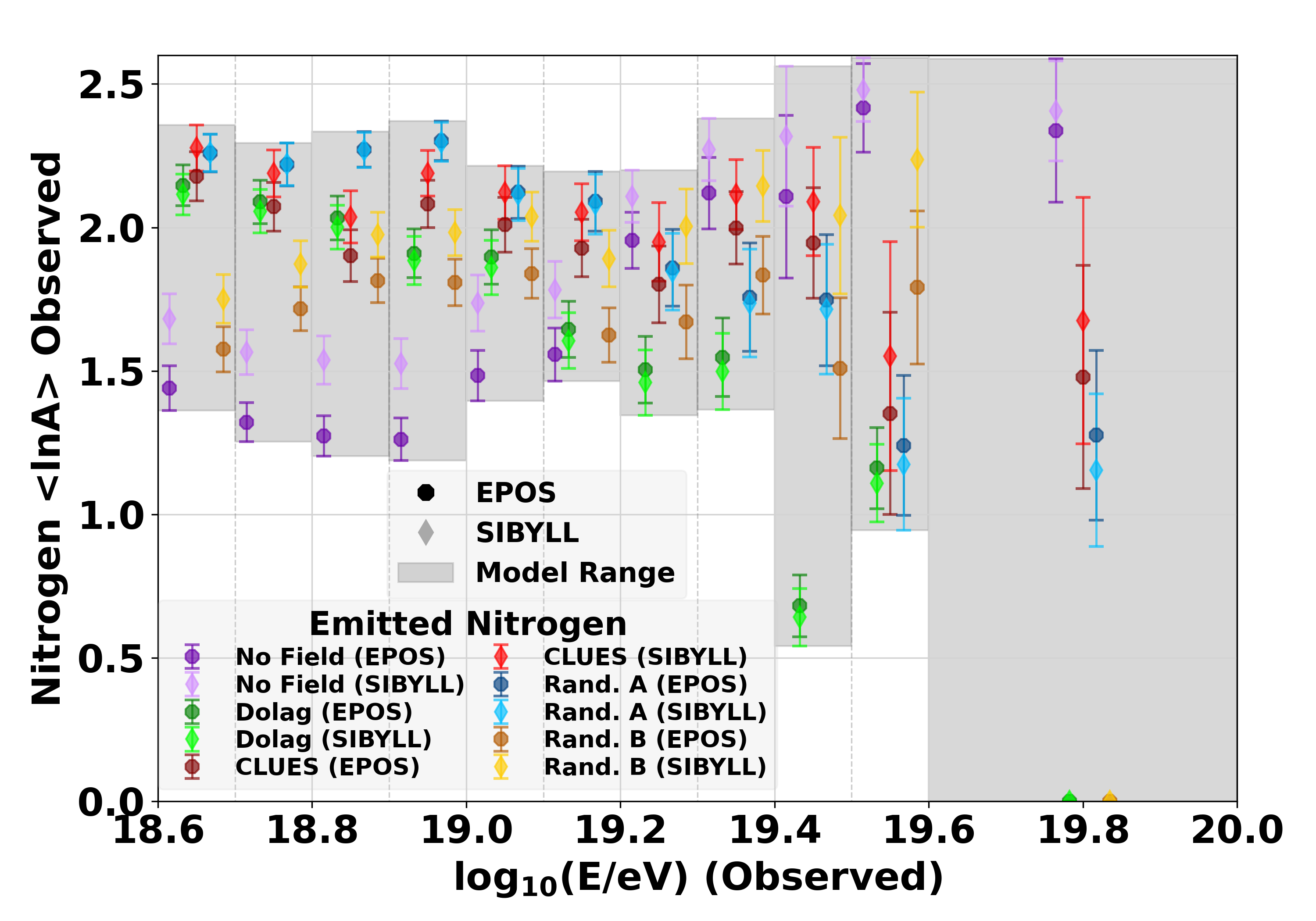}
    \label{fig:lnAN_evolved}}\\
    \subfloat[Subfigure 3][]{
    \includegraphics[width=0.45\textwidth]{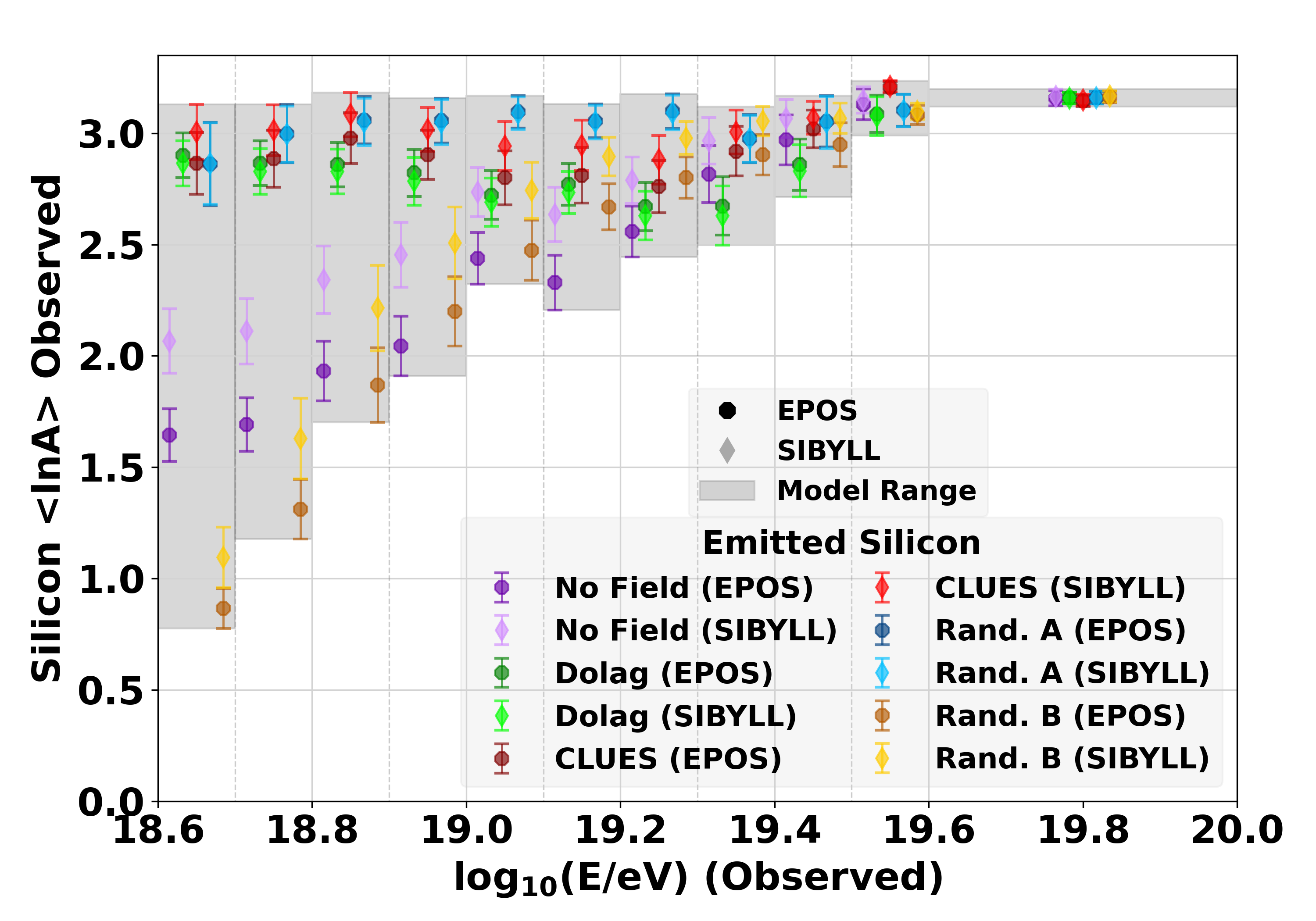}
    \label{fig:lnASi_evolved}}
    \subfloat[Subfigure 4][]{
    \includegraphics[width=0.45\textwidth]{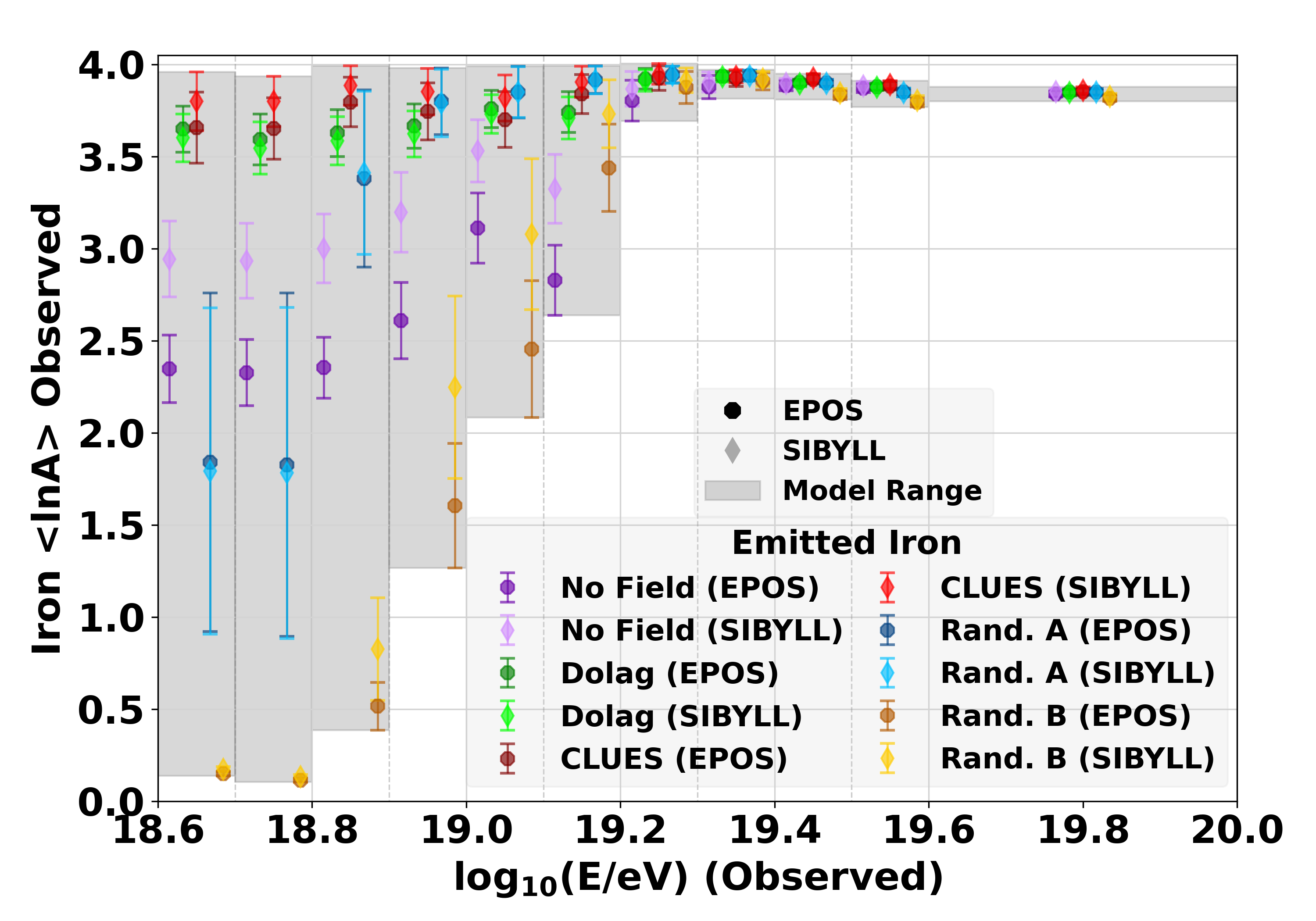}
    \label{fig:lnAFe_evolved}}\\
    \caption{The evolving-fraction fit results observed \lnA versus emitted energy for the ten configurations based on the EPOS-LHC (EPOS) and \Sibyll{} (SIBYLL) hadronic interaction models and all five magnetic fields. Proton has a constant \lnA$=0$ and is not shown. Offsets are applied to simulations within each bin on the x-axis for improved visibility. Grey areas display the $\pm1\sigma$~bounds of all the simulation configurations. Each figure's y-axis upper bounds are approximately the \lnA of the emitted nuclei.}
    \label{fig:lnA_evolved}
\end{figure*}

\begin{figure*}[htb]
    \centering
    \subfloat[Subfigure 1][]{
    \includegraphics[width=0.45\textwidth]{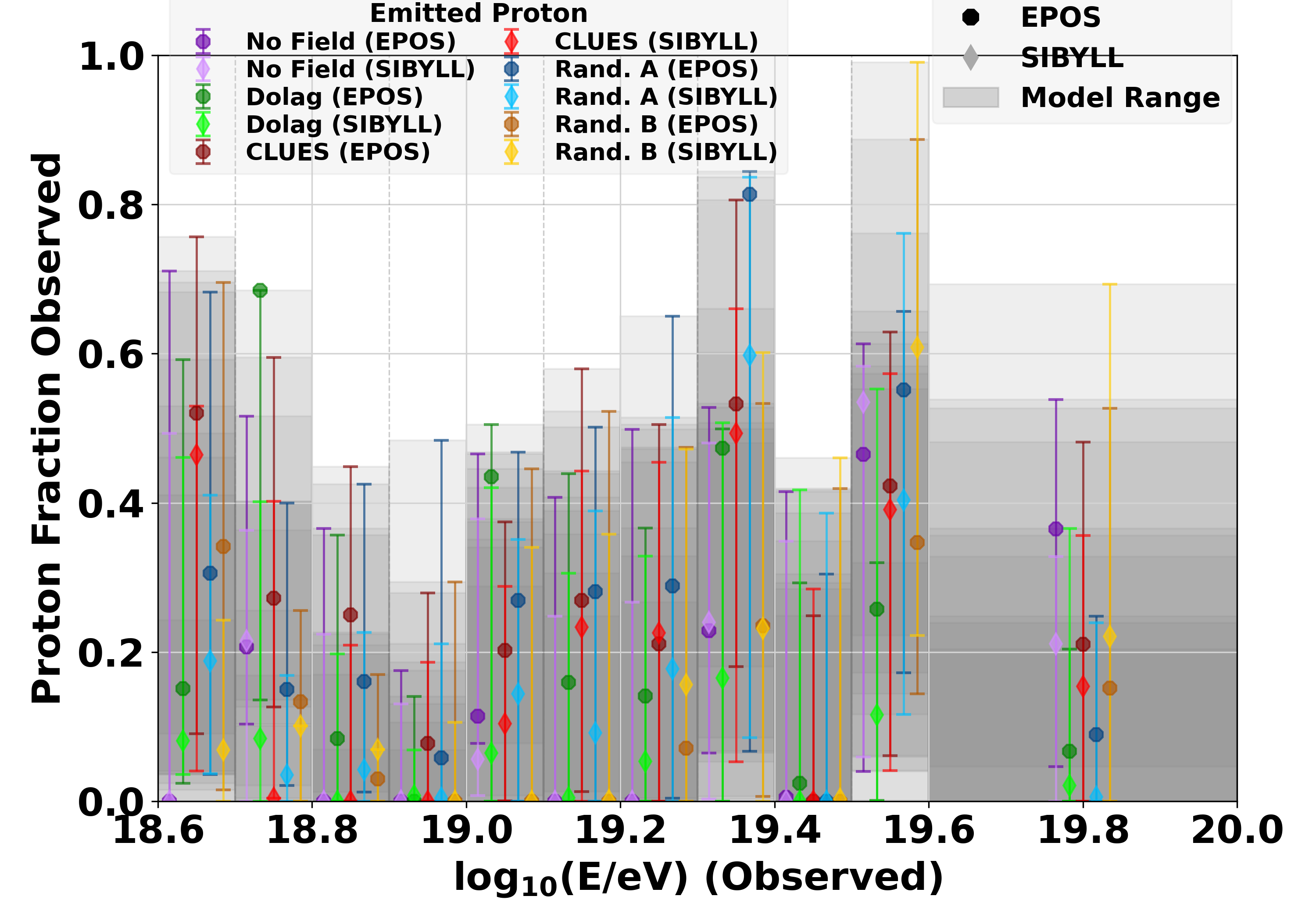}
    \label{fig:ObsH_evolved}}
    \subfloat[Subfigure 2][]{
    \includegraphics[width=0.45\textwidth]{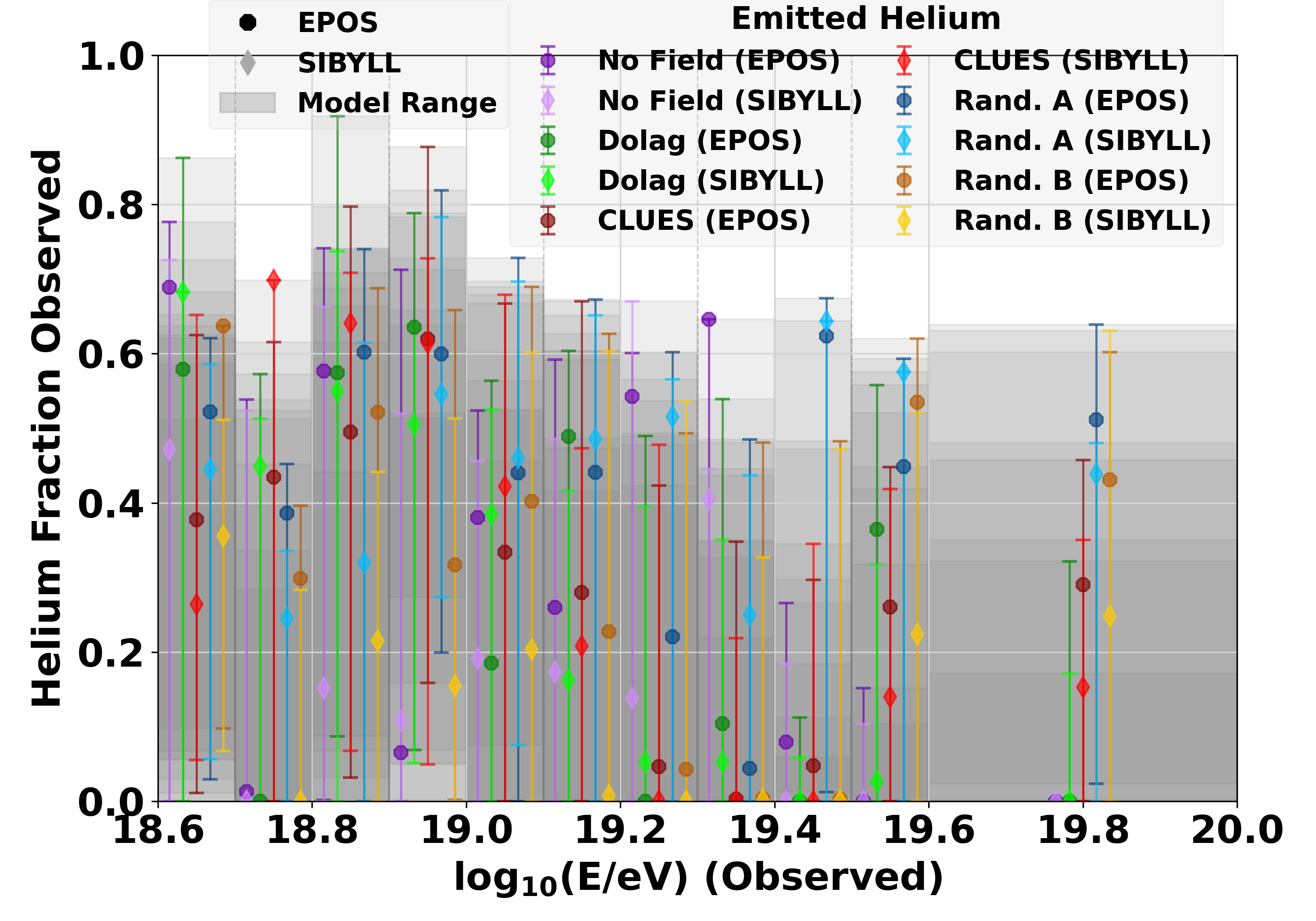}
    \label{fig:ObsHe_evolved}}\\
    \subfloat[Subfigure 3][]{
    \includegraphics[width=0.45\textwidth]{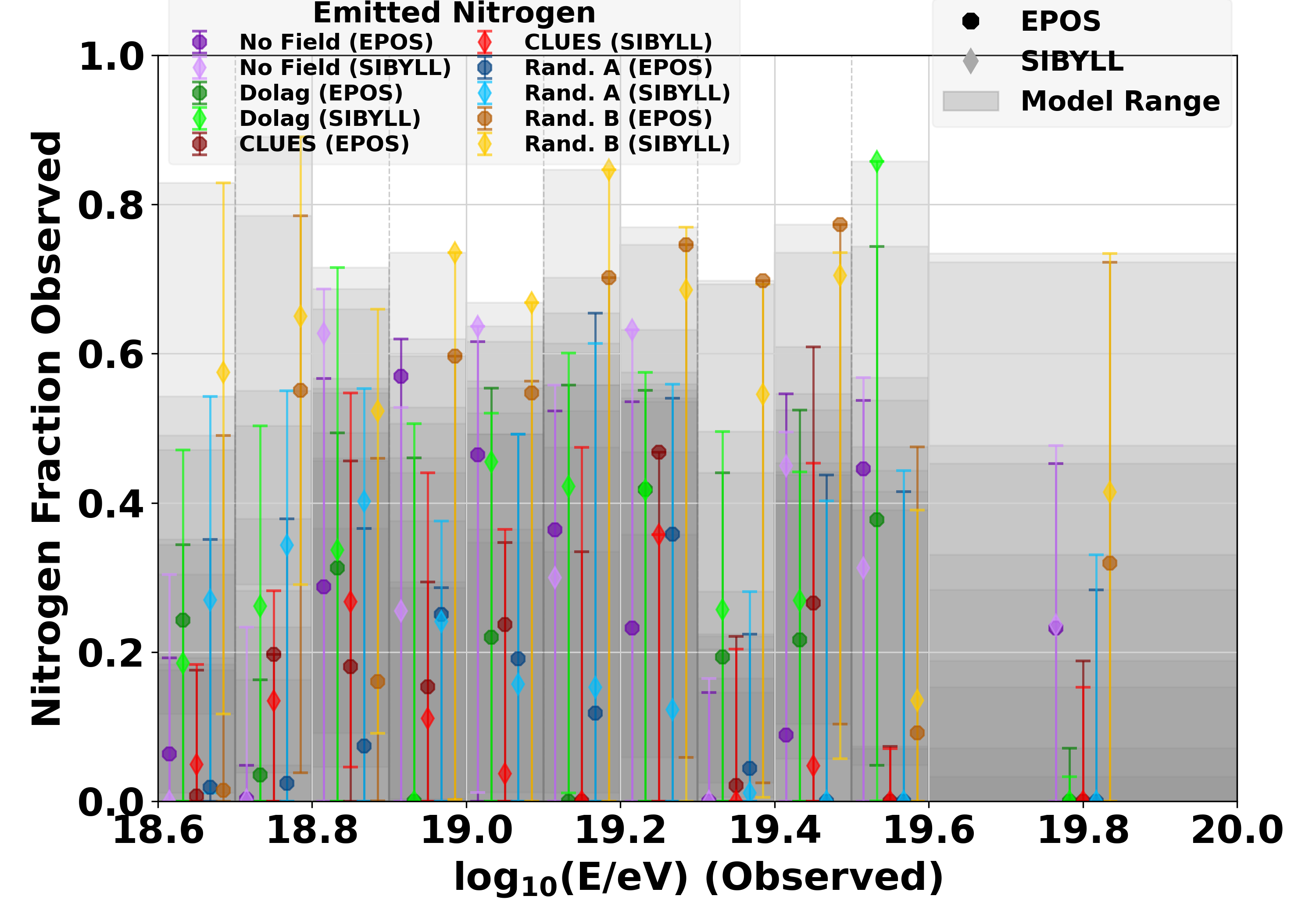}
    \label{fig:ObsN_evolved}}
    \subfloat[Subfigure 4][]{
    \includegraphics[width=0.45\textwidth]{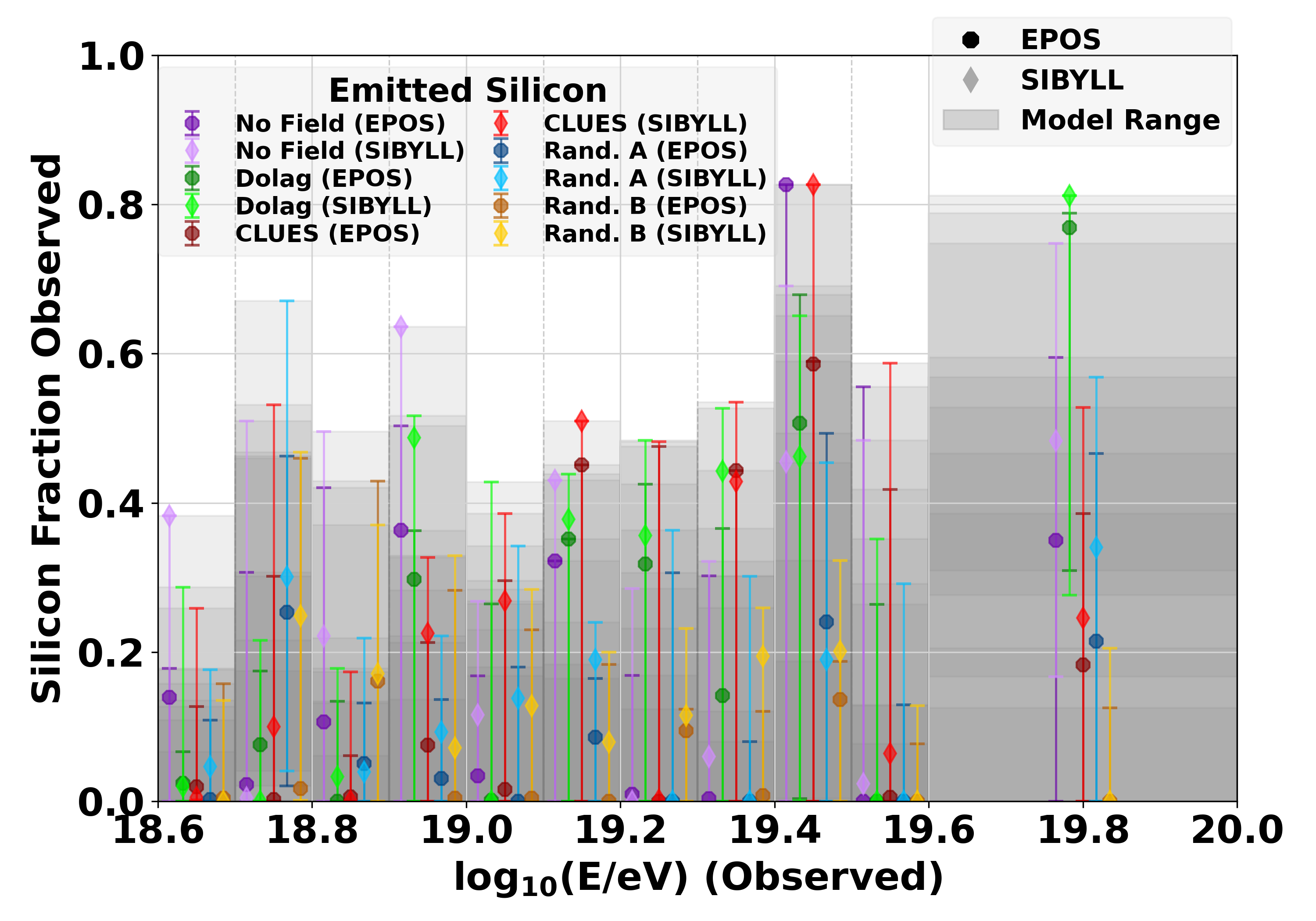}
    \label{fig:ObsSi_evolved}}\\
        \subfloat[Subfigure 5][]{
    \includegraphics[width=0.45\textwidth]{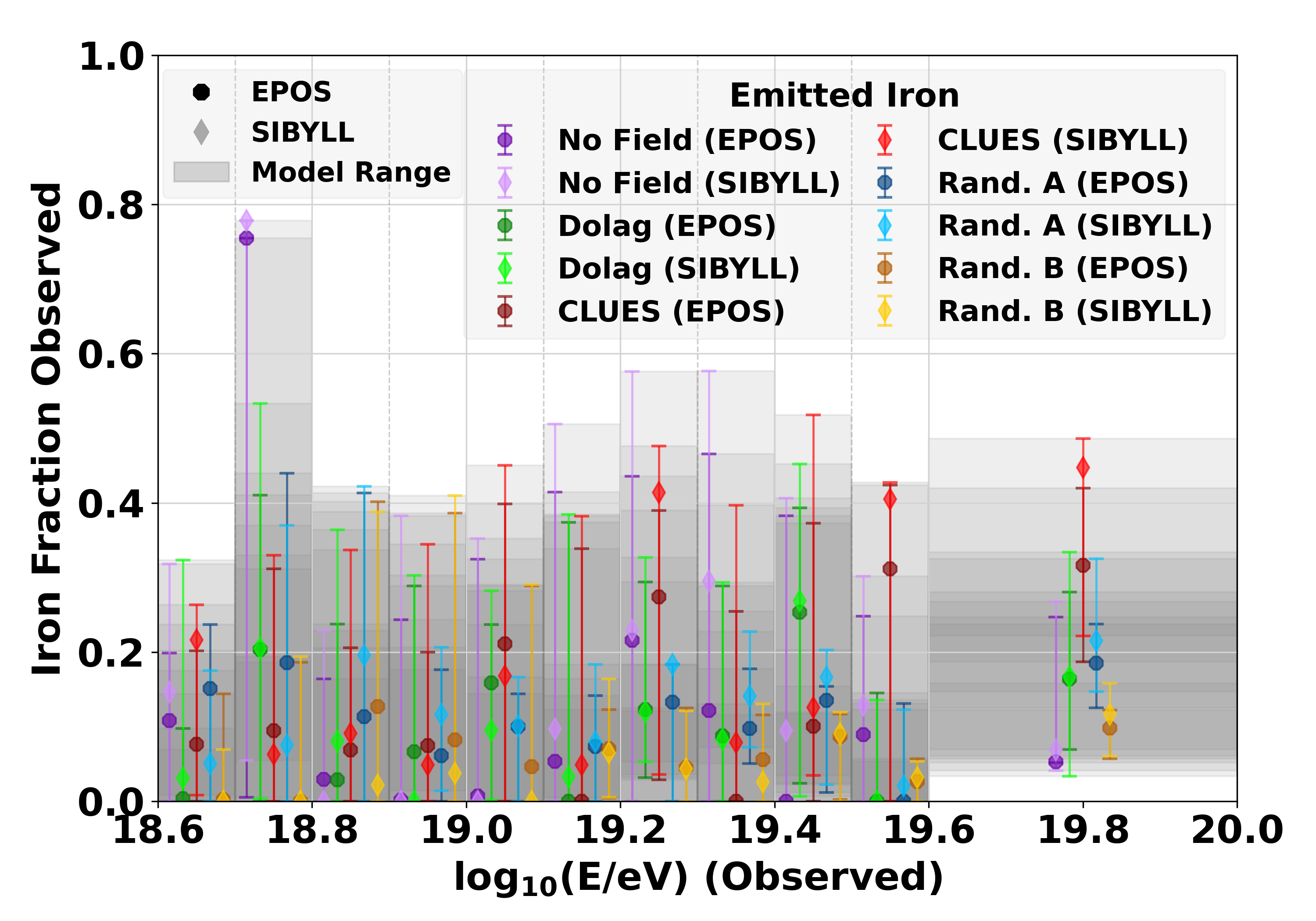}
    \label{fig:ObsFe_evolved}}\\
    \caption{The evolving-fraction fit FR0 observed nuclei fractions versus emitted energy necessary to best fit the data for the ten configurations based on the EPOS-LHC (EPOS) and \Sibyll{} (SIBYLL) hadronic interaction models and all five magnetic fields. These observed fractions are the free composition parameters in the fit. Offsets are applied to simulations within each bin on the x-axis for improved visibility. Grey bands extend from $\pm1\sigma$ for each configuration and is darkest where the ranges overlap.}
    \label{fig:Obs_evolved}
\end{figure*}

\begin{figure*}[htb]
    \centering
    \subfloat[Subfigure 1][]{
    \includegraphics[width=0.45\textwidth]{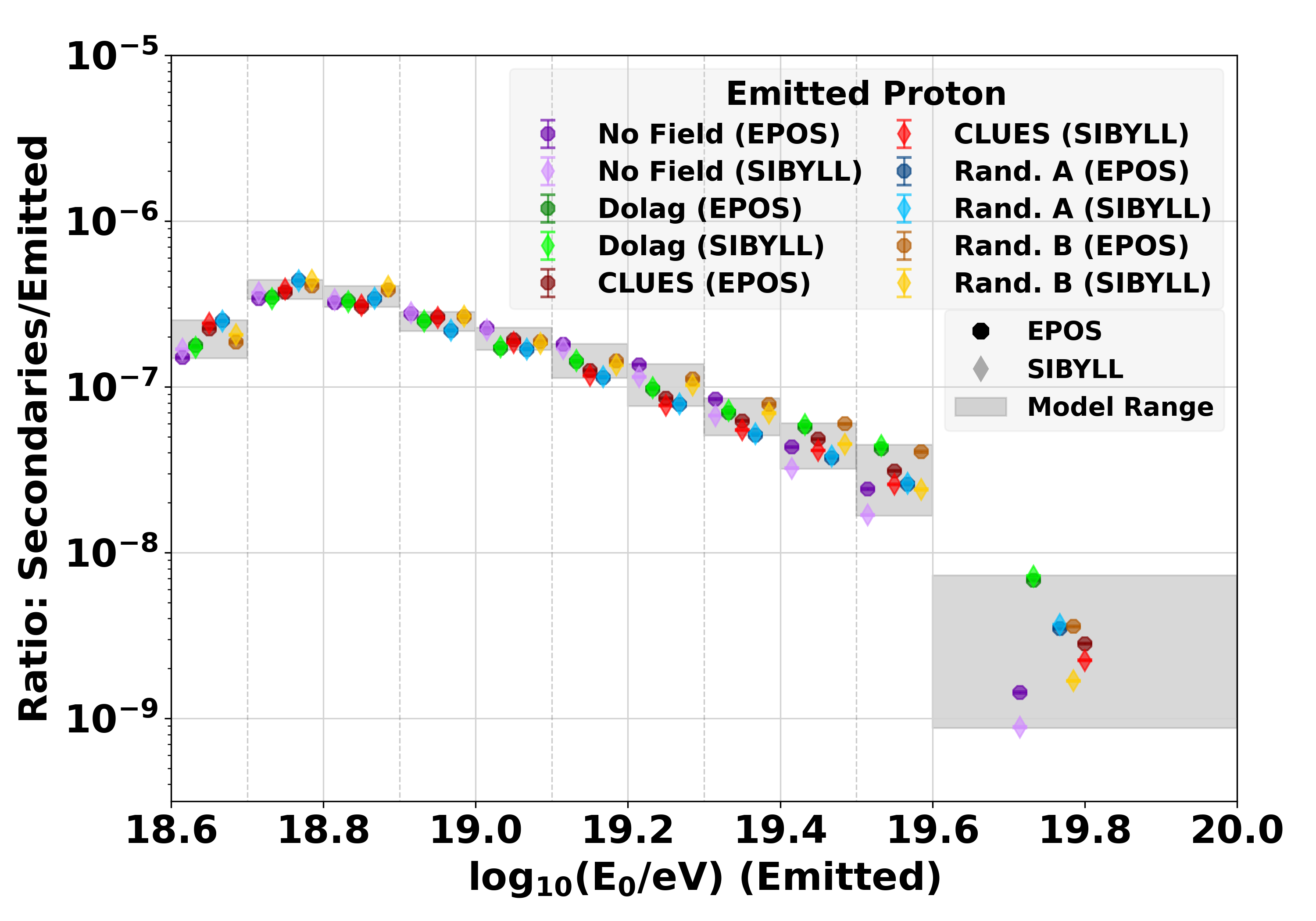}
    \label{fig:RatioH_evolved}}
    \subfloat[Subfigure 2][]{
    \includegraphics[width=0.45\textwidth]{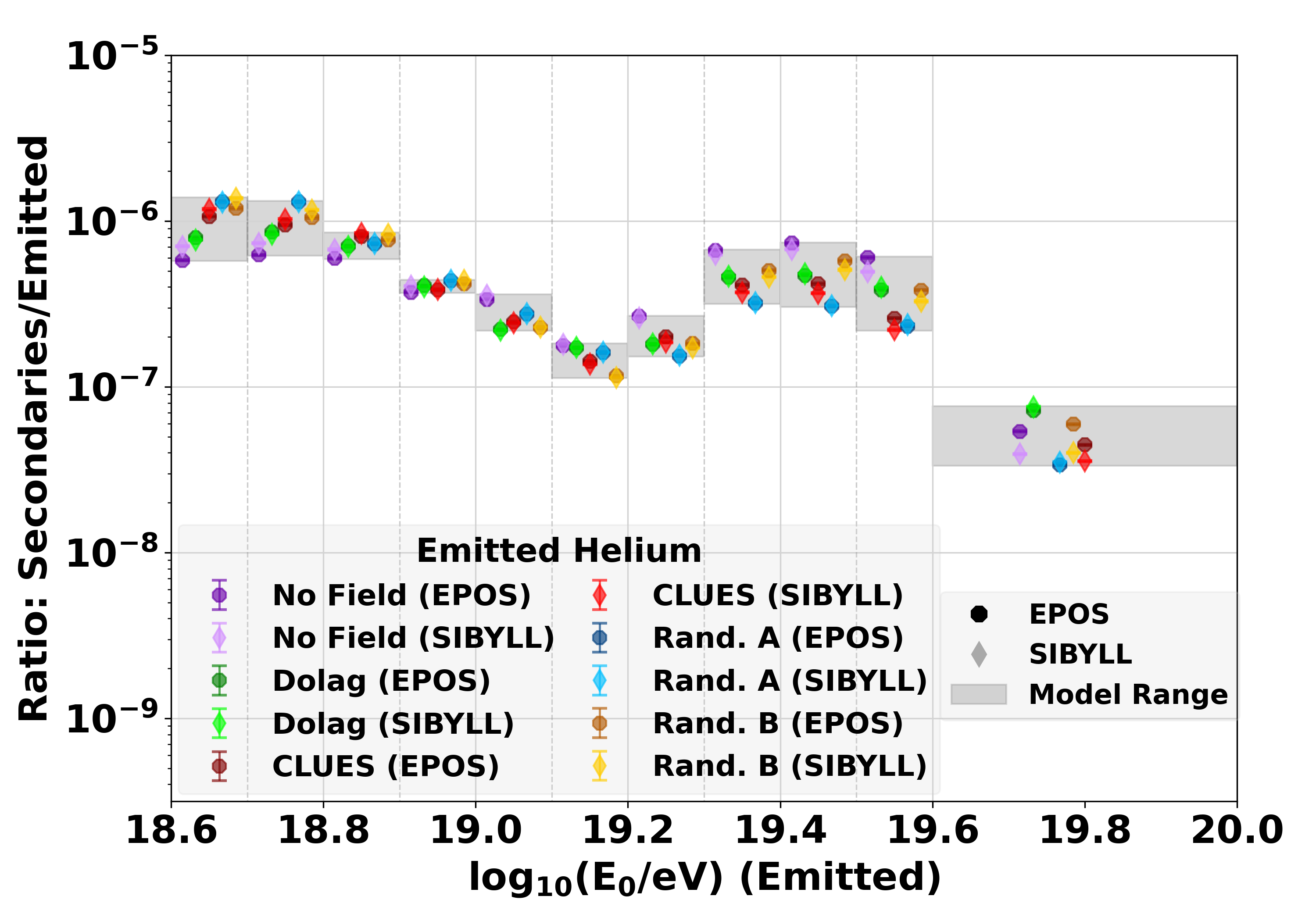}
    \label{fig:RatioHe_evolved}}\\
    \subfloat[Subfigure 3][]{
    \includegraphics[width=0.45\textwidth]{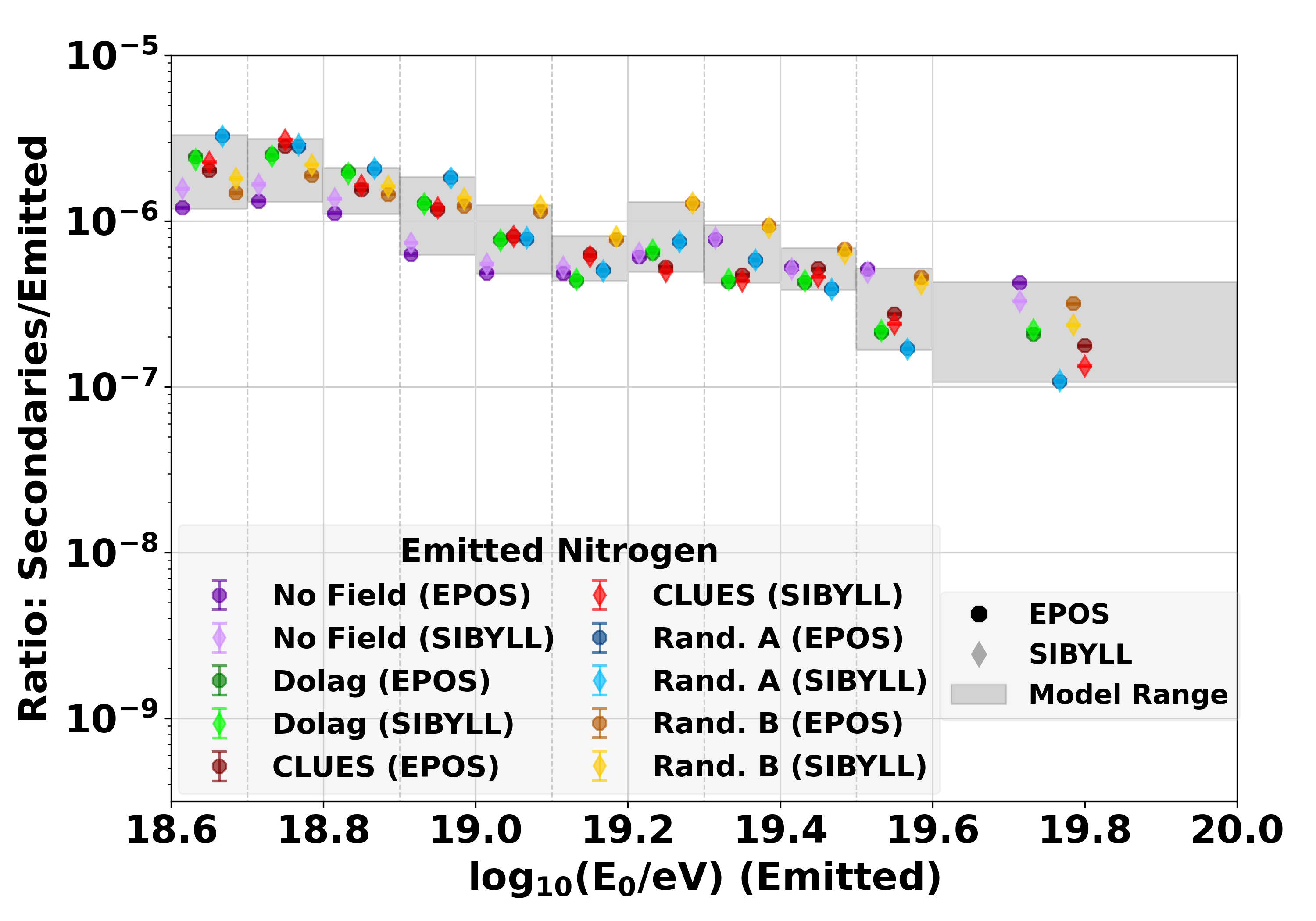}
    \label{fig:RatioN_evolved}}
    \subfloat[Subfigure 4][]{
    \includegraphics[width=0.45\textwidth]{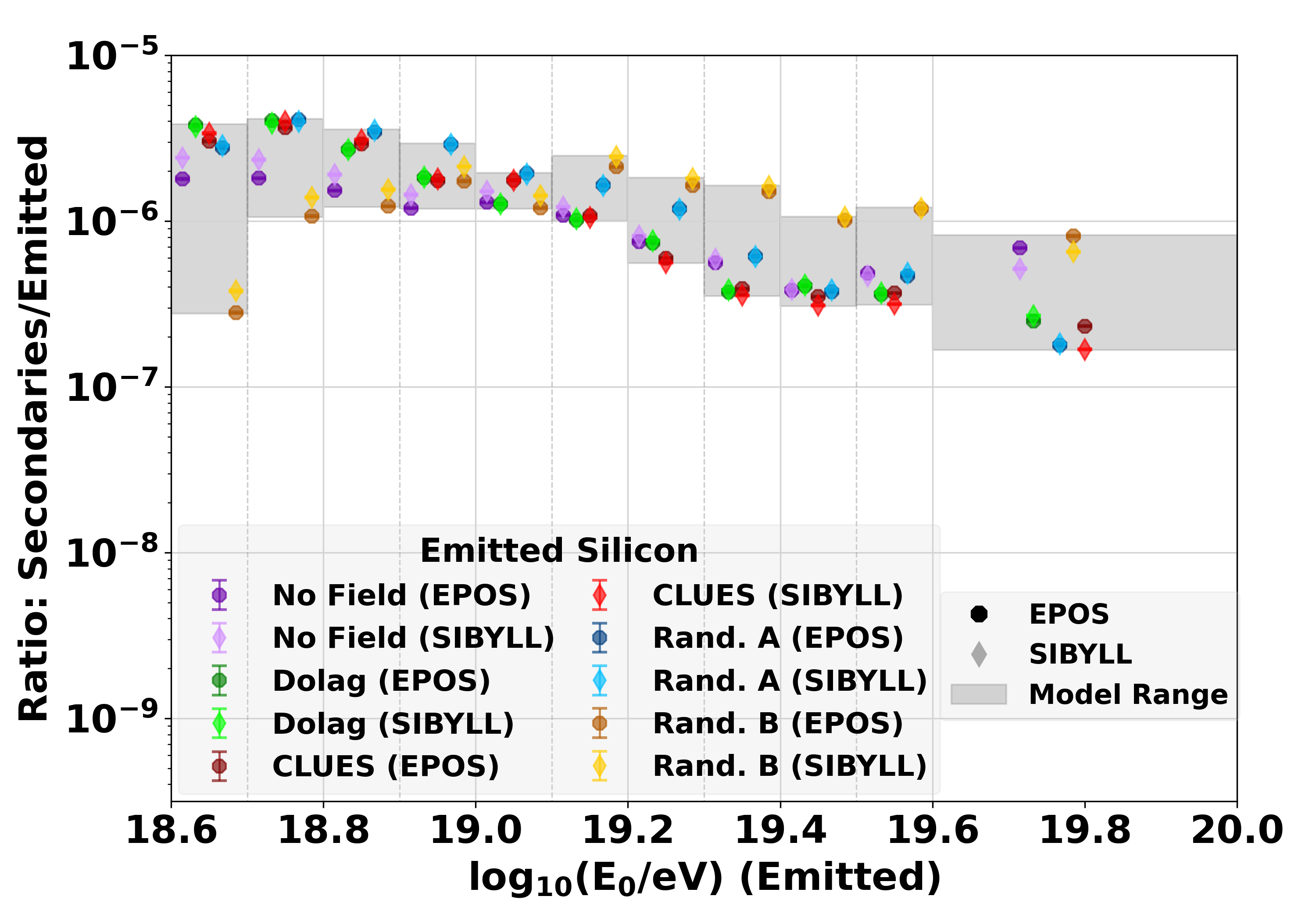}
    \label{fig:RatioSi_evolved}}\\
        \subfloat[Subfigure 5][]{
    \includegraphics[width=0.45\textwidth]{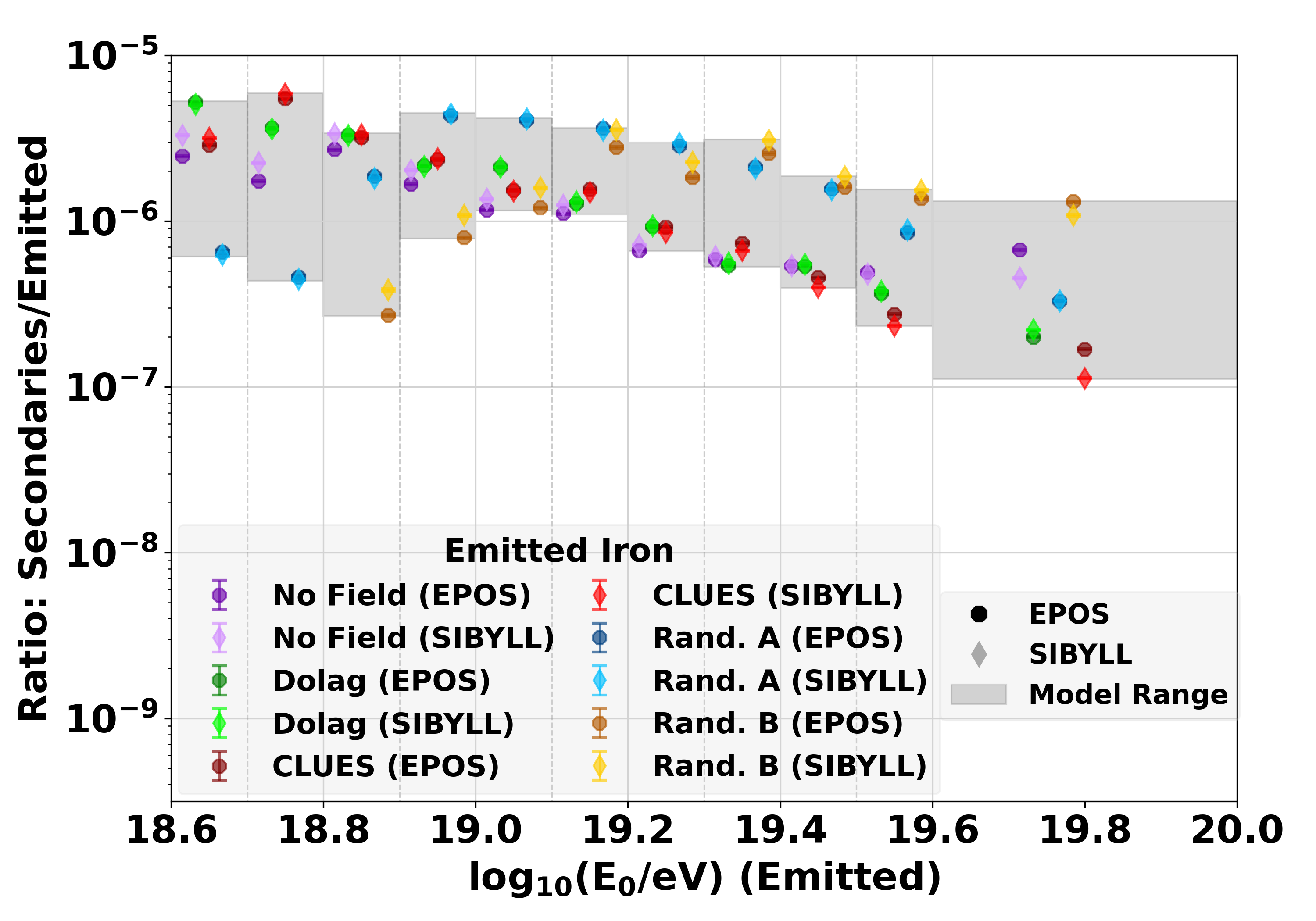}
    \label{fig:RatioFe_evolved}}\\
    \caption{The evolving-fraction fit ratio of secondaries to emitted nuclei for all ten configurations versus emitted energy. The average observed nuclei fractions (Figure~\ref{fig:Obs_evolved}) in emitted-energy bins are converted to the emission fractions using these ratios (shown in Figure~\ref{fig:nuclei_evolved}). Offsets are applied to simulations within each bin on the x-axis for improved visibility. Grey areas display the $\pm1\sigma$~bounds of all the simulation configurations.}
    \label{fig:ratios_evolved}
\end{figure*}

In this appendix are figures for intermediate results of the evolving-fraction fits. Figure~\ref{fig:lnA_evolved} shows the resulting \lnA after propagation for each emitted nuclei species for all ten configurations of magnetic field and EAS model. These are the values used to fit the composition data and have some dependence on the energy-spectrum parameters. Figure~\ref{fig:Obs_evolved} shows the 44 free composition observed fraction parameters that were fit to the data. The final emission fractions (Figure~\ref{fig:nuclei_evolved}) are the average of the observed fractions in emitted-energy bins with the ratio of observed particles to those emitted taken into account (Shown in Figure~\ref{fig:ratios_evolved}). Each figure in this appendix has two legends---the first emphasizes figure element shapes, while the second includes marker colors for all fits, where each base color represents a magnetic field model.

\end{document}